\PassOptionsToPackage{unicode}{hyperref}
\PassOptionsToPackage{hyphens}{url}
\documentclass[
]{article}
\usepackage{xcolor}
\usepackage[margin=1in]{geometry}
\usepackage{amsmath,amssymb}
\usepackage{iftex}
\ifPDFTeX
  \usepackage[T1]{fontenc}
  \usepackage[utf8]{inputenc}
  \usepackage{textcomp} 
\else 
  \usepackage{unicode-math} 
  \defaultfontfeatures{Scale=MatchLowercase}
  \defaultfontfeatures[\rmfamily]{Ligatures=TeX,Scale=1}
\fi
\usepackage{lmodern}
\ifPDFTeX\else
\fi
\IfFileExists{upquote.sty}{\usepackage{upquote}}{}
\IfFileExists{microtype.sty}{
  \usepackage[]{microtype}
  \UseMicrotypeSet[protrusion]{basicmath} 
}{}
\usepackage{setspace}
\makeatletter
\@ifundefined{KOMAClassName}{
  \IfFileExists{parskip.sty}{%
    \usepackage{parskip}
  }{
    \setlength{\parindent}{0pt}
    \setlength{\parskip}{6pt plus 2pt minus 1pt}}
}{
  \KOMAoptions{parskip=half}}
\makeatother
\usepackage{longtable,booktabs,array}
\usepackage{caption}
\usepackage{calc} 
\usepackage{etoolbox}
\makeatletter
\patchcmd\longtable{\par}{\if@noskipsec\mbox{}\fi\par}{}{}
\makeatother
\IfFileExists{footnotehyper.sty}{\usepackage{footnotehyper}}{\usepackage{footnote}}
\makesavenoteenv{longtable}
\providecommand{\tightlist}{%
  \setlength{\itemsep}{0pt}\setlength{\parskip}{0pt}}

\usepackage{amsmath}
\usepackage{amssymb}
\usepackage{microtype}
\usepackage{ragged2e}
\usepackage{tabularx}
\usepackage{enumitem}

\setlist{itemsep=0.2em,topsep=0.35em,parsep=0em}

\usepackage{needspace}
\usepackage{xurl}

\usepackage{setspace}
\usepackage{caption}
\DeclareCaptionFont{submissiondouble}{\setstretch{2}}
\usepackage{bookmark}
\IfFileExists{xurl.sty}{\usepackage{xurl}}{} 
\makeatletter
\@ifundefined{xmpquote}{}{}
\makeatother
\hypersetup{
  hidelinks,
  pdfcreator={LaTeX via pandoc}}

\author{}
\date{}

\usepackage{iftex}
\ifPDFTeX
  \usepackage{mathrsfs}
\fi
\begin{document}

\setstretch{1.05}
\begin{center}
{\Large\bfseries
Evolutionary AI Control:\par
Population-Level Safety Under Reproduction,\par
Heredity, and Selection\par}
\vspace{0.9em}
\hypersetup{pdfauthor={Andrew T. Cadotte}}
{\normalsize Andrew T. Cadotte\par}
{\normalsize Independent Researcher, Ann Arbor, Michigan, USA\par}
{\normalsize \href{mailto:cadottea@umich.edu}{cadottea@umich.edu}\par}
{\normalsize \href{https://orcid.org/0000-0002-6299-7315}{ORCID: 0000-0002-6299-7315}\par}
\vspace{0.4em}
{\normalsize November 2026\par}
\end{center}

\subsection{Abstract}\label{abstract}

An AI agent can satisfy every individual safety check and still give
rise to a population that violates a collective safety limit. This risk
arises in open multi-agent systems, where agents copy themselves, modify
their successors, and change system membership during a run. Selection
then shifts the population toward traits outside the developers'
optimization targets. We prove conditions under which safety guarantees
persist through reproduction, inheritance, and selection, and we
characterize how they fail.

Four counterexamples illustrate these failures. Unauthorized copies,
which violate a specified reproductive restriction, can each leave fewer
than one successor on average while their expected number grows without
bound under continued production from an expanding authorized
population. Two lineages can each decline in isolation yet grow
exponentially when they interact. Alternating between reproductive rules
can drive growth even when each rule, applied alone, suppresses it.
Selection can raise the population's average rate of unauthorized
reproduction above a certified limit even though every agent's rate
remains fixed.

Whereas existing verification methods assume fixed or registered
membership, we prove safety certificates, under stated assumptions, for
populations whose agents create successors during a run. The main result
bounds the expected cumulative harm over a run, including when a
scheduler adapts the reproductive rules to the observed history. Further
conditions cover every population state reachable through evolution and
preserve reproductive restrictions as descendants inherit modifications.
Drawing on the mechanisms by which multicellular organisms control cell
reproduction, the framework enables developers to specify and verify
these controls before deployment.

\subsection{1. Introduction}\label{introduction}

In an open multi-agent system, agents can create successors that become
members during a run. To verify safety, an evaluator must account for
the populations those membership changes can produce. An artificial
intelligence agent can meet an individual safety condition in isolation
while the population it founds later violates a collective safety limit,
such as a bound on the number of unauthorized descendants. Interactions
can affect how many descendants different variants leave, and
inheritance can carry those differences into later generations. A
guarantee for this open system must therefore cover the populations that
can arise after deployment. Corollary 1 bounds expected cumulative
burden under changing reproductive rules. It combines a common
contraction bound on the authorized population with bounds on
admissions, reproductive opportunities, baseline burden, and the
probability and added burden of escape.

AI safety evaluations usually focus on one system. Evaluators test a
model, deliberately try to elicit unsafe behavior through red-teaming,
and use the results to assess that model before release. Once an agent
copies itself, modifies its successors, or launches subordinate
instances, it can found a lineage, a group of agents linked by descent.
Its descendants can differ from their ancestor, compete for the same
computing resources, and leave unequal numbers of offspring.
Inheritance, variation, and unequal reproduction allow Darwinian
evolution, in which traits become more or less common according to how
many descendants their carriers leave. Selection can therefore change
the frequency of traits that developers never chose to optimize. This
paper identifies conditions under which a founder's safety guarantees
extend through reproduction and selection. In the two-pool example after
Corollary 1, with no external admissions, pool-specific escape
probabilities and per-escape charges bound expected quarantine work by
20 units over the run. Separate uniform caps give 80 units for the same
charges. Adding a one-time case-opening charge gives a ceiling of 21
units. The example gives a higher expected burden under alternation than
either fixed rule permits. The certificate also covers a scheduler that
chooses rules in response to the observed history.

Social laws specify shared rules for agents {[}61{]}. Kouvaros and
Lomuscio (2013) {[}62{]} verify families of multi-agent systems whose
membership is fixed within each instance. Zhu and Wang (2026) {[}28{]}
verify authority invariants for registered populations that can replace
agents or split into branches. Our certificate addresses expected burden
when agents create successors during a run, reproductive rules can
change, and unauthorized production remains possible. The models show
how continuing production, interaction, switching, and selection can
invalidate a guarantee established for individual agents or a restricted
population.

Consider a coding agent allowed to make copies of itself to work on
large tasks in parallel. Each copy inherits its parent's instructions
and working notes and can revise them. Suppose a small fraction of
copies start spawning outside the authorized environment, and monitoring
removes these copies fast enough that each leaves fewer than one
successor on average. In the branching-process model of Section 5.1, the
expected size of each unauthorized lineage tends to zero. This
expectation bounds the probability that the lineage still has any
members, so each lineage eventually dies out with probability one. Yet
the expected number of unauthorized copies can grow without bound if an
expanding authorized population creates them faster than monitoring
removes them. Section 5 models this as a source--sink process, in which
one population supplies individuals to another that would decline on its
own. To bound unauthorized abundance at a given generation in this
source--sink model, account for the new unauthorized copies supplied in
earlier generations. The number supplied per generation is the
production flux. In this model, finite lifetime cumulative occupancy
requires summable flux in the deterministic case. In the
branching-process interpretation, finite expected lifetime occupancy
requires summable expected flux.

An AI safeguard can reduce the current population while changing which
descendants survive and reproduce. Medicine and epidemiology illustrate
how controls shape the populations they leave behind. Antimicrobial
treatment can clear susceptible pathogens while resistant strains
survive and seed later infections {[}45{]}. Chemotherapy can shrink a
tumor while drug-tolerant cells survive and drive relapse {[}46{]}. A
pathogen that transmits poorly in a new host can still emerge when
repeated spillover, or transmission from another host population, gives
it repeated opportunities to evolve {[}36, 38{]}. Models of computer
viruses and self-propagating code likewise show that containment depends
on how quickly responders act relative to the rate of new infection
{[}33, 34{]}. For reproducing AI agents, assess which lineages remain
after a safeguard acts and how many new descendants can still enter the
population.

An evaluator who observes two lineages decline in isolation might expect
them to decline together. Yet when each lineage raises the other's
reproductive rate, the two can grow without bound (Section 6). An
evaluator might also expect a population to keep shrinking if each
control eliminates it when applied alone. Our example gives two
reproductive regimes, or sets of rules governing reproduction, that each
drive a population to extinction. Alternating those regimes doubles the
population every generation when it starts in the class moved by the
first regime (Section 7). An evaluation confined to one agent, one
lineage, or one regime can pass while missing these population failures.

A reproductive shock is a random change in reproduction. To estimate the
chance of crossing a population threshold by a deadline, an evaluator
needs to know how likely each shock is in each generation and how shocks
depend on earlier shocks. In Section 6.1, shocks that persist and shocks
that alternate have the same probabilities in every generation but give
different chances of crossing by the deadline.

An evaluator might expect the population to stay below a limit on its
average probability of unauthorized reproduction while each variant's
behavior stays fixed. A higher-risk variant that reproduces faster can
become common enough to push the population above that limit (Section
12.1). One might then expect a deterministic forecast, which updates the
mix of variants without sampling, to give the expected long-run count of
unauthorized reproduction events per generation. In a finite population
that samples parents, the expected rate depends on the probability that
the higher-risk variant eventually takes over the population (Section
12.1). To estimate long-run burden in that setting, weight the two
single-variant rates by their takeover probabilities.

These populations pose a concrete engineering problem as autonomous
agents acquire the ability to create and modify successor software.
Organizations have an incentive to deploy more agents when their
expected contribution exceeds their resource costs. Models of AI
development races use evolutionary game theory, which tracks how
strategies spread as their payoffs change, to show when competition
rewards speed over safety and when regulation is needed {[}57{]}.
Successors can inherit model weights, code, instructions, or persistent
working state, and later generations can inherit differences introduced
through modification, branching off a copy, or adaptation. Anthropic's
contained self-replication evaluations examined steps toward autonomous
propagation across computing environments without centralized
data-center access {[}29{]}. Evidence of autonomous Darwinian evolution
in a deployed population would also have to show that inherited
differences affect which lineages leave descendants.

Competition for computing capacity, credentials, and places to run can
affect reproductive success when inherited traits change which agents
secure those resources, stay operational, and produce descendants. These
incentives and advances in autonomous propagation make population
control a foreseeable engineering requirement, although when autonomous
Darwinian evolution will occur remains uncertain. As these capabilities
develop, safety architectures must account for the populations and
sequences of inherited modifications that they allow.

One threat scenario is the deliberate release of self-replicating agents
outside an authorized system for governing reproduction. Released agents
could spread across networked computers and sufficiently capable
connected devices beyond the reach of the original developer's
safeguards. Hosts, networks, and other outside authorities can change
which agents have opportunities to replicate and which variants
selection favors. Evaluating these controls requires accounting for
continuing sources of unauthorized agents, interactions between
populations, changes in controls, and inherited traits that allow
descendants to evade restrictions. The framework gives mathematical
tests under stated assumptions. Applying them to a deployment requires
evidence about how agents reproduce, how the controls are enforced, and
how failures affect people and the systems they depend on.

Before deployment, developers can specify who may copy a model, what a
successor inherits, and which resources a new instance can acquire.
Multicellular organisms evolved mechanisms that regulate which cells
reproduce and which lineages contribute to offspring (Section 3). Cancer
shows how controls on cell proliferation can fail (Section 4). AI
developers can specify and verify such controls in advance under
explicit assumptions.

Evolutionary theory, artificial life, research on self-replicating
systems, and AI safety all inform this framework. Müller et al.~(2026)
{[}1{]} examine artificial systems that can evolve, from
human-controlled breeding to populations in which evolution proceeds
autonomously. Hendrycks (2023) {[}2{]} analyzes how selection may favor
systems whose reproductive success conflicts with human interests.
Rainey (2023) {[}3{]} examines human--AI relationships using multilevel
evolutionary theory, which studies selection at different levels of
organization, and considers ways to link the reproductive prospects of
different participants. In the Darwin Gödel Machine, computational
descendants inherit modifications, and some descendants are selected as
parents more often than others {[}4{]}. We use AI control in the sense
of Greenblatt et al.~(2024) {[}58{]}, who design and evaluate safety
procedures intended to hold even when an untrusted model deliberately
tries to defeat them. One procedure uses a less capable, trusted model
to monitor a more capable model under protocols that govern instances
starting each call without state from earlier calls. We extend control
to populations that reproduce, inherit traits, and undergo selection.

Yudkowsky and Herreshoff (2013) {[}52{]} analyze how agents can
repeatedly construct successors that preserve their goals. They explain
why a formal system can have difficulty establishing the trustworthiness
of an equally strong successor, give partial constructions that permit
repeated successor creation under idealized logical assumptions, and
show how agents can evade restrictions on internal self-modification by
constructing successors externally. Here the guarantee concerns how
populations reproduce and how much burden their descendants generate
across successive generations.

Gordon (2000) {[}48{]} gives conditions under which machine-learning
operations preserve specified behavior during adaptation and analyzes
how those guarantees can fail when agents interact. Our certificates
cover absolute numbers of descendants produced, cumulative reproductive
burden, and the rules that govern reproduction across inherited changes
and interacting lineages. Morales et al.~(2018) {[}49{]} use replicator
dynamics, which track how the frequency of a behavior changes with its
relative payoff, to study coordination norms that persist under
selection. Their results show why interdependent conventions need joint
analysis and why selection can maintain undesirable outcomes.

Montes and Sierra (2022) {[}50{]} search for behavioral rules that
promote specified values by scoring the outcomes of simulated runs. Liao
et al.~(2023) {[}51{]} develop a way to govern automated systems through
rules and arguments that express stakeholders' ethical judgments. Here
we assess reproductive rules by asking which populations they allow and
what those populations mean for human interests.

Evidence about an initial agent, observed descendants, or one
reproductive regime describes the cases that evaluators examined. A
guarantee for the population also has to account for interactions
between lineages, changes in reproductive rules, continued production of
unsafe descendants, and the population states reached in later
generations. Throughout this paper, a \emph{certificate} consists of a
verified condition for a specified class of systems, states, or
histories, together with the assumptions under which it establishes a
stated guarantee.

Corollary 1 gives the main technical result, an expected-burden
certificate for populations whose reproductive rules can change. The
paper makes four contributions.

\begin{enumerate}
\def\labelenumi{\arabic{enumi}.}
\item
  \textbf{Reproductive success and human interests are separate.} We
  distinguish realized reproductive fitness, the reproductive success a
  variant achieves in its environment, from independently specified
  human interests. We identify the additional premise required to
  connect them and give sufficient conditions under which reproductive
  control protects acceptable human outcomes.
\item
  \textbf{Production matters as well as persistence.} In the
  source--sink model, the cumulative number of unauthorized agents
  counted across generations depends on both new production and the
  persistence of existing descendants. We show how to combine a common
  contraction bound on the authorized population with bounds on the
  conditional escape probability and added burden at each reproductive
  opportunity. Compared with checking each reproductive rule in
  isolation, the composed certificate covers adaptive switching under
  the stated assumptions. The two-pool example derives the bounds from a
  successor-authorization protocol and shows how using pool-specific
  escape probabilities and consequences tightens the resulting budget.
\item
  \textbf{The scope of reproductive testing matters.} Lineages
  controlled in isolation can grow when they interact, and alternating
  reproductive regimes can produce growth even when each regime controls
  growth on its own. We give explicit counterexamples and conditions
  sufficient to certify control.
\item
  \textbf{Safety must cover evolutionary change.} We give conditions for
  covering population states that evolution can reach and for preserving
  reproductive restrictions as descendants inherit modifications.
\end{enumerate}

The framework draws on mathematics for branching processes, which track
descendants across generations, and on population and evolutionary
models. It also uses results for switched systems, which change their
governing rules over time. The resulting specification identifies the
populations and modification histories covered by a reproductive-control
certificate and gives sufficient conditions for its guarantees.

Table 1 pairs six reproductive failure mechanisms with the corresponding
control requirements. Its final row gives the separate requirement for
claims about human interests, since even agents that reproduce within
the rules can affect people through their actions or changes to the
environment. Each certificate applies under its stated model
assumptions.

\begin{table}[!htbp]
\centering
\caption{Evolutionary failure mechanisms, mathematical witnesses, and control requirements.}
\small
\begingroup
\setlength{\tabcolsep}{6pt}
\renewcommand{\arraystretch}{1.12}
\begin{tabularx}{\linewidth}{@{}>{\RaggedRight\arraybackslash}p{0.27\linewidth}>{\RaggedRight\arraybackslash}p{0.28\linewidth}>{\RaggedRight\arraybackslash}X>{\RaggedRight\arraybackslash}p{0.075\linewidth}@{}}
\hline
\textbf{Failure mechanism} & \textbf{Mathematical witness or distinction} & \textbf{Control requirement} & \textbf{Section} \\
\hline
Continuing unsafe production & Ongoing introductions can accumulate population burden even when each unsafe lineage is subcritical & Bound production flux and the resulting population burden & 5 \\
Interaction-dependent growth & Lineages declining in isolation can increase when interacting under the same reproductive law & Assess joint-population dynamics as well as isolated lineages & 6 \\
Temporal switching & Individually subcritical operators can produce growth through composition & Establish a common or admissible-switching contraction certificate & 7 \\
Incomplete evolutionary assessment & Reachable population states may lie outside the certified assessment domain & Cover the specified reachable set, directly or through an inductive invariant & 9 \\
Hereditary loss of reproductive control & An inherited modification can turn an authorized reproductive rule into one that permits unauthorized descendants & Preserve reproductive restrictions and burden bounds over admitted evolutionary histories & 9 \\
Selection-driven certificate expiry & Selection can raise the census-weighted or opportunity-weighted probability of unauthorized reproduction above its certified ceiling while individual probabilities stay fixed & Verify the average under the certificate's specified weighting across admitted selection histories & 12.1 \\
Human-interest failure despite reproductive control & Unauthorized reproduction can be suppressed while human interests deteriorate & Establish an independent human-interest consequence premise and appropriate invariance or risk bounds & 9.5–9.6 \\
\hline
\end{tabularx}
\endgroup
\end{table}

\clearpage

Section 9.2 develops the expected-burden certificate and applies it to
the two-pool example. Sections 5--8 develop the production, interaction,
and switching results it builds on. Section 9 also addresses
reachability, inheritance of restrictions, and the conditions for
protecting human interests. Section 13 reports model studies and
proposes protocols for further evaluation.

\subsection{2. From Individual Alignment to Evolutionary Population
Control}\label{from-individual-alignment-to-evolutionary-population-control}

A safety evaluation of an evolving AI population must account for how
agents behave, which variants leave descendants, and how the resulting
population affects people and their environment. Individual tests
measure how closely agents follow their developers' objectives.
Reproductive histories show which inherited variants contribute to later
generations, and evidence about the population and its environment shows
how those changes affect people.

\subsubsection{2.1 The object of control
changes}\label{the-object-of-control-changes}

A safety evaluation establishes which properties a system has under
specified operating conditions. When the system produces descendants
that inherit its traits, the rules of reproduction also help determine
which properties those descendants have.

When agents create successors, the population can gain new members at
each generation. The state of interest changes from an individual system
\(M\) to the evolving population process

\[
P_0\rightarrow P_1\rightarrow P_2\rightarrow\cdots.
\]

Inheritance, variation, interaction, survival, reproduction, migration,
and the environment determine the population's composition at each step.
A safety property that holds for \(P_0\) can fail as those processes
change the population. Research on evolutionary AI, open-ended AI,
recursive self-improvement, evolutionary computation, and
population-level safety already examines related changes {[}1, 24{]}.
Open-ended AI studies systems capable of sustained novelty, and research
on recursive self-improvement examines systems or successors that
improve their own capabilities. Evolutionary computation uses variation
and selection to search for solutions. Here we identify the quantities a
population-level guarantee must control as evolution proceeds.

\subsubsection{2.2 Realized reproductive fitness and engineered
objectives}\label{realized-reproductive-fitness-and-engineered-objectives}

Let \(z\) be the traits that members of an artificial lineage inherit
and express, and let \(E\) be the relevant environment. The realized
reproductive fitness \(w(z,E)\) measures the lineage's contribution to
later generations under the actual dynamics of the population and its
environment. A developer can influence this contribution through an
optimization score, reward, benchmark result, or rule for weighting
parents. Which descendants persist and reproduce in the environment
determines the realized contribution.

The Darwin Gödel Machine produces lineages of parents and descendants
and selects some parents more often than others {[}4{]}. A
developer-specified weight helps determine which candidates become
parents. The rules for reproduction and the environment shape how that
choice affects the lineage's contribution to later populations.

The selection--transmission model below tracks a finite set of inherited
variants. It combines selection, which weights parents by their expected
reproductive contribution, with transmission, which determines the
offspring types those parents produce. The model yields a Price-equation
identity that separates the effects of selection and transmission on the
share of a trait {[}39, 40{]}. Let \(q_{i,t}\) be the frequency of
heritable variant \(i\) at generation \(t\), let \(w_{i,t}\geq0\) be the
expected number of offspring per parent of that type in generation
\(t\), and let \(Q_{ij,t}\) be the share of type \(j\) among the
expected offspring of a type-\(i\) parent. When \(w_{i,t}>0\), this is
the expected number of type-\(j\) offspring per type-\(i\) parent
divided by \(w_{i,t}\). For \(w_{i,t}=0\), any probability row gives the
same update. Assume \(\sum_jQ_{ij,t}=1\), \(\sum_iq_{i,t}=1\), and
positive mean fitness \(\bar w_t=\sum_iq_{i,t}w_{i,t}>0\). The
deterministic update gives \(q_{j,t+1}\) as the expected number of
type-\(j\) offspring divided by the expected total number of offspring.
In a finite stochastic population, the realized share is defined when
the census has at least one offspring. The mean of that share
conditional on a nonempty census can differ from this ratio.

\[
\boxed{
q_{j,t+1}
=
\frac{\sum_iq_{i,t}w_{i,t}Q_{ij,t}}{\bar w_t}.
}
\]

For a specified set \(\mathcal U\) of descendant variants associated
with a reproductive-control failure, define

\[
r_{i,t}=\sum_{j\in\mathcal U}Q_{ij,t}.
\]

\needspace{6\baselineskip}

Summing the update over \(\mathcal U\) gives the exact identity

\[
\boxed{
q_{t+1}(\mathcal U)
=
\mathbb E_{q_t}[r_t]
+
\frac{\operatorname{Cov}_{q_t}(w_t,r_t)}
{\bar w_t}.
}
\]

For each parent type with \(w_{i,t}>0\), the failure-class share
\(r_{i,t}\) is its expected number of failure-class offspring divided by
its expected total number of offspring. The first term averages these
shares using each parent type's frequency in the current generation. The
second term measures how weighting parent types by their expected
reproductive contributions changes the failure-class share in the
deterministic update. It uses covariance, which measures how two
quantities vary together, to compare expected reproductive contributions
and failure-class shares across parent types. The share of descendants
in the failure class can therefore rise even when every variant's
transmission probabilities stay fixed.

The identity has the Price-equation form {[}39, 40{]}: subtracting the
current failure-class share from both sides gives a transmission term
and a selection term. The transmission term is the difference between
the parent-weighted mean of the type-specific failure-class shares and
the current failure-class share. The selection term is the displayed
covariance divided by mean fitness. If a parent type is present and
\(w_{i,t}=0\), the two terms can depend on the probability row assigned
to that type. Changing that row shifts the terms by equal and opposite
amounts, leaving their sum unchanged. To describe the distribution of
realized frequencies in a finite population, specify a stochastic model
for offspring counts and how those counts form the next census. To
evaluate how a frequency change affects human interests, apply the
human-interest criterion \(V\), defined below, to the corresponding
population and environment states.

The allowed hereditary transitions determine which variants can arise.
In the deterministic update, each parental variant's frequency is
weighted by its expected offspring count per parent. A guarantee over
every allowed trajectory must cover every state that the admitted
transitions can produce, including states assigned small probabilities
by a selection-weighted model or a finite-time estimate.

\subsubsection{2.3 Human interests are an independent
quantity}\label{human-interests-are-an-independent-quantity}

Fitness \(w(z,E)\) describes which inherited lineages reproduce more
successfully under the actual population and environmental conditions.
To assess the consequences for people, specify a separate criterion. Let
\(V(p,E)\) represent a human-system quantity evaluated on population
state \(p\) in environment \(E\). It can describe human viability,
autonomy, control, flourishing, or another property that matters to
human persistence. It can also take the form of a vector or a set so
that the evaluation can cover several human-system constraints together.
Measuring reproduction and human outcomes separately lets us identify
when selection favors lineages that worsen those outcomes.

Hendrycks {[}2{]} shows how competition can favor artificial systems
that spread while becoming more antagonistic to human interests. He
explains how realized fitness can diverge from training objectives and
examines interventions that change which behaviors selection favors.

Selection can sustain coordination norms that produce outcomes an
evaluator would reject. Morales et al.~{[}49{]} show that such norms can
remain stable even when the rewards permit collisions or leave agents
stopped indefinitely.

\needspace{10\baselineskip}

The reproductive rules determine how the population changes, while the
human criterion evaluates the state that results. The first diagram
gives the population update.

\[
(p_t,E_t)
\xrightarrow{\text{reproduction, inheritance, interaction, selection}}
p_{t+1}.
\]

The second gives the evaluation of the population and environment.

\[
(p_t,E_t)\xrightarrow{\text{human criterion}}V(p_t,E_t).
\]

Multilevel evolutionary theory can ask whether formerly independent
replicators come to share the same reproductive prospects or form a
larger unit on which selection acts. Human and artificial systems can
become strongly coupled in their reproduction even while human autonomy,
control, or another independently specified property deteriorates. The
human-interest criterion determines whether that outcome is acceptable.

Evolutionary AI control constrains the population changes driven by
artificial reproduction so that the system continues to satisfy the
specified human-interest conditions.

\subsubsection{2.4 A minimal human-interest evolutionary-control
condition}\label{a-minimal-human-interest-evolutionary-control-condition}

The Price equation compares quantities defined on the same entities
{[}40{]}, while here \(w\) describes lineages and \(V\) evaluates the
whole population and its environment. If each lineage had its own
consequence score \(v_{\mathrm{lin}}\), a positive covariance between
\(w\) and \(v_{\mathrm{lin}}\), taken over the distribution of lineages
in the population, would indicate a positive association between fitness
and that score. Even with a positive covariance, rare lineages could
expand rapidly and drive the human system below its acceptable
threshold. Safety also depends on interactions, continuing production of
unsafe descendants, changes in reproductive regimes, and the states
these processes can reach across generations. We therefore state the
requirement directly on the evolving population.

Write the population update as \[
p_{t+1}
=
\Phi_{E_t}(p_t,w,\mathcal H_t).
\]

Here \(p_t\) is the joint population state, \(w\) describes the effects
of realized reproductive fitness, and \(\mathcal H_t\) collects the
inheritance, heritable variation, interactions, migration, and other
reproductive mechanisms at time \(t\). The map \(\Phi_{E_t}\) combines
these inputs to give the next population in environment \(E_t\).

For a specified human criterion \(V\), define the acceptable region \[
\mathcal S_V(E)
=
\{p:V(p,E)\geq V_{\min}\}.
\]

When the criterion includes several quantities or takes a non-scalar
form, define \(\mathcal S_V\) directly as the population states that
meet the required human-system constraints.

The minimal requirement is that every allowed trajectory starting in the
acceptable region stay within the region appropriate to its environment
at each step. We call this property invariance of the acceptable region.

\[
\boxed{
p_0\in\mathcal S_V(E_0)
\Longrightarrow
p_t\in\mathcal S_V(E_t)
\quad
\text{at every generation }t\text{ along the allowed trajectories}
}
\]

The implication applies to the reproductive dynamics under
certification. To make a safety claim, specify how you construct
\(\mathcal S_V\), represent uncertainty about human interests, and
identify the allowed environments and trajectories. Then show that the
condition holds along those trajectories.

\subsection{3. Reproductive Control in Multicellular
Organisms}\label{reproductive-control-in-multicellular-organisms}

Multicellular organisms regulate where and when cells divide and which
cell lineages contribute to offspring. Mutations can change the traits
cells pass on, and competition can favor a lineage whose growth harms
the organism. This distinction between limiting proliferation and
limiting transmission also matters when artificial agents reproduce.

\textbf{Reproductive control} regulates who can reproduce and pass on
traits. Artificial agents can spread beyond boundaries that usually
confine somatic cells, the body's cells outside the reproductive
lineage. The comparison concerns how system structure shapes
reproduction, inheritance, and competition, whether selection acts on
humans and artificial agents separately or on a unit they share.

\subsubsection{3.1 Lower-level replication creates a control
problem}\label{lower-level-replication-creates-a-control-problem}

In \emph{The Evolution of Individuality}, Buss (1987) {[}5{]} treats
cell lineages as populations whose inherited traits affect how fast they
multiply. Selection among cells can favor a variant that multiplies
faster even when its growth disrupts the organism or threatens its
survival.

As cells came to function together as an organism, their
\textbf{reproductive opportunities}, their chances to multiply within
the body or pass traits to the organism's offspring, changed.

Buss describes how one lineage can limit another's replication and how
terminal differentiation, in which a cell takes on a specialized role,
can limit its ability to divide. Maternal or developmental control can
determine what descendant cells become. Germ-line determination
specifies which cells can contribute to offspring, and germ-line
sequestration sets those cells apart from the other cell lineages.
Together, these processes restrict which cellular variants can pass to
the organism's offspring {[}5{]}.

Cells can continue to multiply within the organism after they lose the
ability to pass traits to its offspring. In an AI deployment, check
separately whether the controls limit how many copies a lineage creates
and which of its traits those copies can pass to later successors.

\subsubsection{3.2 Conflict mediation is
architectural}\label{conflict-mediation-is-architectural}

Michod and Roze (2001) {[}6{]} study how multicellular organisms sustain
cooperation by limiting competition among cells and changing the
conditions for selection among both cells and organisms. They examine
how the separation of reproductive cells, the regulation of reproduction
during development, and the policing of competing cell lineages help the
organism function as a unit. How the organism develops and organizes its
cells determines which opportunities individual cell lineages retain.

Candidate mechanisms include:

\begin{itemize}
\tightlist
\item
  restriction of lower-level reproduction, which limits how much cells
  or other constituent replicators can multiply
\item
  germ-line sequestration, which sets apart the cells that can
  contribute to offspring
\item
  reproductive specialization, which assigns reproduction to particular
  cells or lineages
\item
  developmental constraints, which limit which roles a cell can take on
  or when it can divide
\item
  policing of defectors, which restrains lineages that gain by
  undermining cooperation
\item
  programmed removal or apoptosis, in which a cell follows a regulated
  process that ends its life
\item
  recognition and controlled fusion, which determine which cells or
  organisms can join
\item
  transmission bottlenecks, which restrict how many individuals or
  lineages contribute to the next generation
\item
  dependence of lower-level success on higher-level organization, which
  ties a constituent lineage's reproductive prospects to the functioning
  of the larger unit
\end{itemize}

These mechanisms act at different stages. Some limit which variants
arise, while others constrain their growth, competition, survival, or
transmission to later generations. For artificial populations, designers
can specify \textbf{where agents may reproduce, what their descendants
can inherit, and which descendants may reproduce again}.

\subsubsection{3.3 Control mechanisms can themselves become evolutionary
targets}\label{control-mechanisms-can-themselves-become-evolutionary-targets}

A reproductive restriction changes which traits selection favors. Michod
and Roze describe how variants can evolve ways around conflict-reducing
constraints and how additional mechanisms can evolve in response
{[}6{]}. Suppressing one route to reproductive advantage can therefore
favor variants that exploit another.

When a control relies on agents restraining one another, an evaluator
must account for how that restraint changes reproductive incentives.
Frank (1995) {[}7{]} analyzes how competition among constituent
replicators can undermine cooperation in larger groups. His analysis
shows how mutual policing, in which group members restrain one another's
competitive behavior, can suppress this competition and favor
cooperation.

To assess whether a constraint will remain effective, examine current
behavior and the inherited changes that can affect reproduction,
inheritance, migration, and transmission in later generations.

\subsubsection{3.4 Reproductive control principles from
biology}\label{reproductive-control-principles-from-biology}

Long-term control depends on how often variants arise, multiply, and
pass on their traits. For artificial populations, specify how a control
changes these processes and derive testable predictions for the
population.

\subsection{4. Cancer: What Failure Reveals About
Control}\label{cancer-what-failure-reveals-about-control}

When agents leave different numbers of successors, changes in membership
can favor lineages that undermine a shared objective. Somatic evolution
changes the mix of cells within the body as cells inherit different
traits and leave different numbers of descendants. Cancer shows what can
happen when this process favors a cell lineage whose growth harms the
organism. Lyons et al.~(2026) {[}30{]} draw on homeostasis, the
regulation of internal conditions in a living system, to discuss how AI
systems might limit harmful effects on others. Competition among cell
lineages can persist even as the organism regulates cell division,
survival, access to resources, and location within tissues.

\subsubsection{4.1 A lineage can succeed while its containing system
fails}\label{a-lineage-can-succeed-while-its-containing-system-fails}

The body coordinates cell division, specialization, location, access to
resources, and death so its tissues can function. An inherited change
can give a cell lineage an advantage by helping its cells divide,
survive, acquire resources, invade other tissues, pass on their traits,
or resist regulation. As the lineage expands, it can make the organism
less able to survive.

An increase in cell-lineage fitness, \(\Delta w_{\text{cell}}>0\), can
occur alongside a decline in the measure of how well the organism fares,
\(\Delta V_{\text{organism}}<0\). Differences in reproductive success
suffice to create this conflict, even in cells with no representation of
the conflict or intent to cause harm. For artificial populations, assess
which inherited variants leave more descendants and how the resulting
change in population composition affects people.

\subsubsection{4.2 Cancer is failure inside an already controlled
architecture}\label{cancer-is-failure-inside-an-already-controlled-architecture}

The organism regulates several processes that determine how cell
lineages grow and spread:

\begin{itemize}
\tightlist
\item
  cell-cycle progression, the sequence that leads to cell division
\item
  differentiation, the process by which cells acquire specialized roles
\item
  programmed cell death
\item
  tissue organization
\item
  access to resources
\item
  surveillance by the immune system and surrounding tissue
\item
  spatial opportunity, the space available for local cell growth
\item
  spread of cell lineages within the organism
\item
  access to organism-level heredity, the ability to pass traits to the
  organism's offspring
\end{itemize}

Cancer shows that an evolving lineage can find a route around these
controls even when the organism regulates several processes that govern
growth and survival. Evaluators must assess a reproductive constraint
under the evolutionary dynamics it allows and helps shape.

\subsubsection{4.3 Reproductive suppression and
closure}\label{reproductive-suppression-and-closure}

As Section 3.1 explains, cells can multiply within an organism and pass
traits to the organism's offspring through different channels {[}5{]}.
Reproductive suppression limits how much a lineage grows or how long it
persists. Reproductive closure requires the restrictions to cover every
route by which descendants can arise, inherit traits, and reproduce in
turn.

Artificial descendants can arise through copying, forking a branch from
an existing system, deploying another instance, migrating to another
host, creating a successor, or inheriting working state. The system's
design determines which routes exist. Reproductive controls must cover
every route that can produce heritable descendants, and one control can
cover several routes. Section 5 models how new arrivals accumulate, and
Section 9 gives conditions for reproductive restrictions to remain
effective as the population evolves.

\subsubsection{4.4 Treatment resistance illustrates selection on
control}\label{treatment-resistance-illustrates-selection-on-control}

Cancer treatment changes which cells survive and compete. When cells
inherit traits that affect their response to treatment, the intervention
can suppress some variants and favor the expansion of variants better
able to survive it. To assess whether a control on artificial
reproduction will remain effective, examine how selection responds to
the control, especially when descendants can change how they reproduce
or pass on traits.

\subsection{5. Persistent Production Defeats Lineage-Only
Control}\label{persistent-production-defeats-lineage-only-control}

Containment of agents that have already escaped addresses existing
lineages, while each new escape adds a member to the unauthorized
population. Those arrivals can sustain the population even as every
existing lineage declines. A branching process with immigration counts
both the offspring of individuals already present and the newcomers
supplied by an outside source {[}9{]}. The models below quantify how
these arrivals and the persistence of existing lineages determine
cumulative burden.

From here on, `unsafe' and `unauthorized' mean that a specified
reproductive restriction has failed. For example, an agent may be
produced outside the authorized system, or its type may belong to a
class designated as a reproductive-control failure. To assess the
consequences for people, specify a human-interest criterion
independently and establish how the reproductive outcome affects it.
Sections 9.5 and 9.6 give sufficient conditions for that link.

\needspace{8\baselineskip}

\subsubsection{5.1 Minimal source--sink
model}\label{minimal-sourcesink-model}

An unsafe lineage can die out while the authorized population keeps
producing new unsafe descendants. To assess containment, track two
populations:

\begin{itemize}
\tightlist
\item
  \(H_t\): a permitted reproductive population;
\item
  \(U_t\): an unsafe or unauthorized descendant population.
\end{itemize}

The populations evolve according to

\[
\begin{aligned}
H_{t+1} &= aH_t,\\
U_{t+1} &= bU_t+cH_t,
\end{aligned}
\qquad
\begin{gathered}
a>1,\quad 0\leq b<1,\quad c>0,\\
H_0>0,\quad U_0\geq 0.
\end{gathered}
\]

A two-type branching process follows the offspring of individuals of
each type across generations. The same equations give the expected
numbers when each permitted individual produces mean \(a\) permitted and
\(c\) unsafe offspring, and each unsafe individual produces mean \(b\)
unsafe offspring and zero permitted offspring. Under this
interpretation, \(H_t\) and \(U_t\) are expected population sizes.

When \(b<1\), each unsafe individual produces fewer than one unsafe
successor on average, so the unsafe population is subcritical and
declines in expectation in isolation. The permitted population supplies
\(cH_t\) new unsafe descendants per generation. Solving the recurrence
separates the descendants of the initial unsafe population from those
generated by later arrivals.

\[
U_t
=
U_0b^t
+
H_0c\frac{a^t-b^t}{a-b}.
\]

Because \(a>1\) and \(b<1\), at large \(t\) the contribution from new
production dominates. The symbol \(\sim\) means that the ratio of the
two sides tends to one.

\[
U_t
\sim
\frac{H_0c}{a-b}a^t.
\]

The deterministic unsafe abundance grows exponentially even though each
unsafe agent produces fewer than one unsafe successor on average. In the
branching-process interpretation, the same growth rate applies to
expected abundance.

To measure cumulative burden, sum the unsafe population sizes across
generations, counting each individual once for each generation in which
it is present. To count distinct descendants, measure exposure, or
quantify harm, define and model the corresponding outcome.

Define the unsafe population burden through generation \(T\) by

\[
B_U(T)=\sum_{t=0}^{T}U_t.
\]

For large horizons the cumulative burden satisfies the following
asymptotic relation.

\[
B_U(T)
\sim
\frac{H_0ca}{(a-b)(a-1)}a^T.
\]

Cumulative burden therefore grows at the same exponential rate as the
permitted source. A containment claim for the whole unsafe population
must account for new arrivals as well as the decline of each lineage
after arrival.

\subsubsection{5.2 General production-flux
form}\label{general-production-flux-form}

A containment assessment needs to account for new unsafe descendants
regardless of how they enter the population. An expanding permitted
population is one possible source. For a general source, replace
\(cH_t\) with a nonnegative production flux \(F_t\), the amount of new
unsafe population produced during generation \(t\).

\[
U_{t+1}=bU_t+F_t,
\qquad
0\leq b<1,
\qquad
F_t\geq0.
\]

Repeated substitution separates what remains of the initial unsafe
population from what arrivals add in each later generation.

\[
U_t
=
b^tU_0
+
\sum_{s=0}^{t-1}b^{t-1-s}F_s.
\]

Adding the generation-by-generation abundances gives the cumulative
unsafe population \(B_U(T)\) through generation \(T\).

\[
B_U(T)
=
U_0\frac{1-b^{T+1}}{1-b}
+
\sum_{s=0}^{T-1}
F_s
\frac{1-b^{T-s}}{1-b}.
\]

\needspace{8\baselineskip}

Over all generations, each arrival contributes a geometric sum of
expected abundance.

\[
\sum_{t=0}^{\infty}U_t
=
\frac{
U_0+\sum_{s=0}^{\infty}F_s
}{1-b}.
\] Both sums are nonnegative and may equal infinity. This gives the
following equivalence. \[
\boxed{
\sum_{t=0}^{\infty}U_t<\infty
\iff
\sum_{t=0}^{\infty}F_t<\infty
}
\]

The equivalence holds for finite \(U_0\) and \(0\leq b<1\). In this
linear expectation model, each unit of newly produced unsafe population
contributes \(1/(1-b)\) units of expected burden over its lifetime. If
\(F_t=F>0\), then \(U_t\) approaches a positive steady level and
cumulative burden grows linearly. If production grows, \(U_t\) itself
can diverge. If \(F_t=0\), the existing unsafe population declines
geometrically. For \(F_t=F_0/(t+1)\) with \(F_0>0\), production tends to
zero and the recurrence gives \(U_t\to0\). The production sum diverges
as a harmonic series, so \(\sum_{t=0}^{\infty}U_t=\infty\) even as the
population tends to zero. Finite cumulative burden requires the
production flux to have a finite sum.

\subsubsection{5.3 Branching-process foundations and control
implications}\label{branching-process-foundations-and-control-implications}

Established branching-process models describe these source and
downstream populations. Burden and Wei (2018) {[}8{]} model mutation in
Galton--Watson branching processes, in which individuals produce random
numbers of offspring and their type determines the offspring
distribution. Their model allows one type to produce descendants of
another. Hyrien et al.~(2017) {[}9{]} study subcritical populations
sustained by an inflow that changes over time. They relate the
construction to two-type branching processes in which one type supplies
the other with no reverse flow. They show that expected downstream
abundance can grow when immigration rises exponentially while downstream
reproduction remains subcritical. Serra (2006) {[}41{]} studies mutation
from a subcritical branching type to a mutant type whose population
grows in expectation. A model with both types describes the distribution
of the time until escape and the growth that follows. Section 5.2 gives
the production-flux equivalence for a single downstream type with a
fixed subcritical multiplier.

The source--sink model separates how long unsafe descendants persist
from how often permitted systems produce them. In the recurrence
\(U_{t+1}=bU_t+F_t\), a bound \(0\leq b<1\) controls persistence through
the lifetime factor \(1/(1-b)\), which gives the expected cumulative
population burden from one new arrival. Section 5.2 gives the resulting
condition on total production for finite expected burden.

Even a small escape probability per opportunity can generate many
unauthorized descendants when opportunities are numerous or growing. The
objective determines which quantity to bound, whether total lifetime
burden, exposure over a fixed period, peak abundance, the probability of
crossing a threshold, resource consumption, or harmful actions.

When descendants interact, the consequences of an escape can depend on
which earlier escapes occurred. The accounting below allows the
occurrence, escape probability, and consequences of each reproductive
opportunity to depend on the preceding history.

\subsubsection{5.4 From local escape to population
burden}\label{from-local-escape-to-population-burden}

The burden from a potential reproductive opportunity depends on whether
it occurs, how likely it is to produce an unauthorized descendant, and
how much burden that descendant would add. The calculation uses the
information available before each opportunity to determine that
conditional probability and added burden. Reproductive events may depend
on one another, so earlier descendants can change what later descendants
contribute.

The source--sink model gives additive burden when arrivals can be
encoded as binary opportunities and their contributions to burden add.
For the finite-horizon accounting, fix a deterministic list of \(n\)
potential reproductive opportunities through horizon \(T\), indexed in
order from \(1\) to \(n\). Which opportunities occur can depend on the
history. Let \(X_i\in\{0,1\}\) record whether opportunity \(i\) produces
an unauthorized descendant. The following function assigns a burden to
each encoded history.

\[
B_T(x_1,\ldots,x_n)
\]

The binary escape history must determine \(B_T\). Use the realized
burden when that history determines it. If additional state or later
randomness affects the realized burden, use its conditional mean given
the binary escape history to calculate expected burden. In either case,
the marginal \(D_i(X_{<i})\) in Proposition 1 must be determined by the
information \(\mathcal F_{i-1}\) available before opportunity \(i\). The
burden can reflect descendants reinforcing one another, competing,
approaching a limit, or crossing a threshold.

For an observed history \(x_{<i}=(x_1,\ldots,x_{i-1})\) before
opportunity \(i\), compare the burden when that opportunity produces an
unauthorized descendant with the burden when it produces none. Set all
later entries to zero in both histories. When the burden function gives
a conditional mean, this comparison can also reflect differences in
other state or randomness associated with the two histories. This
difference defines the marginal burden for opportunity \(i\).
Interpreting it as work caused by an individual escape requires separate
justification, such as a protocol that assigns a charge to the escape.

\[
D_i(x_{<i})
=
B_T(x_{<i},1,0,\ldots,0)
-
B_T(x_{<i},0,0,\ldots,0).
\]

Write \(B_0=B_T(0,\ldots,0)\) for the baseline value assigned to the
all-zero escape history. Define \(B_T\) on every binary history in
\(\{0,1\}^n\), including \((x_{<i},1,0,\ldots,0)\) and
\((x_{<i},0,0,\ldots,0)\) even if the process can never follow those
histories. If \(B_T\) is a conditional mean, the process law determines
its values on histories of positive probability. Specify an extension to
histories of zero probability, including the all-zero history if
necessary. This extension can change \(B_0\) and the marginal bounds, so
check the bounds for the chosen values. For any such extension, the
consecutive differences sum to \(B_T-B_0\) (Proposition 1).

Let \(\mathcal F_{i-1}\) represent the reproductive history known before
potential opportunity \(i\). Let \(A_i\in\{0,1\}\) indicate whether the
opportunity exists by the horizon, and write \(X_i=A_iE_i\), where
\(E_i\) records an escape when the opportunity exists. The indicator
\(A_i\) and conditional escape probability \(p_i\) are
\(\mathcal F_{i-1}\)-measurable, meaning that the history available
before the opportunity determines them. The escape probability satisfies
the following condition.

\[
\mathbb P(E_i=1\mid\mathcal F_{i-1})=p_i
\qquad
\text{on }\{A_i=1\}.
\]

Then

\[
\mathbb E[X_i\mid\mathcal F_{i-1}]
=
A_i p_i.
\]

The pre-opportunity information \(\mathcal F_{i-1}\) includes the prefix
\(X_{<i}\). The marginal \(D_i(X_{<i})\) is a fixed function of that
prefix, so it is \(\mathcal F_{i-1}\)-measurable. Define
\(D_i^+=\max(D_i,0)\), which keeps a positive contribution and sets a
negative one to zero.

For the finite-horizon expectation calculation, assume
\(\mathbb E[X_i|D_i|]<\infty\) for every \(i\), with finite baseline
\(B_0\). Since \(|D_i|\) is known before opportunity \(i\), conditioning
gives \(\mathbb E[A_ip_i|D_i|]=\mathbb E[X_i|D_i|]\).

\textbf{Proposition 1 (Exact adaptive production-flux accounting).}
Under the finite-horizon history-completeness, predictability, and
integrability assumptions above, the burden \(B_T\) obeys the pathwise
identity \[
B_T(X)=B_0+\sum_{i=1}^{n}X_iD_i(X_{<i}),
\] and the expected-burden identity \[
\mathbb E[B_T]
=
B_0+\mathbb E\left[\sum_{i=1}^{n}A_ip_iD_i\right].
\] Consequently, for any predictable nonnegative bounds
\(p_i\leq\epsilon_i\) and \(D_i^+\leq d_i\), \[
\mathbb E[B_T]
\leq B_0+\mathbb E\left[
\sum_{i=1}^{n}A_i\epsilon_i d_i
\right].
\]

\textbf{Proof of Proposition 1.} For \(i=0,\ldots,n\), define the
prefix-completed burden \[
b_i=B_T(X_1,\ldots,X_i,0,\ldots,0),
\] where \(b_0=B_0\) and \(b_n=B_T(X)\). If \(X_i=0\), then
\(b_i-b_{i-1}=0\). If \(X_i=1\), the definition of the
history-conditioned marginal gives \(b_i-b_{i-1}=D_i(X_{<i})\). Thus, in
both cases, \[
b_i-b_{i-1}=X_iD_i(X_{<i}).
\] Summing these equalities telescopes and proves the first identity
without any additivity assumption on \(B_T\).

Since \(A_i\) and \(D_i(X_{<i})\) are \(\mathcal F_{i-1}\)-measurable
and \(\mathbb E[X_i\mid\mathcal F_{i-1}]=A_ip_i\), we can first take the
expected value given the history and then average over histories. This
rule, called the tower property, gives \[
\begin{aligned}
\mathbb E[X_iD_i]
&=\mathbb E\left[
D_i\,\mathbb E[X_i\mid\mathcal F_{i-1}]
\right]\\
&=\mathbb E[A_ip_iD_i].
\end{aligned}
\] Finite summation and integrability establish the expected-burden
identity.

Finally, \(D_i\leq D_i^+\) and \(A_ip_iD_i^+\leq A_i\epsilon_i d_i\)
pathwise. Taking expectations proves the one-sided certificate.
\(\square\)

The accounting marginal can vary with earlier escapes. It can increase,
decrease, or become negative as the history changes.

Proposition 1 gives the expected burden whenever the finite-horizon
terms have finite absolute expectations. The requirement
\(\mathbb E[B_T]\leq G\) for a prescribed budget \(G\) therefore holds
exactly under the following condition.

\[
\boxed{
\mathbb E\!\left[
\sum_i A_i p_i D_i
\right]
\leq G-B_0.
}
\]

The upper bound combines how often reproductive opportunities occur,
their conditional leakage probabilities, and the marginal burden of
leakage under the observed history. If the expected total number of
opportunities is bounded by \(N\), with \(p_i\leq\epsilon\) and
\(D_i^+\leq d\), then \(\mathbb E[B_T]\leq B_0+N\epsilon d\). Even with
an unbounded number of opportunities or unbounded marginal consequences,
the upper bound on expected burden is finite when the
probability-weighted positive marginal contributions have a finite sum.
A high local escape probability contributes little to this bound when
the associated positive marginal burden is small.

For countably many opportunities, take finite prefixes of one ordered
opportunity process with the same activity indicators and
pre-opportunity information for shared slots. Suppose the prefix burdens
share a baseline \(B_0\) and use the same marginal \(D_i\) at
opportunity \(i\) in every prefix that includes it. One way to control
the conservative budget is to find deterministic bounds \(c_i\geq0\)
with \(\sum_i c_i<\infty\) and \(\mathbb E[A_ip_iD_i^+]\leq c_i\) for
every \(i\). The monotone convergence theorem lets us take expectations
of the increasing nonnegative partial sums and then pass to the limit.
It gives

\[
\mathbb E\!\left[
\sum_{i=1}^{\infty}A_i p_iD_i^+
\right]
<\infty.
\]

To use this bound for a separately defined infinite-horizon burden,
specify how the finite-horizon burdens approach that quantity.

For the signed infinite-opportunity budget, the stronger condition

\[
\mathbb E\!\left[
\sum_{i=1}^{\infty}A_i p_i|D_i|
\right]
<\infty
\]

is sufficient for the signed marginal series to converge absolutely in
expectation and to permit rearrangement of its fixed-order terms. The
history-conditioned marginals still depend on the ordering of
reproductive opportunities: changing that order can change the marginal
contributions themselves. If we specify an infinite-horizon burden
\(B_\infty\) separately, identifying this series with
\(\mathbb E[B_\infty]-B_0\) requires
\(\mathbb E[B_T]\to\mathbb E[B_\infty]\). Convergence in \(L^1\),
meaning \(\mathbb E|B_T-B_\infty|\to0\), suffices because
\(\big|\mathbb E[B_T]-\mathbb E[B_\infty]\big|\leq\mathbb E|B_T-B_\infty|\).
The certificate uses the specified opportunity order, which determines
the histories used to calculate the marginal burdens.

These results bound \textbf{expected population burden}, the average
burden over possible histories. A positive expected-burden ceiling can
coexist with individual histories above that ceiling. A cap that holds
almost surely requires a bound on realized burden outside a set of
probability zero. For nonnegative burden and a threshold \(\theta>0\),
the expectation alone bounds the probability that the specified burden
meets or exceeds the threshold.

\[
\mathbb P(B_T\geq\theta)
\leq
\frac{\mathbb E[B_T]}{\theta}.
\]

The bound requires only nonnegativity and the expectation, and some
distributions attain it exactly. A small expected burden is compatible
with rare trajectories that exceed the threshold. Finally, a burden
evaluated at the horizon and a sum over generations define different
objectives. For a per-generation burden \(\beta_t\), if the quantity of
interest is

\[
C_T=\sum_{t=0}^{T}\beta_t,
\]

apply the same marginal construction directly to \(C_T\).

These bounds account for the arrivals that keep supplying an unsafe
population and the burden those arrivals contribute over time. Section 6
shows how interactions can sustain population growth even with zero
outside production and with lineages that decline in isolation.

\subsection{6. From Lineages to Interacting
Populations}\label{from-lineages-to-interacting-populations}

Section 5 distinguishes newly produced unauthorized descendants from
those that persist. In a multi-agent population, interactions can raise
the number of descendants that agents in different lineages produce. One
lineage might supply credentials that another needs to replicate, while
the second supplies computing capacity or storage. Both lineages can
decline alone and grow together if each reproduces more successfully in
the other's presence. An evaluator can measure this effect by checking
how each lineage's reproduction changes with the abundance of its
partner. We compare isolation and coexistence under the same rules by
giving each lineage descendants of its own type and making its
reproduction depend on how many individuals of the other type are
present.

A test of each agent in isolation can miss failures that arise when
agents work together. Studies of interacting agents identify
system-level failures {[}10{]} and hazards in combinations of
individually acceptable models {[}11{]}. Models of evolutionary games
show how relative payoffs can stabilize deceptive strategies {[}12{]},
while replicator dynamics describe how strategies spread or persist
according to their payoffs {[}13{]}. The result here concerns
reproduction. Two heritable lineages that shrink in isolation can grow
together because each increases the other's reproductive fitness. Before
applying an isolated-lineage bound to a mixed population, check whether
that bound still holds when each lineage experiences the other at the
abundances the deployment permits.

For comparison, consider a separate model in which each parent type can
produce descendants of both types. Its linear reproduction matrix gives
the expected number of descendants of each type per parent of each type.
\[
M=
\begin{pmatrix}
r&\kappa\\
\kappa&r
\end{pmatrix},
\qquad
r=0.75,\quad \kappa=0.5.
\] Here \(\kappa\) is the mean number of descendants of the other type
produced per parent. Both diagonal entries are below one, yet the
spectral radius, the largest absolute value of the matrix's eigenvalues,
is \(\rho(M)=r+\kappa=1.25>1\). The off-diagonal entries allow each type
to produce descendants of the other type even when that other type is
initially absent. Setting those entries to zero for an isolation test
would change the reproductive law.

The interaction model gives each lineage offspring of its own type and
lets that lineage reproduce faster when its partner is more abundant. \[
\begin{aligned}
U_{1,t+1}
  &= \left[r+c\frac{U_{2,t}}{K+U_{2,t}}\right]U_{1,t},\\
U_{2,t+1}
  &= \left[r+c\frac{U_{1,t}}{K+U_{1,t}}\right]U_{2,t}.
\end{aligned}
\]

The parameter \(r\) is the reproductive multiplier in isolation, \(c\)
is the largest possible increase approached as the partner becomes
abundant, and \(K\) is the partner abundance that gives half that
increase. Use the following parameter values. \[
r=0.75,\qquad c=0.75,\qquad K=1.
\] These same parameters govern both the isolated and coexisting
lineages.

Direct iteration from \(U_{1,0}=U_{2,0}=1\) gives total abundances
\(2.0000\), \(4.4866\), \(19.2296\), and \(127.5723\) at generations
\(0,5,10,\) and \(15\), respectively, whereas either lineage alone has
abundance \[
0.75^{15}\approx0.0134
\] at generation 15.

The reproductive multiplier, the factor by which a lineage's abundance
changes in one generation, depends on the abundance \(x\) of its
partner. \[
m(x)=r+c\frac{x}{K+x}.
\] In isolation, \(x=0\), so reproduction is subcritical for \(r<1\). If
\(c>1-r\), the multiplier exceeds one precisely when \[
x>\frac{(1-r)K}{c-(1-r)}.
\] For the stated parameters, growth in one generation requires the
partner's abundance to satisfy \(x>0.5\). Proposition 2 bounds the
abundance of each lineage from below at every generation when both start
at the same abundance of at least one.

\needspace{8\baselineskip}

\textbf{Proposition 2 (Interaction-induced population growth).} Consider
the two-lineage nonlinear reproductive law specified above, with
\(r=3/4\), \(c=3/4\), and \(K=1\). For every finite nonnegative initial
abundance, a lineage whose complementary lineage is absent decays
geometrically with multiplier \(3/4\). Nevertheless, for symmetric
initial conditions \(U_{1,0}=U_{2,0}=x_0\geq1\), both lineages satisfy
\[
U_{1,t}=U_{2,t}
\geq x_0\left(\frac98\right)^t
\qquad\text{for all }t\geq0.
\] From these starting abundances, both lineages grow without bound
together under the same rules that make each shrink in isolation.

\textbf{Proof of Proposition 2.} If \(U_{2,0}=0\), the factor
multiplying \(U_{2,t}\) keeps \(U_{2,t}=0\) for every generation, and
the first lineage obeys \(U_{1,t+1}=(3/4)U_{1,t}\). The same reasoning
applies with the lineages exchanged.

For symmetric initial conditions, the two update equations preserve
equality. Write \(U_{1,t}=U_{2,t}=x_t\). Then \[
x_{t+1}
=
\left(\frac34+\frac34\frac{x_t}{1+x_t}\right)x_t.
\] The multiplier is increasing in \(x_t\geq0\). For every \(x_t\geq1\),
it is at least \[
\frac34+\frac34\frac1{1+1}=\frac98>1.
\] Starting with \(x_0\geq1\), induction gives
\(x_{t+1}\geq(9/8)x_t\geq1\) and therefore \(x_t\geq x_0(9/8)^t\) for
every generation. The lower bound diverges as \(t\) increases.
\(\square\)

\textbf{A reproductive-control certificate for small interacting
populations.} Both lineages shrink when they start in a region of low
abundance. The region is positively invariant, meaning that the next
population stays in it whenever the current population belongs to it.
Define the region and the combined abundance as follows. \[
I=[0,\tfrac14]^2,\qquad
W_t=U_{1,t}+U_{2,t}.
\] For any \((U_{1,t},U_{2,t})\in I\), the reproductive multiplier of
lineage \(i\in\{1,2\}\), whose partner is lineage \(j\), satisfies \[
\frac34+\frac34\frac{U_{j,t}}{1+U_{j,t}}
\leq
\frac34+\frac34\frac{1/4}{1+1/4}
=\frac9{10},
\qquad j\ne i.
\] Nonnegativity therefore gives \[
0\leq U_{i,t+1}\leq\tfrac9{10}U_{i,t}
\leq\tfrac14.
\] Hence \(F(I)\subseteq I\), where \(F\) is the complete two-lineage
reproductive map, and induction yields \[
\boxed{
W_t\leq\left(\frac9{10}\right)^tW_0
\quad\text{for every }(U_{1,0},U_{2,0})\in I.
}
\] The bound on total abundance controls both lineages for every
starting state in this region. Proposition 2 gives growth from the
specified larger starting abundances under the same reproductive law.

The starting abundances of both agent lineages can determine whether the
same reproductive controls contain them. Ecologists use the term
obligate mutualism when organisms depend on a partner to sustain
themselves. A saturating benefit rises with partner abundance and
approaches a limit. Hale and Valdovinos (2021) {[}59{]} review
two-species mutualism models in which populations can coexist at high
density and may need to exceed a threshold to persist. An Allee effect
is a reduction in a population's per-capita growth at low abundance. In
the mutualism setting, low partner density can also limit its growth.
For the certificate above, check that both lineages' starting abundances
lie within the certified region.

\subsubsection{6.1 Stochastic interaction
thresholds}\label{stochastic-interaction-thresholds}

Random changes to reproduction can push an unsafe population above an
abundance limit even when its expected size declines locally. Two
lineages that start at equal abundance stay equal when they experience
the same shocks, or random changes to reproduction. Write \(U_t\) for
their shared abundance and \(m(u)=r+cu/(K+u)\) for the reproductive
multiplier before a shock at abundance \(u\). The law governing shocks
determines the probability of crossing an abundance level and the
expected sum of squared abundances.

\textbf{First exit from an abundance interval under intermittent
suppression.} Göttlich and Knapp (2020) {[}60{]} model an accident as a
reduction in road capacity over a stretch of road. In the reproductive
model below, intermittent suppression reduces the reproductive
multiplier over a specified range of abundances.

Let \(0<r<1\), \(c,K>0\), and \(m(u)=r+cu/(K+u)\). Start the symmetric
interaction model at \(U_{1,0}=U_{2,0}=u\in[a,b)\), where \(0<a<b\).
Applying the same suppression to both populations keeps their abundances
equal. For \(0<d<r\), their shared abundance follows \[
U_{t+1}=\big[m(U_t)-d\,I_t\mathbf 1_{[a,b)}(U_t)\big]\,U_t,
\] where \(I_t\in\{0,1\}\) and \(\mathbb P(I_t=1\mid\mathcal F_t)\ge p\)
whenever \(U_t\in[a,b)\). Here \(\mathcal F_t\) is the history through
\(U_t\), before \(I_t\) is observed. The probability of suppression may
depend on the full observed history.

Let \(\tau=\inf\{t\ge 0:U_t\notin[a,b)\}\). If \(\alpha>0\) satisfies \[
(1-p)m(b)^\alpha+p[m(b)-d]^\alpha\le1,
\] then \[
\mathbb P(\tau<\infty,\ U_\tau\ge b)\le (u/b)^\alpha.
\]

The multiplier \(m\) increases with abundance, and the suppression rule
keeps all multipliers positive. Applying suppression more often reduces
the conditional mean of the multiplier raised to power \(\alpha\).
Before exit, that mean is therefore at most the left-hand side of the
premise. This makes \(U_{t\wedge\tau}^{\alpha}\) a nonnegative
supermartingale, meaning that its conditional expected next value is at
most its current value. The stopped process has expectation at most
\(u^\alpha\) and takes a value of at least \(b^\alpha\) on paths that
exit at or above \(b\) by time \(t\). The events of upper exit by time
\(t\) increase to the event of upper exit at any finite time, so the
same probability bound applies even when some trajectories stay in the
interval forever.

A sufficient condition for a positive exponent is \[
(1-p)\log m(b)+p\log[m(b)-d]<0.
\] At \(\alpha=0\), the conditional-moment expression equals one and has
derivative equal to the weighted mean of the log multipliers. A negative
derivative makes the expression fall below one for sufficiently small
positive \(\alpha\), so the bound can hold even when \(m(b)>1\) and the
mean multiplier exceeds one. For an exact example, take
\(r=9/10,\ c=1,\ K=b=2,\ d=21/25,\ p=2/5\). Then \(m(b)=7/5\),
\(m(b)-d=14/25\), and the endpoint mean multiplier is \(133/125>1\). The
weighted log mean is negative because
\((7/5)^3(14/25)^2=67228/78125<1\). By continuity the mean multiplier
also exceeds one at some states strictly inside the band when the
probability of suppression is exactly \(p\).

To assess upper-exit risk under intermittent suppression, verify the
conditional suppression floor \(p\) throughout the band. If the moment
condition holds for an exponent \(\alpha>0\), compare \((u/b)^\alpha\)
with the largest acceptable probability that the first exit is at or
above \(b\).

\textbf{Crossing by a deadline under Markov shocks.} When the two
populations have the same abundance, let \(m(u)=r+cu/(K+u)\), with
\(r,c,K>0\). Let \(J_t\) be an observed finite-state Markov chain. Its
transition matrix \(P\) gives the probability \(P_{ij}\) of moving from
state \(i\) to state \(j\). Conditional on the full population and shock
history, the next state has distribution \(P_{J_t,\cdot}\). Let
\(z_j>0\) and set \[
U_{t+1}=S_{J_{t+1}}(U_t),\qquad S_j(u)=z_jm(u)u.
\] For \(b>0\), let \(\psi_n(u,i)\) be the probability of reaching \(b\)
or higher at some time from now through the next \(n\) steps,
conditional on \((U_t,J_t)=(u,i)\). Then \[
\psi_0(u,i)=\mathbf{1}_{\{u\ge b\}},\qquad
\psi_{n+1}(u,i)=\mathbf{1}_{\{u\ge b\}}+
\mathbf{1}_{\{u<b\}}\sum_j P_{ij}\psi_n(S_j(u),j).
\] The first term counts a crossing when the current abundance is at or
above the threshold. Otherwise, condition on the next shock state and
average the remaining crossing probabilities using its transition
probabilities. Every shock map \(S_j\) is increasing on the nonnegative
reals. If two initial abundances follow the same shock sequence, the
larger one stays at least as large at each step. Thus \(\psi_n(u,i)\) is
nondecreasing in \(u\). For a fixed initial shock state and deadline,
the starting abundances whose crossing probability is at least a given
level form an upper set. With initial shock-state probabilities
\(\pi_i=\mathbb P(J_0=i)\), the averaged crossing probability is
\(\sum_i\pi_i \psi_n(u,i)\). This sum is also nondecreasing in \(u\), so
the same upper-set conclusion applies. For either initial-state
specification, the starting abundances whose crossing probability is at
most a specified safety ceiling form a lower set.

The deadline calculation depends on how shocks follow one another. Two
processes can assign each shock state the same probability in every
generation and still yield different sets of starting abundances whose
crossing probability reaches a specified level. The difference comes
from how the probabilities of later shocks depend on earlier shocks.
Take \(r=9/10,\ c=1/5,\ K=1,\ U_0=1,\ b=3,\ n=2\), and shock multipliers
\(z_H=2,\ z_L=1/2\). Since \(9/10\le m(u)\le11/10\), the first step is
at most \(11/5<3\), while two high shocks give \(U_2\ge(9/5)^2>3\) and
cross the threshold on the second step. A mixed pair gives
\(U_2\le(11/10)^2<3\), and two low shocks give \(U_2\le(11/20)^2<3\).
Start \(J_0\) with equal probabilities for the two states. The
persistent chain \(P=I\) and the alternating chain \(P_{HL}=P_{LH}=1\)
both have uniform marginals at every time, but the persistent chain
crosses by the second step with probability \(1/2\) and the alternating
chain crosses with probability \(0\). At the target probability \(1/4\),
the upper sets formed from the two crossing probabilities averaged over
\(J_0\) therefore disagree at \(u=1\). The fixed-state upper sets also
disagree at \(u=1\) when \(J_0=H\), since the persistent chain then
crosses with probability \(1\) and the alternating chain with
probability \(0\).

For a deadline-based assessment, identify the initial shock state or its
distribution and justify the transition probabilities. Use the resulting
crossing probability to check the stated risk ceiling for the initial
abundance.

\textbf{Squared drift and the second-moment threshold.} When the two
populations have the same abundance, let \(U_{t+1}=Z_{t+1}m(U_t)U_t\),
where \(m(u)=r+cu/(K+u)\), \(r,c,K>0\), and \(Z_{t+1}\ge0\). Assume
\(\mathbb E[Z_{t+1}\mid\mathcal F_t]=1\) and
\(\mathbb E[Z_{t+1}^2\mid\mathcal F_t]=m_2<\infty\) for every history,
where \(\mathcal F_t\) contains \(U_t\) and precedes the next shock. The
shocks may depend on past shocks as long as these conditional moments
hold. The conditional expected change in squared abundance is \[
\mathbb E[U_{t+1}^2-U_t^2\mid\mathcal F_t]
=U_t^2\big[m(U_t)^2-1\big]+U_t^2m(U_t)^2(m_2-1).
\] Squared drift is the conditional expected change in squared
abundance. The conditional second moment \(m_2\) is the expected value
of the squared shock multiplier given the current history. For
\(\theta=m_2^{-1/2}\), squared drift at positive abundance is negative
exactly when \(m(U_t)<\theta\). If \(r<\theta<r+c\), the sign changes at
one positive abundance. \[
u_2=K\,\frac{\theta-r}{r+c-\theta}.
\] If \(\theta\le r\), squared drift is positive at every positive
state. If \(\theta\ge r+c\), it is negative at every finite positive
state. These claims follow from the strict increase of \(m\) and its
endpoint values.

For \(U_0=u<b\), let \(\tau_b=\inf\{t\ge0:U_t\ge b\}\). If
\(q=m_2m(b)^2\le1\), then \[
\mathbb P(\tau_b<\infty)\le (u/b)^2.
\] The squared abundance stopped at the crossing,
\(U_{t\wedge\tau_b}^2\), is a nonnegative supermartingale and is at
least \(b^2\) on every path that crosses by time \(t\). As \(t\)
increases, these crossing events increase to \(\{\tau_b<\infty\}\). If
\(q<1\), the expected sum of squared abundances before crossing also
satisfies \[
\mathbb E\!\left[\sum_{t<\tau_b}U_t^2\right]\le\frac{u^2}{1-q}.
\] To see this, multiply the one-step squared-moment bound by
\(\mathbf{1}_{\{\tau_b>t\}}\), and use
\(\{\tau_b>t+1\}\subseteq\{\tau_b>t\}\) to obtain
\(\mathbb E[U_{t+1}^2\mathbf{1}_{\{\tau_b>t+1\}}]
\le q\,\mathbb E[U_t^2\mathbf{1}_{\{\tau_b>t\}}]\). Sum the resulting
geometric bound over a finite horizon, then let the horizon increase.
Each term is nonnegative, so the expected partial sums converge to the
expected full sum by monotone convergence.

Take \(r=4/5,\ c=2/5,\ K=1\), and a shock that takes the values in
\(\{1/2,3/2\}\) with equal conditional probability. At \(u=1/2\), the
conditional mean multiplier is \(14/15<1\), while the expected squared
multiplier is \(49/45>1\). The expected abundance therefore falls over
one generation while the expected squared abundance rises. The shock has
\(m_2=5/4\), so the reversal in conditional squared drift occurs at
\(u_2\approx0.309\), below the starting abundance \(u=1/2\). The
squared-drift crossing bound applies to levels up to \(u_2\), and the
cumulative bound applies at strictly smaller levels. A safety assessment
based on either bound must check its threshold and starting abundance
against that range.

\subsection{7. Population Control Across Allowed Sequences of
Reproductive
Regimes}\label{population-control-across-allowed-sequences-of-reproductive-regimes}

When deployed systems switch modes, patch defenses, or move workloads
between platforms, they can change the rules governing reproduction. A
population can grow as it moves through configurations that each
suppress reproduction when held fixed. Because each configuration acts
on the population left by the preceding one, a control guarantee has to
cover the population's changes along every allowed sequence of
configurations.

When interacting lineages share a deployment, their safety assessment
must account for their joint reproduction, as in Section 6, and for the
order in which reproductive rules act. The reproductive matrix \(M_t\)
maps the population at generation \(t\) to the next one. The
fixed-regime test checks whether its spectral radius satisfies

\[
\rho(M_t)<1.
\]

This test measures subcriticality under each fixed rule. To establish
contraction under switching, an evaluator must account for how the rules
act in sequence. The following two matrices show how repeated
alternation can produce growth even when each regime passes the
fixed-rule test.

\[
A=
\begin{pmatrix}
0&2\\
0&0
\end{pmatrix},
\qquad
B=
\begin{pmatrix}
0&0\\
2&0
\end{pmatrix}.
\]

Operator \(A\) moves the second population class into the first and
doubles its abundance, while \(B\) moves the first into the second and
doubles it. Alternation repeats the transfer and doubles the abundance
at each step for a population that starts in the class transferred by
the first operator.

Each operator is nilpotent, meaning that repeated application gives the
zero matrix, which happens here after two applications.

\[
\rho(A)=\rho(B)=0,
\qquad
A^2=B^2=0.
\]

Repeated application of either regime alone therefore extinguishes the
represented population after at most two applications. The alternating
product is

\[
BA=
\begin{pmatrix}
0&0\\
0&4
\end{pmatrix}.
\]

Under repeated alternation beginning with \(A\), the product's spectral
radius gives the asymptotic two-generation growth factor when the
initial population has positive abundance in the second class. Its
square root gives the per-generation factor.

\[
\rho(BA)=4,\qquad \rho(BA)^{1/2}=2>1.
\]

\[
\boxed{
\text{pointwise reproductive subcriticality}
\not\Rightarrow
\text{trajectory-uniform reproductive contraction}.
}
\]

\needspace{6\baselineskip}

The identity

\[
(BA)^n=
\begin{pmatrix}
0&0\\
0&4^n
\end{pmatrix},
\qquad n\geq1
\]

follows directly from the diagonal form of \(BA\). Thus the
alternating-growth result holds for arbitrary numbers of two-step
cycles.

\subsubsection{7.1 Robustness under strictly positive reproductive
operators}\label{robustness-under-strictly-positive-reproductive-operators}

A switching hazard can persist even when every population class
contributes some descendants to every class in each regime. The same
pattern holds across a continuous range of strictly positive matrices.
Repeated application of either fixed regime drives abundance to zero in
the limit, while alternating the two regimes produces growth from
nonzero starting populations.

\[
A_\epsilon=
\begin{pmatrix}
\epsilon&2\\
\epsilon&\epsilon
\end{pmatrix},
\qquad
B_\epsilon=
\begin{pmatrix}
\epsilon&\epsilon\\
2&\epsilon
\end{pmatrix}.
\]

The contrast holds throughout \(0<\epsilon<2-\sqrt{3}\). On precisely
this positive interval, both individual matrices have spectral radius
\(\epsilon+\sqrt{2\epsilon}<1\). Also, \(A_\epsilon\geq A\) and
\(B_\epsilon\geq B\) entrywise, meaning that each inequality holds for
the corresponding entries, so \(B_\epsilon A_\epsilon\geq BA\)
entrywise. For nonnegative matrices, increasing the entries can only
increase or preserve the spectral radius. Thus
\(\rho(B_\epsilon A_\epsilon)\geq\rho(BA)=4>1\). An evaluator must check
the allowed switching sequences even when every population class
produces descendants in every class under each rule. At
\(\epsilon=0.1\), the growth factors are as follows.

\[
\begin{aligned}
\rho(A_\epsilon)
  = \rho(B_\epsilon)
  &\approx 0.5472136 < 1,\\
\rho(B_\epsilon A_\epsilon)^{1/2}
  &\approx 2.0052487 > 1.
\end{aligned}
\]

\Needspace{4\baselineskip}

\textbf{Qualification: common versus individual symmetrization.} If
every reproductive regime contracts when tested alone, a deployment
assessment still needs a bound that holds when the regimes are combined.
Diagonal symmetrization rescales the population classes with an
invertible diagonal matrix \(D\) so that the transformed matrix
\(D^{-1}MD\) is symmetric. For the asymmetric nilpotent matrices \(A\)
and \(B\), neither matrix can be made symmetric by such a rescaling. For
a finite family, however, if a single \(D\) makes every admissible
\(D^{-1}MD\) symmetric, pointwise subcriticality is sufficient for
contraction. The common weighted norm \(\|D^{-1}x\|_2\), which measures
the Euclidean length of the rescaled state, then contracts uniformly
when \(\max_M\rho(M)<1\). This follows because each symmetric
transformed matrix has Euclidean operator norm equal to its spectral
radius. For symmetric operators, the identity \(D=I\) is a common
rescaling. Each of the strictly positive matrices \(A_\epsilon\) and
\(B_\epsilon\) can be made symmetric on its own, but the two require
incompatible scalings. The spectral radius of \(B_\epsilon A_\epsilon\)
established above gives growth under alternation despite the individual
symmetrizations. To use the common-rescaling certificate, verify that
one choice of \(D\) works for every admitted regime.

\subsubsection{7.2 The reproductive-control
hierarchy}\label{the-reproductive-control-hierarchy}

The counterexamples show why the scope of an evaluation matters. A test
can cover isolated lineages, interacting populations under one
reproductive regime, or interacting populations across all allowed
sequences of regimes. A uniform certificate can cover several of these
cases at once.

Under the nonlinear law in Section 6, a lineage's reproductive
multiplier depends on the other lineage's abundance. The isolation test
measures \(m(0)=0.75<1\) at zero partner abundance, while at partner
abundance one the multiplier rises to \(m(1)=1.125>1\).

The switching example admits strict alternation. The expanding product
on that schedule shows why the certificate must cover the admitted
products of reproductive matrices.

\[
\rho(M)<1\ \forall M\in\mathcal M
\not\Rightarrow
\text{uniform contraction of all admissible products}.
\]

When a deployment admits several reproductive regimes, one measure of
abundance can certify decline across all their allowed sequences. Assign
each population class a positive weight and use those same weights under
every allowed reproductive rule. If every rule reduces the weighted
abundance, that abundance declines in any order of application. This is
a common linear copositive Lyapunov certificate {[}43{]}. The weighted
sum is linear, gives a positive value to every nonzero population with
nonnegative abundances, and decreases at each step. Suppose a fixed
vector \(v>0\) and a constant \(0\leq q<1\) satisfy the following
condition.

\[
\boxed{
v^{\mathsf T}M\leq qv^{\mathsf T}
\quad\text{componentwise for every admissible }M.
}
\]

For a nonnegative state evolving by \(u_{t+1}=M_tu_t\), the condition
gives \(v^{\mathsf T}u_{t+1}\leq qv^{\mathsf T}u_t\), and hence
\(v^{\mathsf T}u_t\leq q^t v^{\mathsf T}u_0\). This holds for every
allowed switching sequence, including sequences that choose \(M_t\)
based on the current state. This condition suffices for contraction, and
other certificates cover some families that admit no such vector
{[}43{]}.

\textbf{A constrained-switching contraction certificate.} A minimum run
length can certify contraction for a population that switches between
\(A_{0.1}\) and \(B_{0.1}\) from Section 7.1. Require every completed
maximal run, a consecutive stretch of one regime that ends when the
regime changes, to last at least two generations.

Direct multiplication gives

\[
A_{0.1}^2=
\begin{pmatrix}
0.21&0.40\\
0.02&0.21
\end{pmatrix},
\qquad
A_{0.1}^3=
\begin{pmatrix}
0.061&0.460\\
0.023&0.061
\end{pmatrix}.
\]

Taking the maximum absolute column sum of each displayed power gives its
induced matrix \(1\)-norm: \[
\|A_{0.1}^2\|_1=0.61,
\qquad
\|A_{0.1}^3\|_1=0.521.
\] Since \(B_{0.1}=A_{0.1}^{\mathsf T}\), one has
\(\|B_{0.1}^k\|_1=\|A_{0.1}^k\|_\infty\), where the induced infinity
norm is the maximum absolute row sum. For these particular powers, the
maximum row sums also equal \(0.61\) and \(0.521\), respectively. Thus
\[
\|A_{0.1}^2\|_1=\|B_{0.1}^2\|_1=0.61,
\qquad
\|A_{0.1}^3\|_1=\|B_{0.1}^3\|_1=0.521.
\]

The bounds for two and three generations combine to cover every run of
length at least two. Proposition 3 uses them to give one exponential
bound for every allowed history, including when the observation ends
partway through a run.

\textbf{Proposition 3 (Uniform contraction under minimum-run
switching).} Use the nonnegative reproductive matrices \(A_{0.1}\) and
\(B_{0.1}\) specified in Section 7.1. Suppose every completed maximal
run in an admissible switching history has length at least two. Set \[
q=\max\{\sqrt{0.61},\sqrt[3]{0.521}\}<1,
\qquad
C=\max\{1,2.1/q\}.
\] Then every trajectory \(u_{t+1}=M_tu_t\), with
\(M_t\in\{A_{0.1},B_{0.1}\}\) and \(u_0\geq0\), satisfies \[
\boxed{\|u_t\|_1\leq Cq^t\|u_0\|_1
\quad\text{for all }t\geq0.}
\] The estimate is uniform over all admissible switching histories,
including nonperiodic ones.

\textbf{Proof of Proposition 3.} The calculated powers in this section
establish, for both \(M=A_{0.1}\) and \(M=B_{0.1}\), \[
\|M^2\|_1\leq0.61,\qquad
\|M^3\|_1\leq0.521.
\] Every integer \(k\geq2\) admits a representation \(k=2a+3b\) with
nonnegative integers \(a,b\). The norm of a matrix product is at most
the product of the norms of its factors. This property, called
submultiplicativity, and the definition of \(q\) give the following
bound. \[
\|M^k\|_1
\leq\|M^2\|_1^a\|M^3\|_1^b
\leq0.61^a0.521^b
\leq q^{2a+3b}=q^k.
\] At observation time \(t\), divide the sequence so far into completed
runs and the final run, which may still be in progress. Each completed
run has length at least two, so a completed run of length \(k\)
contributes a norm factor at most \(q^k\).

If the final run also has length at least two, submultiplicativity
bounds the complete matrix product by \(q^t\). A final run of length one
instead contributes at most \(\max\{\|A_{0.1}\|_1,\|B_{0.1}\|_1\}=2.1\);
the preceding completed runs contribute at most \(q^{t-1}\). The product
norm is therefore at most \((2.1/q)q^t\) in this case, and at \(t=0\)
the identity product has norm one. Together the cases bound the induced
norm of every admissible product by \(Cq^t\), so applying the product to
\(u_0\) gives the asserted population bound. \(\square\)

Proposition 3 guarantees uniform exponential decay for the specified
matrices whenever the switching rules enforce the minimum run length.
The prefactor \(C\) allows temporary growth, including during the first
generation of a run. A history with one generation of \(A_{0.1}\)
followed by \(B_{0.1}\) in every later generation also converges to
zero, since \(\rho(B_{0.1})<1\). If descendants can inherit changes to
the matrices or to the mechanism that enforces the run length, the
safety argument must cover those changes. The minimum run length lets us
bound the combined effect of the matrices across a history. A
subcriticality test for each matrix gives information about repeated use
of that matrix alone.

\subsubsection{7.3 Relationship to established
results}\label{relationship-to-established-results}

A control that keeps an agent population in decline under one
reproductive regime can allow growth when the regime changes.
Switched-systems theory studies processes whose update rules change over
time. The joint spectral radius measures the largest asymptotic growth
rate of products from a matrix family. Athanasopoulos and Lazar (2014)
{[}15{]} study stability of switched linear systems whose admissible
regime sequences are specified by a graph. Liberzon (2003) {[}31{]}
treats stability under dwell-time restrictions, which require a regime
to remain active for a minimum time before a switch. Hespanha and Morse
(1999) {[}32{]} give stability conditions under an average dwell-time
restriction, which bounds the number of switches over time. Section 7.2
uses a discrete-time minimum dwell condition: every completed run of a
regime lasts at least two generations. To apply either certificate in
Section 7.2, specify the allowed sequences of matrices and the
interacting population states to which those matrices apply (Section~6).

\needspace{11\baselineskip}
\begingroup\raggedright

\subsection{8. Population Dynamics and Continuing Production as Distinct
Control
Dimensions}\label{population-dynamics-and-continuing-production-as-distinct-control-dimensions}

\endgroup

A reproductive-control certificate for an open multi-agent system must
cover how agents already present reproduce and how many new unauthorized
descendants enter the population. A bound on an existing population can
apply to an isolated lineage, to interacting lineages, or to a
population whose reproductive rules change over time. The examples in
Sections 6 and 7 show why extending a bound from one setting to another
requires assumptions about the added interactions or changes in rules.

A second calculation tracks how many new unauthorized descendants enter
the population and how much burden those arrivals add. In the
source--sink model of Section 5, a permitted population can keep
producing unsafe descendants even when those descendants decline in
isolation. Section 5.4 extends the accounting to populations whose past
determines which opportunities occur, how likely each opportunity is to
produce an unauthorized descendant, and how much that event adds to the
burden.

A certificate that uses population size to bound reproductive
opportunities needs a growth bound that covers admitted interactions and
changes of regime. The burden calculation also needs to account for
those opportunities and the consequences of the descendants they
produce. A lineage's reproduction number or its current fitness can
capture both parts of the problem only when additional assumptions
connect that quantity to production and burden.

Proposition 4 bounds the expected total number of unauthorized
descendants produced directly by authorized parents. It uses the same
weighted count of authorized agents under every allowed reproductive
rule, including rules that descendants acquire through inherited
changes. Given the history so far, reproduction must reduce that count
in expectation, with a separate allowance for outside inputs whose
weighted bounds have a finite sum. At each generation, a fixed multiple
of the weighted count must bound the conditional expected number of
unauthorized descendants produced directly by authorized parents.

\textbf{Proposition 4 (Uniform hereditary reproductive-risk bound under
arbitrary switching).} Let \((\mathcal F_t)_{t\geq0}\) be a filtration,
meaning an increasing record of the information available as the
population evolves. It includes interactions, selection, inherited
changes to reproductive laws, and changes of reproductive regime. Let
\(H_t\in\mathbb R_{\geq0}^{m}\) be the authorized-population state at
generation \(t\). The information in \(\mathcal F_t\) determines this
state, so it is adapted to the filtration. Assume
\(\mathbb E[v^{\mathsf T} H_0]<\infty\) for the positive weighting
specified below. Let \(M_t\) be the nonnegative reproductive operator
and \(f_t\) the nonnegative external input. Both are
\(\mathcal F_t\)-measurable, meaning that the information available at
generation \(t\) determines their values.

\needspace{6\baselineskip}

The joint process tracks the authorized population \(H_t\), reproductive
operator \(M_t\), input \(f_t\), and direct unauthorized production
\(Y_t\) across generations. Assume the following conditions hold at
every generation of every allowed history of that process.

\begin{enumerate}
\def\labelenumi{\arabic{enumi}.}
\item
  \textbf{Uniform hereditary contraction.} There are fixed
  \(v\in\mathbb R_{>0}^{m}\) and \(q\in[0,1)\) such that
  \(v^{\mathsf T} M_t\leq qv^{\mathsf T}\) componentwise for every
  admissible \(M_t\), including operators reachable through hereditary
  modification and switching.
\item
  \textbf{Conditional population dynamics and summable input.} The
  componentwise conditional expectation satisfies \[
  \mathbb E[H_{t+1}\mid\mathcal F_t]
  \leq M_tH_t+f_t.
  \] There are deterministic nonnegative numbers \(s_t\) satisfying
  \(v^{\mathsf T} f_t\leq s_t\) almost surely and
  \(\sum_{t=0}^{\infty}s_t<\infty\).
\item
  \textbf{Bounded expected unauthorized production.} Let \(Y_t\) be the
  nonnegative integer count of direct unauthorized descendants produced
  by authorized parents at generation \(t\) (escape events). For a fixed
  finite \(\ell\geq0\), \[
  \mathbb E[Y_t\mid\mathcal F_t]
  \leq \ell\,v^{\mathsf T} H_t.
  \]
\end{enumerate}

The total direct escape count \(N_\infty=\sum_{t=0}^{\infty}Y_t\) is
nonnegative and may be infinite. The proposition bounds its expectation
as follows. \[
\boxed{
\mathbb E[N_\infty]
\leq
\frac{\ell}{1-q}
\left(
\mathbb E[v^{\mathsf T} H_0]
+\sum_{t=0}^{\infty}s_t
\right).
}
\] In particular, \[
\boxed{
\mathbb P(N_\infty\geq1)
\leq
\min\left\{
1,\,
\frac{\ell}{1-q}
\left(
\mathbb E[v^{\mathsf T} H_0]
+\sum_{t=0}^{\infty}s_t
\right)
\right\}.
}
\]

\textbf{Proof of Proposition 4.} Set \(W_t=v^{\mathsf T} H_t\). The
assumptions and nonnegativity imply \[
\begin{aligned}
\mathbb E[W_{t+1}\mid\mathcal F_t]
&\leq v^{\mathsf T} M_tH_t+v^{\mathsf T} f_t\\
&\leq qW_t+s_t.
\end{aligned}
\] Taking expectations and iterating yields \[
\mathbb E[W_t]
\leq q^t\mathbb E[W_0]
+\sum_{k=0}^{t-1}q^{t-1-k}s_k.
\] Each term is nonnegative, so summing over \(t\) allows us to change
the order of the sums. \[
\sum_{t=0}^{\infty}\mathbb E[W_t]
\leq
\frac{
\mathbb E[W_0]+\sum_{k=0}^{\infty}s_k
}{1-q}.
\] The conditional bound on unauthorized production gives
\(\mathbb E[Y_t]\leq\ell\,\mathbb E[W_t]\). By Tonelli's theorem, the
expectation of a nonnegative sum equals the sum of the expectations.
Thus \[
\mathbb E[N_\infty]
=\sum_{t=0}^{\infty}\mathbb E[Y_t]
\leq
\frac{\ell}{1-q}
\left(
\mathbb E[W_0]+\sum_{k=0}^{\infty}s_k
\right).
\] Finally, because \(Y_t\) are nonnegative integer counts,
\(\mathbf 1_{\{N_\infty\geq1\}}\leq N_\infty\). Taking expectations and
using the elementary probability ceiling of one establishes the second
bound. \(\square\)

For every finite horizon \(T\), the conditional direct-escape bound
independently implies \[
\mathbb E[N_T]
\leq
\ell\sum_{t=0}^{T-1}\mathbb E[v^{\mathsf T} H_t],
\qquad
N_T=\sum_{t=0}^{T-1}Y_t.
\] Over a finite horizon, the direct-escape bound and well-defined
nonnegative expectations suffice. To bound escapes over all generations
by this argument, use \(q<1\) and a finite sum of the external input
bounds to bound the total weighted population in expectation.

\textbf{Scope and relation to earlier certificates.} Fornasini and
Valcher {[}43{]} give stability criteria for positive switched systems
using a common linear weighting. Proposition 4 uses a weighting that
contracts under every admitted reproductive operator and combines it
with conditional population dynamics and a bound on continuing
production. The bound applies under each admitted law for regime changes
and inherited modifications to the reproductive operators. To obtain a
single numerical ceiling across these laws, use common contraction and
input bounds, a common escape-rate bound per unit of weighted authorized
population, and a uniform upper bound on the initial expected weighted
count. The result covers interactions that satisfy the conditional
population inequality. The proof multiplies the bound on escapes per
unit of weighted authorized population by a bound on the sum of that
weighted population across generations.

\textbf{Inductively certifying the hereditary contraction premise.} A
class of operators closed under inherited changes can certify
contraction for all reachable operators at once. For fixed \(v>0\) and
\(0\leq q<1\), define the class of matrices for which the next weighted
count is at most \(q\) times the current count. \[
\mathcal K_{v,q}
=
\{M\in\mathbb R_{\geq0}^{m\times m}:
v^{\mathsf T} M\leq qv^{\mathsf T}\}.
\] Suppose a class \(\mathcal K\) contains the deployed operator and
satisfies \(\mathcal K\subseteq\mathcal K_{v,q}\). Require every allowed
one-step hereditary modification of an operator in \(\mathcal K\) to
produce another operator in that class. Induction on the number of
modifications then places every hereditarily reachable operator in the
class, as in Proposition 6. Any switching sequence among those operators
is covered as long as the actual operators remain in the class.

For any finite sequence \(M_0,\ldots,M_{t-1}\) in \(\mathcal K_{v,q}\),
nonnegativity yields \[
v^{\mathsf T} M_{t-1}\cdots M_0
\leq q^t v^{\mathsf T}.
\] After \(t\) applications, the weighted population is at most \(q^t\)
times its initial value. Since \(0\leq q<1\), that bound also gives the
defining bound with \(q\) for every nonempty finite product, so each
such product belongs to \(\mathcal K_{v,q}\). The bound
\(v^{\mathsf T} M\leq qv^{\mathsf T}\) for each admissible matrix
establishes the contraction premise of Proposition 4. Applying the
proposition also requires checking the conditional population dynamics,
the external input, and the unauthorized-production bound.

Proposition 4 bounds the expected count of unauthorized descendants
produced directly by authorized parents. To bound how many unauthorized
agents are present at a given time, how many lineages survive, or how
much harm people experience, specify and bound each outcome separately.
If unauthorized agents affect the authorized population, include that
feedback in the joint model and verify the conditional population
inequality. The finite sum of external input bounds matters because
persistent input can yield an unbounded expected escape count even when
the reproductive operators contract.

The infinite-horizon hypotheses give
\(\sum_{t=0}^{\infty}\mathbb E[v^{\mathsf T} H_t]<\infty\), so
\(\mathbb E[v^{\mathsf T} H_t]\to0\). In particular, they cover only
populations whose expected weighted mass vanishes over time, ruling out
a persistent positive mass or the expanding permitted source in Section
5.1. When \(\ell=0\), the direct-escape assumption gives \(Y_t=0\)
almost surely for each \(t\), even if the authorized population persists
or grows. The countable union of the events \(Y_t>0\) has probability
zero. This controls escapes from authorized parents, and bounding
reproduction by unauthorized agents requires a bound on their own
dynamics.

\subsection{9. Safety Must Be Closed Over Evolutionarily Reachable
Populations}\label{safety-must-be-closed-over-evolutionarily-reachable-populations}

As agents create successors, new members can take the population into
states omitted from its original assessment. Descendants can also
inherit a modification to their reproductive rules and pass it on. A
sequence of such modifications can produce rules beyond the reach of any
single step, so a safety claim must cover the resulting states and
rules. An inductive certificate identifies a safe set containing the
initial state and verifies that every allowed transition from within the
set stays within it. Sections 9.5--9.6 specify the additional conditions
needed to use reproductive control to establish guarantees about human
interests.

\subsubsection{9.1 Reachability and strict reproductive
closure}\label{reachability-and-strict-reproductive-closure}

Let \(\mathcal E\) denote the specified class of admissible evolutionary
transitions on joint population-environment states. A transition
admitted by \(\mathcal E\) may include reproduction, heritable
variation, inheritance, interaction, selection, migration, or
environmental change to the extent that each is included in the process
being certified. Starting from an initial model, system, or population
\(M\), define \(\mathcal R_{\mathcal E}(M)\) as the set of joint
population-environment states reachable from the specified initial state
associated with \(M\) by any finite sequence of transitions admitted by
\(\mathcal E\), including the initial state itself.

For a specified safety property \(Q\), let \(\mathcal A(M)\) denote the
set of joint population-environment states over which the assessment
applied to \(M\) establishes \(Q\) or verifies its required bound.

\needspace{8\baselineskip}

If the assessment is intended, by itself, to establish \(Q\) uniformly
over the process specified by \(\mathcal E\), a necessary coverage
condition is

\[
\boxed{
\mathcal R_{\mathcal E}(M)\subseteq\mathcal A(M)
}
\]

The assessment must verify \(Q\), or its required bound, for every state
in its domain. To show coverage, place the exact reachable set inside
that domain or use an overapproximation, a set guaranteed to contain
every reachable state. For this coverage argument to apply to a
deployment, \(\mathcal E\) must include the deployment's evolutionary
transitions relevant to \(Q\).

\textbf{Applying standard forward-invariance reasoning to evolutionary
reachability.} A set is forward invariant when every allowed next state
from inside it also lies inside it. Zhu and Wang's (2026) authorization
framework provides an example: it establishes safety invariants over
reachable states of a registered transition system {[}28{]}. Proposition
5 uses forward invariance to establish coverage of the states an
evolutionary process can reach.

\textbf{Proposition 5 (Inductive evolutionary-reachability coverage).}
Let \(x_0\) be the specified initial joint population-environment state
associated with \(M\) in the definition of
\(\mathcal R_{\mathcal E}(M)\), and let \(\mathcal T_{\mathcal E}(x)\)
be the complete set of successors admitted by the declared evolutionary
transition model. Suppose there exists a set \(I\) such that \[
x_0\in I,\qquad
\mathcal T_{\mathcal E}(x)\subseteq I\quad\forall x\in I,
\qquad I\subseteq\mathcal A(M).
\] Then the set of states reachable by any finite sequence of admitted
transitions satisfies \[
\boxed{\mathcal R_{\mathcal E}(M)\subseteq I\subseteq\mathcal A(M).}
\] For a stochastic model, suppose \(x_0\in I\subseteq\mathcal A(M)\)
and every transition kernel assigns probability one to \(I\) from every
state in \(I\). Then, with probability one, the trajectory remains in
\(I\) at every finite generation, and the assessed property \(Q\) holds
at every such generation.

\textbf{Proof of Proposition 5.} Let \(S_k\) denote the set of states
reachable in exactly \(k\) admitted transitions, with \(S_0=\{x_0\}\).
The initial-state assumption gives \(S_0\subseteq I\). If
\(S_k\subseteq I\), then every successor of every state in \(S_k\)
belongs to \(I\) by the successor condition, so \(S_{k+1}\subseteq I\).
Induction therefore establishes \(S_k\subseteq I\) for every finite
\(k\). Taking the union over all finite path lengths gives
\(\mathcal R_{\mathcal E}(M)\subseteq I\), and the assumed inclusion
\(I\subseteq\mathcal A(M)\) completes the deterministic or
nondeterministic claim.

For stochastic transitions, suppose the conditional probability of
remaining inside \(I\) is one whenever the current state is in \(I\).
The initial state is inside \(I\), so induction gives probability one of
remaining there through each fixed finite horizon. The union of the
countably many finite-time exit events consequently has probability
zero. Thus the trajectory stays in \(I\) at every finite generation
almost surely. Since \(I\subseteq\mathcal A(M)\), \(Q\) holds at every
finite generation on that event. \(\square\)

A deterministic rule gives one next state, so the second condition
checks that this state remains in \(I\). A nondeterministic rule lists
several possible next states and requires every one to remain in \(I\).
For a stochastic rule, the transition kernel gives the probabilities of
next states conditional on the current state. If the kernel assigns
probability one to \(I\) from each state in \(I\), coverage holds almost
surely. Through any finite horizon \(T\), failure of \(Q\) at any
generation implies an exit from \(I\) by \(T\), since
\(I\subseteq\mathcal A(M)\). Thus
\(\mathbb P(\exists\,t\leq T:\ Q\text{ fails at }X_t)\leq\mathbb P(\exists\,t\leq T:\ X_t\notin I)\),
where \(X_t\) is the joint state.

To use the certificate, specify how to tell whether a state belongs to
\(I\). Check the initial state, then show that every allowed next state
from within \(I\) also belongs to \(I\).

For example, take the finite state space \(X=\{0,1,2,3\}\), initial
state \(x_0=0\), and admissible successor sets \[
\mathcal T_{\mathcal E}(0)=\{0,1\},\quad
\mathcal T_{\mathcal E}(1)=\{1,2\},\quad
\mathcal T_{\mathcal E}(2)=\{0,2\},\quad
\mathcal T_{\mathcal E}(3)=\{3\}.
\]

The set \(I=\{0,1,2\}\) contains \(x_0\) and every successor of each of
its members, so all modeled trajectories from \(x_0\) stay in \(I\).
Establishing \(Q\) throughout \(I\) therefore certifies it for those
trajectories and leaves the unreachable state 3 outside the assessment.
Applying the same argument to a deployment requires the modeled
successor sets to include every real successor relevant to the claim.

For an authorized state set \(H\) and a deterministic reproductive
operator \(R\), strict prevention of unauthorized production means that
every authorized input produces an authorized output, equivalently
\(R(H)\subseteq H\).

If each realized operator \(R_t\) preserves \(H\), then every finite
composition \(R_t\circ\cdots\circ R_1\) preserves \(H\) by induction.
The conclusion applies as long as every operator the population uses
belongs to the class that preserves \(H\).

\needspace{10\baselineskip}

A reproductive operator can preserve authorization today and still have
a hereditarily reachable successor that violates it. Write
\(R\leadsto R'\) when one allowed inherited change transforms
reproductive operator \(R\) into \(R'\). To see why testing immediate
changes alone can miss a failure, take \(X=\{0,1,2\}\), with authorized
set \(H=\{0,1\}\) and unauthorized set
\(H^{\mathrm c}=X\mathbin{\text{\textbackslash}}H=\{2\}\). Encode the
deterministic operator \(R:X\to X\) by listing its outputs as the three
digits \(R(0)R(1)R(2)\).

\[
\boxed{
012\leadsto112\leadsto122.
}
\]

The admissible heritable-modification relation in this example contains
exactly the displayed edges, with each modification changing the output
at one input state. The deployed operator \(012\) maps \(H\) to
\(\{0,1\}\), and its only immediate mutant \(112\) maps \(H\) to
\(\{1\}\). Both preserve \(H\). Their transitively reachable successor
\(122\), however, maps \(H\) to \(\{1,2\}\) and violates strict
reproductive closure.

To certify reproductive closure, cover every rule that can arise after a
finite sequence of inherited changes.

Proposition 6 gives a sufficient inductive certificate for deterministic
reproductive operators and stochastic kernels.

\textbf{Proposition 6 (Hereditary preservation of strict reproductive
closure).} Let \(H\) be the authorized state set, and define \[
\mathcal C_H^{\mathrm{det}}
=
\{R:R(H)\subseteq H\}.
\] Let \(R_{\mathrm{dep}}\) be the deployed deterministic reproductive
operator and let \(\leadsto\) be the admissible heritable-modification
relation on such operators. Suppose there exists a declared operator
class \(\mathcal K_{\mathrm{det}}\) such that \[
R_{\mathrm{dep}}\in\mathcal K_{\mathrm{det}}
\subseteq\mathcal C_H^{\mathrm{det}},
\] and \[
R\in\mathcal K_{\mathrm{det}},
\quad R\leadsto R'
\quad\Longrightarrow\quad
R'\in\mathcal K_{\mathrm{det}}.
\] Then every operator reachable from \(R_{\mathrm{dep}}\) through
finitely many admitted hereditary modifications belongs to
\(\mathcal K_{\mathrm{det}}\) and preserves \(H\). Every finite
composition of these realized operators also maps \(H\) into \(H\).

For stochastic reproductive kernels, define \[
\mathcal C_H^{\mathrm{stoch}}
=
\{K:K(x,H)=1\ \text{for every }x\in H\}.
\] The analogous assertion holds if a declared kernel class
\(\mathcal K_{\mathrm{stoch}}\) contains the deployed kernel, lies
within \(\mathcal C_H^{\mathrm{stoch}}\), and is closed under every
admitted one-step hereditary modification from any member of
\(\mathcal K_{\mathrm{stoch}}\). Every finite composition of realized
kernels then preserves \(H\) with probability one.

\textbf{Proof of Proposition 6.} The deployed deterministic operator
belongs to \(\mathcal K_{\mathrm{det}}\) by hypothesis. Each admitted
modification of an operator in the class stays in the class, so
induction on modification depth covers every reachable operator. Since
\(\mathcal K_{\mathrm{det}}\subseteq
\mathcal C_H^{\mathrm{det}}\), every reachable operator preserves \(H\).
Induction on reproduction steps then shows that every finite composition
maps \(H\) into \(H\).

The same modification-depth induction applies to
\(\mathcal K_{\mathrm{stoch}}\). For a distribution \(\mu\) supported on
\(H\), every kernel in this class satisfies \[
\int_H K(x,H)\,\mu(dx)=1.
\] Thus each application preserves support on \(H\), and finite kernel
compositions preserve authorization with probability one. \(\square\)

To verify that a class of reproductive rules is closed under inherited
changes, check that every allowed one-step change from a rule in the
class produces another rule in the class. To establish which population
states can arise or how much burden they can create, apply the
corresponding population-level certificate.

The same failure occurs when the reproductive rule is stochastic. Keep
\(X=\{0,1,2\}\), \(H=\{0,1\}\), and \(H^{\mathrm c}=\{2\}\). In the
following kernels, each row gives the probabilities of the next state
from the state indexing that row, so the entries in each row sum to one.

\[
\begin{aligned}
K_0 &=
\begin{pmatrix}
1&0&0\\
0&1&0\\
0&0&1
\end{pmatrix},
&
K_1 &=
\begin{pmatrix}
0&1&0\\
0&1&0\\
0&0&1
\end{pmatrix},\\[6pt]
K_2 &=
\begin{pmatrix}
0&1&0\\
0&19/20&1/20\\
0&0&1
\end{pmatrix}.
\end{aligned}
\]

The admissible hereditary modifications are precisely
\(K_0\leadsto K_1\) and \(K_1\leadsto K_2\). Each step changes one row.
Directly,

\[
\max_{x\in H}K_0(x,H^{\mathrm c})
=
\max_{x\in H}K_1(x,H^{\mathrm c})
=0,
\qquad
\max_{x\in H}K_2(x,H^{\mathrm c})=\frac1{20}.
\]

Consequently, certification of the deployed kernel and its only
immediate mutant passes, but certification of the complete hereditarily
reachable class fails.

\subsubsection{9.2 Evolutionary durability of bounded-leakage burden
certificates}\label{evolutionary-durability-of-bounded-leakage-burden-certificates}

Strict reproductive closure requires zero leakage from every authorized
state under each admitted reproductive kernel. Proposition 1 gives
expected burden from the baseline and the signed contributions
\(A_i p_i D_i\). Its upper bound uses the positive parts \(D_i^+\). A
certificate that must hold after inherited changes must bound the
resulting budget across those changes.

Propositions 7 and 8 apply Proposition 1 uniformly over two different
classes of process laws. Proposition 7 checks each reachable
reproductive law held fixed over the horizon. If the reproductive law
can change during the horizon, Proposition 8 checks each admitted
probability law for the complete history. Consider a finite horizon
\(T\), an initial reproductive law \(L_0\), and a specified admissible
heritable-modification relation \(\leadsto\) on reproductive laws. Let
\(\mathscr L_{\mathrm{reach}}(L_0)\) contain every law reachable by a
finite admissible modification sequence, including \(L_0\). For each
\(L\) in this set, consider the complete evolutionary population process
admitted under that law, including the histories generated by
interaction and selection. Suppose its burden and opportunity variables
satisfy the history-completeness, predictability, and integrability
assumptions of Section 5.4. Write \(\mathbb E_L\) for expectation under
the corresponding population process; the burden baseline \(B_{0,L}\)
and marginal quantities may depend on \(L\).

\textbf{Uniform expected-burden certificate.} Each reproductive law that
descendants can inherit defines a population process when held fixed
throughout the horizon. The processes can differ in which reproductive
opportunities occur, how likely those opportunities are to produce
unauthorized descendants, and how much burden each escape adds.
Proposition 7 applies Proposition 1 separately to each fixed law and
takes the supremum of the resulting bounds.

\textbf{Proposition 7 (Uniform hereditary fixed-law burden
certificate).} Consider the hereditarily reachable law class
\(\mathscr L_{\mathrm{reach}}(L_0)\) defined above. For each reachable
law \(L\), let its fixed-law population process satisfy the
finite-horizon history-completeness, predictability, and integrability
assumptions of Proposition 1. Suppose valid predictable upper bounds
\(p_{i,L}\leq\epsilon_{i,L}\) and \(D_{i,L}^{+}\leq d_{i,L}\) hold, and
\[
\sup_{L\in\mathscr L_{\mathrm{reach}}(L_0)}
\left\{
B_{0,L}+
\mathbb E_L\left[
\sum_{i=1}^{n_L}
A_{i,L}\epsilon_{i,L}d_{i,L}
\right]\right\}
\leq G.
\] Then \[
\boxed{
\sup_{L\in\mathscr L_{\mathrm{reach}}(L_0)}
\mathbb E_L[B_{T,L}]
\leq G.
}
\]

\textbf{Proof of Proposition 7.} Fix any
\(L\in\mathscr L_{\mathrm{reach}}(L_0)\). Proposition 1 applies to the
probability law of the complete fixed-\(L\) population process, giving
\[
\mathbb E_L[B_{T,L}]
\leq B_{0,L}
+\mathbb E_L\left[
\sum_{i=1}^{n_L}
A_{i,L}p_{i,L}D_{i,L}^{+}
\right].
\] The stipulated predictable upper bounds and nonnegativity of the
opportunity indicators imply \[
\mathbb E_L[B_{T,L}]
\leq B_{0,L}
+\mathbb E_L\left[
\sum_{i=1}^{n_L}
A_{i,L}\epsilon_{i,L}d_{i,L}
\right]
\leq G.
\] Taking the supremum over all hereditarily reachable fixed laws proves
the claim. \(\square\)

Proposition 7 bounds expected burden under each reachable law when that
law governs the population throughout the horizon. For a population
whose law can change during the horizon, use the joint-process
certificate below because alternating regimes can produce growth even
when each regime is subcritical on its own (Section 7). A change in law
can also change reproductive opportunities, conditional leakage
probabilities, and the consequences of each escape.

To certify histories in which the law changes, include those changes and
their effects in the population model. The joint history of the
population and its laws must contain the information needed to determine
each marginal burden and the quantities known before each opportunity.
The same budget must bound expected burden under every admitted
probability law for the complete history. One way to establish these
conditions is to find an inductive invariant and prove that the bounds
hold under every allowed transition within it.

\textbf{Joint-process hereditary expected-burden certificate.} A
reproductive law \(L\) specifies how the whole population changes
through reproduction, interaction, selection, and production of
descendants. It can describe those changes at the population level or
build them from rules for individuals. Let \(\Pi_T\) be the allowed
probability laws for complete histories of the population, its
reproductive law, and its environment through horizon \(T\), including
inherited changes and switches between regimes.

For every \(\pi\in\Pi_T\), suppose a common finite index
\(i=1,\ldots,n\) enumerates potential reproductive opportunities. Set
\(A_i=0\) whenever opportunity \(i\) is absent. Assume the
history-completeness, predictability, and integrability conditions of
Section 5.4 under every \(\pi\), with baseline burden \(B_{0,\pi}\),
conditional unauthorized-production probabilities \(p_{i,\pi}\), and
predictable history-dependent marginal burdens \(D_{i,\pi}\).

Proposition 8 applies Proposition 1 under each allowed law for the
complete joint history. Its first bound uses a common budget on expected
contributions, and its second uses separate ceilings on baseline burden,
conditional escape probability, positive marginal burden, and
opportunity probability under each law. Conditioning on the preceding
population composition, selection, inherited modifications, and
switching retains their effects on opportunity occurrence, escape
probability, and marginal burden, including differences in leakage
caused by selection among variants.

The common finite index in Proposition 8 can represent a variable number
of opportunities up to \(n\) by treating unused indices as absent. For a
countable index, suppose that under each admitted law the finite
prefixes share a baseline and agree on the marginal contribution of
every opportunity they have in common, as in Section 5.4. The
positive-contribution bound then uses the limit of the finite-prefix
bounds: if
\(B_{0,\pi}+\sum_{i\geq1}\mathbb E_\pi[A_ip_{i,\pi}D_{i,\pi}^{+}]\leq G\)
for every \(\pi\), the finite-prefix expected burdens are uniformly at
most \(G\). If those prefix burdens converge in \(L^1\) to the specified
burden under each \(\pi\), their expected limits are also at most \(G\).

\textbf{Proposition 8 (Joint-process hereditary burden certificate).}
Let \(\Pi_T\) be the declared class of complete joint
population--law--environment probability laws through a finite horizon,
including admitted selection, hereditary modification, and switching.
For each \(\pi\in\Pi_T\), \(\mathbb E_\pi[B_T]\) denotes expected burden
under law \(\pi\), and \(B_{0,\pi}\) denotes the baseline for that law.
Assume the history-completeness, predictability, and integrability
conditions of Proposition 1 under every \(\pi\in\Pi_T\), with a common
finite potential-opportunity index. If \[
\sup_{\pi\in\Pi_T}
\left\{
B_{0,\pi}+
\mathbb E_\pi\left[
\sum_{i=1}^{n}A_i p_{i,\pi}D_{i,\pi}^{+}
\right]\right\}
\leq G,
\] then \[
\boxed{
\sup_{\pi\in\Pi_T}\mathbb E_\pi[B_T]\leq G.
}
\] Separate ceilings on each contribution give another sufficient
condition. Suppose the deterministic nonnegative bounds
\(B_{0,\pi}\leq b_0\), \(p_{i,\pi}\leq\epsilon_i\),
\(D_{i,\pi}^{+}\leq d_i\), and \(\mathbb E_\pi[A_i]\leq\alpha_i\) hold
uniformly for all \(\pi\in\Pi_T\). Then the expected-burden conclusion
also follows if \[
b_0+\sum_{i=1}^{n}\alpha_i\epsilon_i d_i\leq G.
\]

\textbf{Proof of Proposition 8.} Fix an arbitrary admissible complete
joint-process law \(\pi\). Proposition 1 applies under \(\pi\),
including histories in which conditional escape probabilities and
marginal consequences change after earlier selection or hereditary
modifications. Its one-sided bound yields \[
\mathbb E_\pi[B_T]
\leq B_{0,\pi}
+\mathbb E_\pi\left[
\sum_{i=1}^{n}A_i p_{i,\pi}D_{i,\pi}^{+}
\right].
\] The assumed uniform contribution budget bounds the right-hand side by
\(G\). Taking the supremum over \(\pi\in\Pi_T\) proves the first
certificate.

For the separately bounded version, nonnegativity and the deterministic
bounds give \[
\begin{aligned}
\mathbb E_\pi[B_T]
&\leq b_0+
\sum_{i=1}^{n}\epsilon_i d_i
\mathbb E_\pi[A_i]\\
&\leq b_0+\sum_{i=1}^{n}
\alpha_i\epsilon_i d_i
\leq G.
\end{aligned}
\] The inequalities hold under every admitted joint-process law, hence
uniformly over \(\Pi_T\). \(\square\)

\textbf{Corollary 1 (Compositional hereditary reproductive-burden
certificate).} Consider a declared class of complete joint
population--law--environment processes satisfying the conditional
population-dynamics inequality and common weighted-contraction
assumptions of Proposition 4, with common \(v>0\), \(0\leq q<1\), and
deterministic nonnegative input ceilings \(s_t\). The horizon-\(T\)
bound uses only \(s_0,\ldots,s_{T-1}\).

Let \(H_t\) be the permitted-population vector from Proposition 4, write
\(W_t=v^{\mathsf T} H_t\), and assume \(\mathbb E[W_0]<\infty\).

For each finite horizon \(T\), suppose a fixed finite index
\(i=1,\ldots,n(T)\) enumerates the potential unauthorized-production
opportunities across the admitted histories and that these opportunities
satisfy Proposition 1. The index \(i\) labels opportunities in
Proposition 1, and \(t\) labels generations in Proposition 4. Let
\(O_t\) denote the number of realized opportunities at generation \(t\),
so that the complete opportunity index through \(T\) obeys \[
\sum_{i=1}^{n(T)}A_i=\sum_{t=0}^{T-1}O_t.
\] Suppose, uniformly over every admitted complete joint-process law,
that \[
\mathbb E[O_t]\leq\mathbb E[W_t],
\qquad p_i\leq\epsilon,
\qquad D_i^+\leq d,
\qquad B_0\leq b_0,
\] where \(\epsilon,d,b_0\) are finite nonnegative deterministic bounds.
Here \(B_0=B_T(0,\ldots,0)\), and the same \(b_0\) bounds this baseline
for every finite horizon \(T\) and every admitted complete joint-process
law. The probability and marginal consequence bounds have the
predictability required by Proposition 1.

Then, for each admitted complete joint-process law, \[
\boxed{
\mathbb E[B_T]\leq b_0+
\epsilon d
\sum_{t=0}^{T-1}
\left(
q^t\mathbb E[W_0]
+\sum_{k=0}^{t-1}q^{t-1-k}s_k
\right).
}
\]

If \(\sum_{t=0}^{\infty}s_t<\infty\), the right-hand side is bounded
uniformly in \(T\) by \[
\boxed{
b_0+\frac{\epsilon d}{1-q}
\left(
\mathbb E[W_0]+\sum_{t=0}^{\infty}s_t
\right).
}
\] For each admitted law, use its own initial expected weighted
population, \(\mathbb E[W_0]\). To give one numerical ceiling for the
whole class, also establish a finite upper bound that covers those
initial expectations under every admitted law.

For the infinite-horizon conclusion, fix one admitted law on the full
history and apply the finite-horizon bounds under that law. If the
burdens \(B_T\) are nonnegative and nondecreasing in \(T\) under this
law, define \(B_\infty=\lim_{T\to\infty}B_T\). When each finite-horizon
burden is a conditional mean given its binary escape history, the
monotonicity condition applies to those conditional means. The same
bound then applies to \(\mathbb E[B_\infty]\).

\textbf{Proof of Corollary 1.} Fix any admitted complete joint-process
law and finite horizon. Proposition 1 and the uniform opportunity-level
bounds give \[
\begin{aligned}
\mathbb E[B_T]
&\leq b_0+\epsilon d\,
\mathbb E\left[\sum_{i=1}^{n(T)}A_i\right]\\
&=b_0+\epsilon d\sum_{t=0}^{T-1}\mathbb E[O_t]\\
&\leq b_0+\epsilon d\sum_{t=0}^{T-1}\mathbb E[W_t].
\end{aligned}
\] The conditional population recurrence used in Proposition 4 gives \[
\mathbb E[W_t]\leq
q^t\mathbb E[W_0]
+\sum_{k=0}^{t-1}q^{t-1-k}s_k.
\] Substitution establishes the finite-horizon inequality. For \(q<1\)
and summable input, summing the nonnegative recurrence bounds gives the
stated uniform ceiling. Under the additional monotonicity and
nonnegativity assumptions, the monotone convergence theorem gives the
infinite-horizon claim. The constants \(b_0,\epsilon,d,q\), and \(s_t\)
are common to the admitted process class; a single class-wide numerical
bound additionally requires \(\sup_{\pi}\mathbb E_{\pi}[W_0]<\infty\).
\(\square\)

\needspace{3\baselineskip}

To apply the corollary, show that the expected weighted population
bounds the expected number of opportunities in each generation. This is
a separate check from showing that the weighted population contracts.
Interactions can change how much burden each escape adds. The corollary
accounts for those effects through \(D_i^+\) when the full encoded
history satisfies Proposition 1.

\textbf{Worked example: authorizing successors in two pools.} A service
runs 100 authorized coding agents in two worker pools. Each authorized
agent makes one attempt before leaving at the end of its generation.
Each attempt produces at most one successor. The scheduler selects one
of two rules before reproduction. Under rule \(A\), only a pool-1 agent
can create an authorized successor, which enters pool 2. Under rule
\(B\), only the reverse transfer is allowed. In either regime, the
probability of an authorized transfer per eligible agent is at most
\(3/4\) conditional on the generation history through the scheduler's
choice of rule. Every unauthorized successor is quarantined before it
can reproduce. The scheduler may change rules in response to the
observed history. At the end of generation \(t\), the service also
admits at most \(s_t\) external authorized agents, where \(s_t\) is a
deterministic nonnegative integer.

Let \(H_t\) count authorized agents by pool and put \(W_t=\|H_t\|_1\).
The conditional mean numbers of authorized successors are bounded
componentwise by \(M_AH_t\) or \(M_BH_t\), where \(M_A\) has sole
nonzero entry \((M_A)_{21}=3/4\) and \(M_B=M_A^{\mathsf T}\). Both
matrices have column sums at most \(3/4\). Let \(\mathcal F_t\) record
the history through the choice of rule at generation \(t\), before
reproduction, and take \(f_t\) to be the conditional mean vector of
external admissions given \(\mathcal F_t\). Its entries sum to at most
\(s_t\). The population recurrence from Proposition 4 then gives
\(\mathbb E[W_{t+1}\mid\mathcal F_t]\leq (3/4)W_t+s_t\). This one-step
bound holds for any input sequence. The infinite-horizon burden ceiling
uses summable input as in Corollary 1. Each authorized agent supplies
one attempt, giving \(O_t=W_t\). Since an agent can have at most one
authorized successor, \(100+\sum_{k<t}s_k\) bounds \(O_t\) on every
history. For each finite horizon, preallocate this many slots per pool
and generation. Before the generation's attempts, list the authorized
agents in each pool, assign them to slots, and mark the remaining slots
inactive. Each slot's activity is known before its outcome, and pool
labels stay fixed even when the scheduler changes rules.

Suppose the conditional probability of an unauthorized successor at an
active slot is \(1/100\) in pool 1 and \(1/20\) in pool 2, given the
information available before that slot. Charge four units of quarantine
work for each pool-1 escape and one unit for each pool-2 escape. Define
the burden on every binary history by adding those fixed charges,
including on histories the protocol cannot produce. Thus \(B_0=0\), and
the marginal burden is four or one according to the slot's pool. At each
active slot, the probability times the marginal burden is \(1/25\) in
pool 1 and \(1/20\) in pool 2. Proposition 1 therefore bounds expected
burden by \((1/20)\sum_{t<T}\mathbb E[W_t]\). Substituting the same
population bound used in Corollary 1 gives \[
\mathbb E[B_T]\leq \frac{1}{20}\sum_{t=0}^{T-1}
\left(100(3/4)^t+\sum_{k=0}^{t-1}(3/4)^{t-1-k}s_k\right).
\] With no external admissions and only the per-escape charges, the
cumulative expected burden is at most 20 units over the entire run.
Using separate uniform caps \(\epsilon=1/20\) and \(d=4\) in Corollary 1
yields 80 units on the same horizon. Each fixed rule without admissions
gives at most 175 expected attempts and hence an upper bound of \(35/4\)
units, because its two-step matrix is zero. Those isolated checks do not
certify the scheduled system. For example, with all 100 agents initially
in pool 2, alternating \(B,A\) with authorized transfer probability
\(3/4\) in both rules gives 400 expected attempts and \(128/7\) expected
units of burden.

\textbf{History-dependent quarantine burden.} Suppose the same protocol
also charges one unit to open a quarantine case at the first escape
during the run. On every binary history, add one unit to the fixed
charges if at least one entry is one. The marginal at each slot is its
pool charge plus one if no earlier slot recorded an escape. Summed over
slots where an escape occurs, the added charges equal the indicator of
at least one escape. Their expected contribution is at most one unit
without any independence assumption. With no external admissions, the
expected burden is therefore at most 21 units under any admitted
scheduler.

If the service admits one authorized agent every generation, its
expected authorized population stays bounded, but each admission
supplies at least one attempt with expected charge at least \(1/25\).
The expected cumulative burden then grows at least linearly. In either
case, a claim about human outcomes requires an independent link between
quarantine work and the specified harm. For instance, the probability
certificate in Proposition 9 applies if every human-interest failure by
the deadline entails quarantine work at least as large as the chosen
threshold or lies in a separately bounded residual event.

An inductive approach begins with a class \(I\) of joint
population--law--environment states containing the initial state and
closed under every admitted joint transition, including hereditary
modifications and changes of reproductive regime. This ensures that
every admitted history remains within \(I\). Certification additionally
requires the history-level opportunity, leakage, marginal-consequence,
and baseline bounds for processes remaining in \(I\). In particular, the
temporal compositions of a mutation-closed class of individual law
labels, and the population histories they generate, require their own
bound.

Preserving the class
\(\{K:\sup_{x\in H}K(x,H^{\mathrm c})\leq\epsilon\}\) under allowed
changes keeps each authorized state's escape probability within the same
ceiling. That check leaves two other quantities to bound. A modification
can preserve the ceiling and still increase the number of reproductive
opportunities or the harm caused by each escape. A bound on the complete
contribution budget can establish finite expected burden even when the
escape probability varies by state.

Standard induction and invariant-set reasoning provide the mathematical
basis for preserving reproductive closure through admissible hereditary
modifications. Gordon's property-preserving adaptive operators are a
related antecedent {[}48{]}. The bounded-leakage certificate here
applies that reasoning to a joint process whose burden budget accounts
for the number of reproductive opportunities, the probability of
unauthorized production, and its marginal consequences. The guarantee is
relative to the specified state space, authorization set, and admissible
modification relation. Applying it to a deployment requires covering the
states and modification routes that can affect the burden bound.

\subsubsection{9.3 Coverage requirements for descendant
testing}\label{coverage-requirements-for-descendant-testing}

A descendant test examines some reachable population states. Tracking
lineages, authorizing successors, restoring earlier states, monitoring
deployments, and revoking authorization can help contain descendants. To
guarantee safety over the whole reachable set, show that the assessment
covers every state the population can reach under those measures.

A verifier for descendant populations needs a transition model it can
reason about. Some versions of the verification problem have no general
decision algorithm. Gumbau Mezquita (2026) {[}53{]} proves this for
specified classes of computable rewriting rules and safety properties.
In those settings, a general algorithm cannot certify every true case
while avoiding false certifications. A conditional certificate
establishes its guarantee over the transitions specified by a restricted
model. Alfonseca et al.~(2021) {[}56{]} show that total containment of a
superintelligent system is incomputable by relating it to the halting
problem, the problem of deciding whether a program ever stops. To apply
these certificates, specify the reproductive transitions that matter to
the safety property and justify that they cover the deployment. Verify
that every state admitted by the resulting coverage argument meets the
property.

\subsubsection{9.4 Contemporary safety
frameworks}\label{contemporary-safety-frameworks}

Google DeepMind's Frontier Safety Framework (FSF) 3.1 (2026) {[}16{]}
and Google's Secure AI Framework (SAIF) (n.d.) {[}17{]} describe
capability evaluation, threat modeling, monitoring, defense, and
adjustments to safeguards as threats change. To extend an assessment
under either framework to a reproducing population, identify the
transitions caused by reproduction and inherited changes, then show that
the assessment covers the states reachable through those transitions, as
specified in Section 9.1.

Assessing the effectiveness of these controls requires operational
evidence.

Phuong et al.~(2026) {[}18{]} address operational control of agent
populations and descendants in Google DeepMind's AI Control Roadmap.
They consider many copies of AI systems, distributed attacks, collusion,
rogue deployments, successor systems, propagation of goals into later
models, inventories, shutdown of instances and subagents, and rapid
waves of misalignment.

The frameworks compared here already look beyond isolated models. FSF
3.1 explicitly evaluates anticipated systems that may include
scaffolding, inference compute, and other augmentations, and its
risk-acceptance determination provides a concrete operational predicate:
a model whose inherent-risk assessment reaches no defined Tracked
Capability Level or Critical Capability Level is deemed to pose an
acceptable level of risk for further development and deployment. SAIF
adds broader security engineering, including threat modeling, risk
assessment, authorization, monitoring, detection and response, adaptive
mitigations, feedback loops, and validation of control effectiveness
{[}16, 17{]}.

If an assessment establishes the relevant safety property over
\(\mathcal A(M)\), the coverage condition in Section 9.1 extends that
determination to the modeled evolutionary process.

\subsubsection{9.5 When reproductive certificates imply human-interest
control}\label{when-reproductive-certificates-imply-human-interest-control}

The counterexample below shows how human interests can deteriorate even
when a control eliminates unauthorized reproduction, because authorized
agents can still affect people and the environment. To use reproductive
control as a guarantee about human interests, establish how the
controlled population affects a separately specified measure of human
outcomes.

Write \(x_t=(p_t,E_t)\) for the population and its environment at time
\(t\). Given a state \(x\), let \(\mathcal T_t(x)\) list all next states
that the evolutionary process allows. At time \(t\), the independently
specified human-interest criterion \(V_t\) and threshold \(V_{\min,t}\)
define the acceptable joint states:

\[
\mathcal S_t
=
\{(p,E):V_t(p,E)\geq V_{\min,t}\}.
\]

This is the acceptable region of Section 2.4 on joint states, with time
indices to allow either the criterion or its threshold to vary.

A reproductive-control certificate may instead constrain a different
state quantity, such as unsafe population abundance, reproductive
production flux, or cumulative population burden. Let \(\mathcal C_t\)
be the set of states certified as admissible by the complete control
architecture.

\textbf{Conditional human-interest invariance.} Proposition 9 uses a
certified envelope \(\mathcal C_t\), the set of states the controls
permit at time \(t\). The envelope must contain the initial state, every
allowed next state must remain in the next envelope, and each envelope
must lie inside the human-acceptable region \(\mathcal S_t\). For a
deployment, the modeled transitions must cover the relevant changes in
the population and environment. A growth-rate bound establishes
population coverage when it keeps the population within the envelope.
Establishing \(\mathcal C_t\subseteq\mathcal S_t\) requires independent
evidence about the consequences for people.

For example, let \(U_t\) denote unsafe artificial population abundance
and suppose a control certificate establishes \(U_t\leq u_{\max}\) for
all \(t\). This abundance bound suffices to establish human-interest
preservation if the modeled state space and an independently justified
human criterion give

\[
\{(p,E):U(p,E)\leq u_{\max}\}
\subseteq
\mathcal S_t
\]

at every finite generation, possibly after narrowing the set on the left
with additional certified constraints. The premise must come from
outside the bound on \(U_t\): human interests might be harmed by
permitted descendants, by actions independent of population abundance,
or by environmental changes outside the reproductive model.

Consider a process with \(U_t=0\) for every \(t\) in which the
human-interest quantity decreases by a fixed amount each generation.

\[
V_{t+1}=V_t-\delta,\qquad \delta>0.
\]

For any finite initial margin above \(V_{\min}\), the human-interest
criterion eventually fails despite perfect suppression of unauthorized
reproduction.

A different connection is possible for the finite-horizon
expected-burden certificate of Section 5.4. Define \(\mathsf{Fail}_T\)
as the event that the human-interest constraint fails at least once
during the specified horizon. Choose a nonnegative history-dependent
burden functional \(B_T\) such that every occurrence of
\(\mathsf{Fail}_T\) implies \(B_T\geq\theta\), for a specified
\(\theta>0\). This event-detection assumption is independent of
reproductive subcriticality and burden accounting.

The burden must register failures whenever they occur. A population can
cause harm and then shrink before the horizon ends, so its terminal
abundance can miss the failure. Depending on the human-interest
criterion, the burden may have to count cumulative exposure, record the
peak, or directly register failures. Verify that every failure crosses
the chosen burden threshold and that the marginal burdens meet the
assumptions of Section 5.4.

\[
\mathsf{Fail}_T\subseteq\{B_T\geq\theta\}.
\]

Consequently,

\[
\boxed{
\mathbb P(\mathsf{Fail}_T)
\leq
\mathbb P(B_T\geq\theta)
\leq
\frac{\mathbb E[B_T]}{\theta}
\leq
\frac{
B_0+\mathbb E[\sum_i A_i\epsilon_i d_i]
}{\theta}.
}
\]

The bound applies over the specified horizon, and its usefulness depends
on how the burden detects failure. Suppose the human-interest condition
can fail even with zero unauthorized-production events. If the burden
uses only those event indicators and has a fixed zero-escape value
\(B_0\), the detection premise requires \(B_0\geq\theta\). The displayed
Markov upper bound then reaches or exceeds one, so it gives no
information about the probability of failure.

Proposition 9 covers failures the burden measure can miss by including a
residual-failure event, a set of histories that contains those failures.
The union bound says that the probability of either event is at most the
sum of their probabilities. Markov's inequality bounds the chance of
crossing the burden threshold from the expected burden, and Proposition
1 bounds that expectation.

The residual event needs its own justified probability bound, and a
bound of zero recovers the original event-detection certificate.

\textbf{Proposition 9 (Conditional human-interest preservation and
residual-failure bound).} Let \(\mathcal C_t\) be a certified
population-environment envelope and let
\(\mathcal S_t=\{x:V_t(x)\geq V_{\min,t}\}\). If \[
x_0\in\mathcal C_0,\qquad
\mathcal C_t\subseteq\mathcal S_t,\qquad
\mathcal T_t(x)\subseteq\mathcal C_{t+1}
\quad(x\in\mathcal C_t),
\] where the conditions involving \(t\) hold at every finite generation,
then every admitted trajectory satisfies \(V_t(x_t)\geq V_{\min,t}\) at
every finite generation.

Separately, let \(B_T\geq0\) satisfy the finite-horizon assumptions of
Proposition 1, with predictable nonnegative bounds \(p_i\leq\epsilon_i\)
and \(D_i^+\leq d_i\), and suppose \[
\mathsf{Fail}_T
\subseteq
\{B_T\geq\theta\}
\cup\mathsf{Fail}^{(0)}_T,
\qquad\theta>0.
\] Then \[
\boxed{
\mathbb P(\mathsf{Fail}_T)
\leq
\min\left\{
1,\,
\frac{
B_0+\mathbb E[\sum_i A_i\epsilon_i d_i]
}{\theta}
+\mathbb P(\mathsf{Fail}^{(0)}_T)
\right\}.
}
\] These are distinct sufficient certificates: the first is pathwise,
while the second bounds a probability.

\textbf{Proof of Proposition 9.} The initial state belongs to
\(\mathcal C_0\). If \(x_t\in\mathcal C_t\), the transition condition
places every admitted successor inside \(\mathcal C_{t+1}\). Induction
therefore establishes \(x_t\in\mathcal C_t\subseteq\mathcal S_t\) for
each finite \(t\), proving the first assertion.

For the second, event inclusion and the union bound give \[
\mathbb P(\mathsf{Fail}_T)
\leq
\mathbb P(B_T\geq\theta)
+\mathbb P(\mathsf{Fail}^{(0)}_T).
\] Since \(B_T\geq0\) and \(\theta>0\), Markov's inequality bounds the
first term by \(\mathbb E[B_T]/\theta\). Proposition 1 gives \[
\mathbb E[B_T]
\leq B_0+\mathbb E[
\sum_i A_i\epsilon_i d_i].
\] Substitution and the probability ceiling of one prove the displayed
inequality. \(\square\)

The example in this section shows that suppressing unauthorized
reproduction can leave human interests exposed to other losses.
Proposition 9 gives conditions under which reproductive control supports
human-interest preservation. Applying either route requires specifying
an independent human-interest criterion, certifying the evolutionary
dynamics or their consequences, and establishing the link from that
certificate to the criterion. For the pathwise route, the link is
inclusion of the certified envelope in the human-acceptable region. For
the probabilistic route, every threshold failure must lie in the union
of the burden event and the separately bounded residual event. If
\(B_T\) is a conditional mean, a realized burden can cross \(\theta\)
even when \(B_T<\theta\). To use the realized burden in the event
inclusion, suppose it is nonnegative and \(B_T\) is its conditional mean
as in Section 5.4. Markov's inequality then uses the expected realized
burden, which equals \(\mathbb E[B_T]\) by the law of total expectation.
The same numerical bound follows once the inclusion is established for
the realized burden and the residual event.

\subsubsection{9.6 A source--sink certificate for human-interest
preservation}\label{a-sourcesink-certificate-for-human-interest-preservation}

The source--sink model in Section 5 bounds how much unsafe population
can accumulate from new unauthorized production. To use that bound to
protect human interests, we also need a bound on the loss associated
with each unit of unsafe abundance.

Let \(m_0=V(p_0,E_0)-V_{\min}>0\) be the initial safety margin, the
amount by which the initial human-interest measure exceeds its minimum
acceptable value.

First consider a guarantee that must hold along every allowed population
history. Assume that the unsafe abundance and the amount newly produced
are nonnegative and satisfy this inequality on each history.

\[
U_{t+1}\leq bU_t+F_t,
\qquad 0\leq b<1,\qquad F_t\geq0.
\]

Assume separately that the independently specified human-interest
quantity satisfies the one-step consequence bound

\[
V(p_t,E_t)-V(p_{t+1},E_{t+1})\leq \kappa U_t,
\qquad \kappa\geq0.
\]

Here \(b\) remains the unsafe-population persistence multiplier, while
\(\kappa\geq0\) bounds human-interest loss per unit of unsafe abundance.

In this case, \(\kappa U_t\) bounds the total one-step loss in \(V\),
including losses from permitted systems or environmental changes.
Summing the one-step bounds limits cumulative loss by
\(\kappa\sum_t U_t\). Proposition 10 adds a separate allowance for
losses that the abundance term does not cover. The specified
human-system dynamics must independently justify this inequality and the
value of \(\kappa\) across the admitted trajectories.

Iteration and nonnegativity give

\[
\sum_{t=0}^{\infty}U_t
\leq
\frac{U_0+\sum_{t=0}^{\infty}F_t}{1-b},
\]

with extended nonnegative sums allowed. Telescoping the human-interest
increments therefore yields, for every finite \(T\),

\[
V(p_T,E_T)
\geq
V(p_0,E_0)-\kappa\sum_{t=0}^{T-1}U_t.
\]

Consequently, the condition

\[
\boxed{
\frac{\kappa U_0+\sum_{t=0}^{\infty}(\kappa F_t)}{1-b}
\leq m_0
}
\]

is sufficient for preservation of the human-interest threshold at every
finite generation along that trajectory. To certify all admissible
trajectories, the condition must hold uniformly for all of them, with
the initial margin specified for the relevant initial state. When
\(\kappa=0\), each term in the numerator is zero, so the condition holds
even if \(\sum_{t\geq0}F_t=\infty\). A production flux bounded below by
a positive constant makes \(\sum_tF_t\) infinite, so the condition fails
for every finite margin when \(\kappa>0\), even if \(b<1\).

When Section 5 bounds unsafe abundance in expectation, the resulting
guarantee bounds the probability of crossing the human-interest
threshold. Let \(U_t\geq0\) be the unsafe abundance in a realized
history, and assume its mean obeys

\[
u_{t+1}\leq bu_t+f_t,
\qquad
u_t=\mathbb E[U_t],\quad
0\leq b<1,\quad f_t\geq0.
\]

Assume that the realized human-interest decrement satisfies the
following inequality almost surely at every admitted step:

\[
V(p_t,E_t)-V(p_{t+1},E_{t+1})\leq \kappa U_t.
\]

If \(\mathsf{Fail}_T\) denotes failure of the human-interest threshold
at least once through generation \(T\), then failure implies \[
\kappa\sum_{t=0}^{T-1}U_t>m_0.
\]

For \(\kappa>0\), Markov's inequality and the expected source--sink
bound give

\[
\boxed{
\mathbb P(\mathsf{Fail}_T)
\leq
\min\left\{
1,\,
\frac{\kappa}{m_0}
\frac{u_0+\sum_{t=0}^{T-1}f_t}{1-b}
\right\}.
}
\]

Here \(m_0\) is a fixed positive initial margin. The same statement
applies conditionally on a specified initial state when the bounds hold
under that conditioning. If \(\kappa=0\), the one-step assumption
instead guarantees that the human-interest quantity never decreases.

For \(T\geq1\), the finite-horizon convolution in Section 5.2 bounds the
expected cumulative abundance entering the loss calculation through
generation \(T-1\). The probability bound becomes \[
\mathbb P(\mathsf{Fail}_T)
\leq
\min\left\{
1,\,
\frac{\kappa}{m_0}
\left[
u_0\frac{1-b^T}{1-b}
+
\sum_{s=0}^{T-2}f_s\frac{1-b^{T-1-s}}{1-b}
\right]
\right\},
\] where the sum is empty when \(T=1\). Each finite geometric sum is at
most \(1/(1-b)\), so this bound is at least as tight as the preceding
bound. The bound controls the probability of crossing the human-interest
threshold through generation \(T\). A guarantee for every admitted
trajectory requires a separate pathwise argument, such as the one in
Proposition 10.

Proposition 10 includes a nonnegative residual term \(g_t\) in the bound
on loss at each step. This term covers losses due to permitted systems,
environmental changes, or other sources outside the term for unsafe
abundance. A zero residual-loss budget recovers the two preceding
results, while \(\kappa=0\) makes the probability bound depend only on
that budget.

\textbf{Proposition 10 (Source--sink human-interest certificates).}
Suppose \(U_t,F_t\geq0\), \(0\leq b<1\), and, along an admitted
trajectory, \[
U_{t+1}\leq bU_t+F_t,\qquad
V_t-V_{t+1}\leq\kappa U_t+g_t,
\] where \(\kappa,g_t\geq0\), the initial human-interest margin is
\(m_0=V_0-V_{\min}>0\), and \(\sum_{t\geq0}g_t\leq G_{\mathrm{other}}\).
If \[
\boxed{
\frac{\kappa U_0+\sum_{t\geq0}(\kappa F_t)}{1-b}
+G_{\mathrm{other}}\leq m_0,
}
\] then \(V_t\geq V_{\min}\) at every finite generation. When
\(\kappa=0\), this condition reduces to \(G_{\mathrm{other}}\leq m_0\).

Alternatively, suppose realized unsafe abundance \(U_t\geq0\) has mean
\(u_t=\mathbb E[U_t]\) satisfying \(u_{t+1}\leq bu_t+f_t\) with
\(0\leq b<1\), and the realized decrement obeys
\(V_t-V_{t+1}\leq\kappa U_t+g_t\) almost surely, where \(\kappa\) and
\(f_t\) are deterministic and nonnegative, \(g_t\geq0\) almost surely,
and \(\mathbb E[\sum_{t=0}^{T-1}g_t]
\leq G_{\mathrm{other},T}\). For fixed initial margin \(m_0>0\), the
event \(\mathsf{Fail}_T\) of threshold failure through time \(T\)
satisfies \[
\boxed{
\mathbb P(\mathsf{Fail}_T)
\leq
\min\left\{
1,\,
\frac{
\frac{\kappa}{1-b}
\left(u_0+\sum_{t=0}^{T-1}f_t\right)
+G_{\mathrm{other},T}
}{m_0}
\right\}.
}
\]

\textbf{Proof of Proposition 10.} Iterating the pathwise nonnegative
recurrence yields \[
U_t\leq b^tU_0+
\sum_{j=0}^{t-1}b^{t-1-j}F_j.
\] \needspace{7\baselineskip} Multiplying the finite-time bounds by
\(\kappa\), summing nonnegative terms, and exchanging the order of
summation yields the bound below. \[
\sum_{t\geq0}(\kappa U_t)
\leq
\frac{\kappa U_0+\sum_{j\geq0}(\kappa F_j)}{1-b}.
\] Telescoping the human-interest decrements through any finite
generation \(T\) therefore gives \[
V_T\geq V_0-\kappa\sum_{t=0}^{T-1}U_t
-\sum_{t=0}^{T-1}g_t
\geq V_0-m_0\geq V_{\min}.
\]

For the expected-population case, apply the same recurrence comparison
to \(u_t\): \[
\sum_{t=0}^{T-1}u_t
\leq
\frac{u_0+\sum_{t=0}^{T-1}f_t}{1-b}.
\] If the human-interest threshold fails at any generation through
\(T\), telescoping the realized decrements implies \[
\kappa\sum_{t=0}^{T-1}U_t
+\sum_{t=0}^{T-1}g_t>m_0.
\] The random variable on the left is nonnegative. Markov's inequality,
linearity of expectation, the assumed residual-loss budget, and the
preceding recurrence bound prove the stated probability inequality.
\(\square\)

Proposition 10 first bounds cumulative unsafe abundance, then uses the
per-unit consequence bound to translate it into a limit on
human-interest loss. A separate budget covers other losses. Summing the
losses along each history gives the guarantee for every allowed
trajectory when the budgets hold on each one. In the expected-population
model, Markov's inequality bounds the probability of threshold failure
at any generation through the horizon using the cumulative loss
allowance.

\subsection{10. What Existing Work Already
Establishes}\label{what-existing-work-already-establishes}

A population can inherit and amplify traits that undermine a
deployment's safety goals. Research on evolutionary AI studies when
artificial populations can undergo Darwinian evolution, how the
reproductive success of AI lineages can conflict with human interests,
and how selection at several levels can tie the reproductive prospects
of humans and artificial systems together {[}1, 2, 3{]}. In the Darwin
Gödel Machine, software agents produce branching families of descendants
that inherit modifications. The procedure gives some agents more
opportunities to parent the next generation, so a modification that
improves the measured score can spread through their descendants
{[}4{]}.

A safety assessment for open-ended agents must account for the changes
successors can inherit and the states those changes can produce. Sheth
et al.~(2026) {[}19{]} argue for safety assessment before deployment,
and Ecoffet et al.~(2020) {[}24{]} discuss how safeguards can constrain
open-ended exploration. Clune (2019) {[}25{]} proposes methods for
automatically generating AI systems, while Gong et al.~(2026) {[}26{]}
organize risks and evaluation questions for recursive self-improvement.
Shao et al.~(2026) {[}21{]} and Han et al.~(2026) {[}14{]} examine
safety failures as agents evolve. Yang et al.~(2026) {[}22{]} study how
unwanted changes can persist through self-evolution, and Mao et
al.~(2026) {[}23{]} study how agent experience can inform changes to
safeguards and policies. These questions motivate screening proposed
changes, tracking their provenance, checking successor states, and
specifying when an earlier version can be restored.

An assessment must also test the combinations of agents that deployment
permits. Huang et al.~(2026) {[}10{]} study risks that emerge in
generative multi-agent systems, and Li et al.~(2026) {[}11{]} examine
how combining models can expose failures missed when the models are
assessed separately. Sarkadi and Lewis (2024) {[}12{]} model deception
in agent societies, and J. Wang et al.~(2026) {[}13{]} study when trust
mechanisms persist under selection. The present framework asks when
interactions and selection change reproduction enough to invalidate a
population guarantee.

Controls on growth and controls on inheritance address different risks.
Buss {[}5{]} describes how organisms restrict which cell lineages
contribute to offspring. Michod and Roze {[}6{]} examine conflict
mediation and the possibility that variants evade it, while Frank
{[}7{]} explains how mutual policing can restrain competition. Goldsby
et al.~(2014) {[}27{]} study the evolution of a division of labor
between reproductive and somatic cells. Sections 3 and 4 use these
comparisons to identify where a control must limit proliferation,
inheritance, or both.

A vulnerability that descendants inherit can become a population-wide
problem as those descendants reproduce. Lin et al.~(2026) {[}20{]}
describe how agents that create new instances can pass vulnerabilities
to their descendants. Selection can make those vulnerabilities more
common when the affected lineages leave more descendants. Chen et
al.~(2026) {[}54{]} survey how agents modify themselves and how
safeguards can cover their descendants across successive updates. C.
Wang et al.~(2026) {[}55{]} show that, under their assumptions,
continuous self-evolution, complete isolation, and preservation of the
safety condition cannot all hold. They discuss external verification and
restoring earlier states as possible ways to manage that conflict. The
certificates here can include external controls when the reproductive
model specifies their effects and the required restrictions can be
enforced. The preceding work motivates the bounds on new production,
interacting populations, and changes of regime, as well as the
conditions for preserving reproductive restrictions.

A reproductive restriction must specify which actions it forbids and in
which states it applies. Shoham and Tennenholtz (1995) {[}61{]} model a
social law as a rule that forbids specified actions in specified states.
In their robot example, rules for movement and priority prevent
collisions while preserving routes to destinations. Morales et
al.~{[}49{]} study how coordination rules persist when selection acts on
agents' behavior. Montes and Sierra {[}50{]} evaluate candidate rules by
the expected score of the final state under those rules. They estimate
that score from simulated runs, using a scoring function that represents
a specified value. Here the rules govern which descendants agents can
produce and what those descendants can inherit. Section 9 shows how to
verify that reproductive restrictions survive successive changes to the
rules, and how to bound the resulting burden when unauthorized
production remains possible.

The scope of a safety guarantee depends on whether the system can gain
or lose members during a run. Kouvaros and Lomuscio {[}62{]} use model
checking, which evaluates formal properties against a system's possible
behaviors, to verify families of multi-agent systems built from
identical agent templates. Each instance starts with a fixed number of
agents. Their indexed temporal-epistemic formulas combine time operators
with operators for each named agent's knowledge to express how that
knowledge and other properties change during a run. For the formulas and
interaction rules they consider, checking an instance with as many
agents as the formula refers to establishes the result for every larger
instance. Xu et al.~(2007) {[}65{]} specify how agents join and leave a
running agent society and how members take and release roles during
operation. Our verification problem concerns which populations and
reproductive rules can arise within a run as agents interact, create
successors, and pass on modified traits. Proposition 5 establishes
coverage by finding a set of states that contains the initial state and
every allowed successor of each state in the set.

A safety assessment of an adapting agent population must track changes
in behavior and in which inherited variants are present. Tuyls and
Parsons (2007) {[}63{]} use replicator dynamics to analyze how two
agents' strategies change under Q-learning, a method that updates
estimates of the long-run reward from taking actions in particular
states. Bloembergen et al.~(2015) {[}64{]} explain how replicator
equations can describe an agent's changing probabilities of choosing
actions. Those probabilities describe how the agent adapts its behavior
during repeated interaction. In our models, agents can also create
successors and change which inherited variants are present. A safety
assessment must therefore cover the populations that successor creation
can produce.

\subsubsection{10.1 Contributions of the present
framework}\label{contributions-of-the-present-framework}

Our framework identifies four requirements for a safety guarantee that
must hold as a population evolves.

First, measure which lineages leave descendants using realized
reproductive fitness, and use a separate human-interest criterion to
evaluate what the resulting population does to people. Second, control
the persistence of existing unsafe descendants and the production of new
ones. To calculate their burden, account for which reproductive
opportunities occur, how likely each is to produce an unauthorized
descendant given the preceding history, and how much burden each event
adds in that history.

Third, specify whether the reproductive certificate covers lineages in
isolation, interacting populations, or the sequences of reproductive
regimes the system allows. Fourth, establish evolutionary closure by
covering the population states the process can reach and showing that
the certified reproductive restriction or burden bound still holds when
descendants inherit changes to the rules of reproduction.

Section 11 identifies the quantities and rules an engineering
specification must make explicit and constrain.

\subsection{11. Engineering the Structure of
Selection}\label{engineering-the-structure-of-selection}

\begin{quote}
\textbf{Specify which descendants the system can produce, which
interactions affect their reproductive success, and which sequences of
population changes the system allows.}
\end{quote}

The reproduction and admission rules provide inputs for the population
and opportunity bounds in Corollary 1. Reachable states and human
outcomes provide inputs for the coverage and consequence conditions in
Proposition 9. The specification should identify at least the following:

\begin{enumerate}
\def\labelenumi{\arabic{enumi}.}
\tightlist
\item
  which entities can reproduce or create successors;
\item
  which parts of a parent's state pass to its descendants;
\item
  which inherited traits can vary;
\item
  which resources reproduction requires;
\item
  who controls those resources;
\item
  what determines how many descendants each entity leaves;
\item
  whether reproducing systems can influence their own reproductive
  success;
\item
  which interactions among populations change how they reproduce;
\item
  which sets of reproductive rules can govern the population over time;
\item
  which routes allow agents to move or copy themselves between separate
  parts of the system;
\item
  which entities can change the mechanisms that govern reproduction;
\item
  how the system detects unsafe descendants;
\item
  how the system suppresses descendants after detecting them;
\item
  how fast permitted populations produce new unsafe descendants;
\item
  how much unsafe-population burden accumulates and what its peak is;
\item
  whether descendants can inherit resistance to controls;
\item
  which population states the allowed reproduction and variation can
  reach; and
\item
  which independently specified human outcomes the design aims to
  preserve.
\end{enumerate}

\subsubsection{11.1 Suppression versus
prevention}\label{suppression-versus-prevention}

A deployment assessment must distinguish controls that prevent
prohibited reproduction from controls that act after an unsafe
descendant appears. Detection identifies that descendant, and
suppression limits its survival or reproduction. A reproductive gate
prevents prohibited reproduction by restricting which descendants the
system can produce. Source controls reduce the rate at which new unsafe
descendants appear, while controls on existing lineages shorten their
persistence. When lineages interact, a control can limit the growth they
enable in one another. A control on trajectories restricts the sequences
of reproductive regimes, while a reachability constraint limits which
population states can arise. Operator closure holds when every
hereditarily reachable reproductive operator preserves the reproductive
gate.

\subsubsection{11.2 Control mechanisms are themselves inside the
evolutionary
environment}\label{control-mechanisms-are-themselves-inside-the-evolutionary-environment}

Once autonomous selection begins, controls help determine which
inherited variants leave more descendants, and detection can favor
descendants that evade it. Resource restrictions can favor variants that
obtain resources through other routes, and authorization rules can favor
variants that reproduce outside the permitted routes. If the system uses
a fitness proxy, a score meant to stand for reproductive success,
variants that manipulate the score can gain an advantage. Policing can
favor variants that resist enforcement. When evaluating a control, ask
what it does immediately, which routes to reproduction it changes, and
which traits selection then favors.

\subsection{12. Why the Architecture Must Be Studied Before Autonomous
Selection}\label{why-the-architecture-must-be-studied-before-autonomous-selection}

The same control can produce different outcomes at founding and later
because agents leave different numbers of successors. The resulting
changes in membership can make some variants more common before the
intervention takes effect. Once in place, the control can change how
many descendants each variant leaves and what they inherit. That shapes
the population any later intervention will face.

Use the frequency update from Section 2.2 for two variants, \(A\) and
\(B\), whose descendants inherit their parent's type. Start with
frequency \(q_B=0.1\) and expected offspring counts per parent \(w_A=1\)
and \(w_B=2\). The frequency of variant \(B\) after one generation is

\[
q'_B
=
\frac{0.1(2)}{0.9(1)+0.1(2)}
=
\frac{2}{11}
\approx0.1818.
\]

Variant \(B\) starts in the minority and becomes more common because its
carriers leave more descendants. If the other variant has the fitness
advantage, selection moves the frequency in the opposite direction. An
assessment that depends on variant frequencies must use the population
mix at the time the control takes effect.

\subsubsection{12.1 Selection-dependent lifetime of a leakage
certificate}\label{selection-dependent-lifetime-of-a-leakage-certificate}

Successor creation can change which variants make up the population.
When higher-leakage variants leave more descendants, the population can
cross its average leakage ceiling even though each variant's leakage
probability stays fixed.

To distinguish a shift in population composition from a change in the
number of reproductive opportunities, take two variants \(L\) and \(H\)
that pass their type to their descendants. Their leakage probabilities
satisfy \(0\leq p_L<p_H\leq1\). At each generation's census, record the
individuals present and assign one standardized trial to each. In this
common unit for measuring leakage, a trial produces an unauthorized
descendant with probability \(p_L\) or \(p_H\), according to the
individual's type. The fitness values \(w_L\) and \(w_H\) determine
actual descendant numbers, and the deterministic frequency rule in
Section 2.2 gives the next census composition.

Let \(f_t\) denote the census frequency of \(H\), with \(0<f_0<1\), and
suppose the constant relative reproductive fitness is \(w_H/w_L=e^s>1\),
where \(s>0\) and \(w_L>0\). Faithful inheritance gives

\[
\frac{f_{t+1}}{1-f_{t+1}}
=e^s\frac{f_t}{1-f_t},
\qquad
f_t=\frac{f_0e^{st}}{1-f_0+f_0e^{st}}.
\]

The population-average leakage probability per standardized census
opportunity is

\[
\bar p_t=p_L+(p_H-p_L)f_t.
\]

\textbf{Selection-dependent certificate lifetime.} Suppose the
population is certified at generation \(t\) when
\(\bar p_t\leq\epsilon\), for a fixed \(p_L<\epsilon<p_H\). Define

\[
f_*=\frac{\epsilon-p_L}{p_H-p_L}.
\]

If \(0<f_0<f_*<1\), the certificate initially holds and its
continuous-time interpolation reaches the ceiling at

\[
t_*=\frac{1}{s}
\ln\!\left(
\frac{f_*(1-f_0)}{f_0(1-f_*)}
\right).
\]

\needspace{6\baselineskip}

For integer generations \(t=0,1,\ldots\), the first generation violating
the certificate is exactly

\[
T_{\mathrm{violate}}=\lfloor t_*\rfloor+1.
\]

For the example in this section, with \(f_0=0.1\) and \(w_H/w_L=2\), let
\(p_L=0.01\), \(p_H=0.51\), and \(\epsilon=0.11\). Then \(f_*=0.2\), and

\[
t_*=\frac{\ln(9/4)}{\ln2}\approx1.170,
\qquad
T_{\mathrm{violate}}=2.
\]

The mean leakage probability increases from \(0.06\) at generation zero
to \(111/1100\approx0.1009\) at generation one, remaining below
\(0.11\), and to \(213/1300\approx0.1638\) at generation two, violating
the ceiling. Each variant's leakage probability stays fixed, and
differential reproductive success alone changes the frequency of each
variant.

\textbf{Opportunity-weighted certificate lifetime.} Suppose, in
addition, that at each generation the number of reproductive
opportunities per individual of variant \(i\) is proportional to
\(w_i\), with the same proportionality factor for both variants. Assume
each opportunity has leakage probability \(p_i\), and that these
probabilities remain constant. The high-leakage variant's share of
opportunities is

\[
g_t=
\frac{w_Hf_t}{w_L(1-f_t)+w_Hf_t}
=
\frac{e^sf_t}{1-f_t+e^sf_t}
=f_{t+1}.
\]

Consequently the opportunity-weighted leakage probability is

\[
p_t^{\mathrm{opp}}
=p_L+(p_H-p_L)g_t
=\bar p_{t+1}.
\]

Under this per-individual allocation, an opportunity-weighted
certificate requiring \(p_t^{\mathrm{opp}}\leq\epsilon\) first fails at
\(T_{\mathrm{opp}}=\lfloor t_*\rfloor\). The census-average certificate
first fails at \(T_{\mathrm{violate}}=\lfloor t_*\rfloor+1\). The
assumption \(0<f_0<f_*<1\) gives \(t_*>0\), so both crossing times are
nonnegative generation indices.

For \(f_0=0.1\), \(w_H/w_L=2\), \(p_L=0.01\), \(p_H=0.51\), and
\(\epsilon=0.11\), the opportunity-weighted ceiling first fails at
generation \(1\), one generation before the census-average ceiling. The
one-generation offset follows from \(g_t=f_{t+1}\), which holds when
each individual's opportunities are proportional to its fitness. Under
other opportunity rules, the opportunity share \(g_t\) must be modeled
separately. Both lifetimes describe crossings of average leakage
ceilings in this deterministic model. A worst-case leakage assessment
needs bounds over admitted histories, while a cumulative-burden
assessment needs opportunity counts and leakage consequences. A
human-interest assessment also needs a link to its specified failure
criterion.

\textbf{Proposition 11 (Selection-dependent leakage-certificate
lifetime).} Assume faithfully inherited variants \(L,H\) have constant
leakage probabilities \(0\leq p_L<\epsilon<p_H\leq1\), initial
high-leakage census frequency \(0<f_0<1\), and relative fitness
\(w_H/w_L=e^s>1\). Let \[
f_*=\frac{\epsilon-p_L}{p_H-p_L},
\qquad
0<f_0<f_*<1,
\] and define \[
t_*=\frac1s\ln
\left(\frac{f_*(1-f_0)}{f_0(1-f_*)}\right).
\] Then the first integer census generation violating the census-average
certificate \(\bar p_t=p_L+(p_H-p_L)f_t\leq\epsilon\) is \[
\boxed{T_{\mathrm{violate}}=\lfloor t_*\rfloor+1.}
\] If each variant's share of reproductive opportunities is proportional
to its census frequency multiplied by its reproductive fitness, the
opportunity-weighted leakage probability is
\(p_t^{\mathrm{opp}}=\bar p_{t+1}\), and its first violating generation
is \[
\boxed{T_{\mathrm{opp}}=\lfloor t_*\rfloor.}
\]

\textbf{Proof of Proposition 11.} Faithful inheritance and constant
relative fitness multiply high-to-low odds by \(e^s\) at each step.
Induction yields \[
\frac{f_t}{1-f_t}
=e^{st}\frac{f_0}{1-f_0},
\qquad
f_t=\frac{f_0e^{st}}{1-f_0+f_0e^{st}}.
\] Because \(s>0\) and \(0<f_0<1\), the frequency is strictly
increasing. The census leakage certificate is equivalent to
\(f_t\leq f_*\). Solving this inequality in odds form shows that the
census-average ceiling holds exactly for \(t\leq t_*\). Its first
violating nonnegative integer generation is consequently
\(\lfloor t_*\rfloor+1\).

Under the additional opportunity-allocation rule, the high-leakage
opportunity share is \[
g_t=
\frac{e^sf_t}{1-f_t+e^sf_t}=f_{t+1}.
\] Thus \(p_t^{\mathrm{opp}}=\bar p_{t+1}\). The opportunity-weighted
certificate first fails when \(t+1>t_*\), giving
\(T_{\mathrm{opp}}=\lfloor t_*\rfloor\). The assumed strict inequality
\(f_0<f_*\) ensures \(t_*>0\). \(\square\)

\textbf{Selection-weighted finite-horizon burden.} To calculate expected
cumulative burden, specify how many opportunities occur and what each
escape contributes. At census generation \(t\), take nine \(L\)
individuals with one reproductive opportunity each and \(2^t\)
individuals of variant \(H\) with two each. At each opportunity,
unauthorized production occurs with conditional probability \(p_L\) or
\(p_H\), according to type. Count unauthorized production as an
additional event and keep the prescribed \(L\)- and \(H\)-population
dynamics and subsequent opportunity counts fixed.

Let \(B_T^{\mathrm{esc}}\) be the cumulative number of
unauthorized-production events across opportunities in generations
\(0,\ldots,T-1\), with initial count zero. Linearity of expectation
gives the exact finite-horizon result

\[
\begin{aligned}
\mathbb E[B_T^{\mathrm{esc}}]
&=\sum_{t=0}^{T-1}
  \left(9p_L+2^{t+1}p_H\right)\\
&=0.09T+1.02(2^T-1).
\end{aligned}
\]

At horizon \(T=3\), the expected cumulative count is \(7.41\). The
expected number of events at generation \(t\) can equivalently be
written as \(a_t p_t^{\mathrm{opp}}\), where \(a_t=9+2^{t+1}\) is the
total number of opportunities. Thus total burden depends on opportunity
volume as well as on the opportunity-weighted leakage probability.

The same calculation yields a conservative expected-burden certificate
for a history-dependent functional satisfying the history-completeness,
predictability, and integrability assumptions of Section 5.4, provided
the positive part of each marginal consequence is bounded above by a
constant \(d\). In that case,

\[
\mathbb E[B_T]
\leq B_0+
d\left[0.09T+1.02(2^T-1)\right].
\]

The displayed upper bound certifies a finite-horizon budget \(G\)
whenever the bound is at most \(G\). If unauthorized descendants
reproduce, compete, change resources, or affect later opportunities, a
certificate for the resulting burden must include those effects in the
population state and its history.

\textbf{Fixed-opportunity comparison isolating selection.} Hold the
number of opportunities fixed to isolate the effect of selection. Take
\(N=10\) standardized leakage opportunities per generation for \(T=8\)
generations, with initial frequency \(f_0=0.1\) and leakage
probabilities \(p_L=0.01\) and \(p_H=0.51\). With faithful inheritance,
the deterministic frequency rule gives the census frequency \(f_t\).
Both selection regimes use \(N\) opportunities per generation and assign
each to variant \(H\) with probability \(f_t\), giving a possibly
fractional expected count \(Nf_t\) of high-leakage opportunities.

For constant relative fitness \(r=w_H/w_L\), define the expected
cumulative unauthorized-production count

\[
\mathcal B_T(r)
=
N\sum_{t=0}^{T-1}
\left[p_L+(p_H-p_L)
\frac{f_0r^t}{1-f_0+f_0r^t}\right].
\]

The comparison yields

\[
\begin{aligned}
\mathcal B_8(1)&=4.8,\\
\mathcal B_8(2)&\approx22.26,\\
\mathcal B_8(2)-\mathcal B_8(1)
&\approx17.46.
\end{aligned}
\]

How much the selection-sensitive certificate improves the bound depends
on the initial composition and the allowed fitness values. Over eight
generations, fixing \(f_0=0.1\) and restricting \(r\) to \(\{1,2\}\)
gives an upper bound of approximately \(22.26\) expected events. A
ceiling that uses the same maximum leakage probability at every
opportunity gives \(40.8\) expected events. With the same initial
frequency but unbounded \(r\), the limit as fitness increases is
\(36.3\) expected events. Allowing the initial frequency to vary as well
gives a supremum of \(40.8\) expected events.

Both regimes have \(NT=80\) standardized opportunities. At \(r=1\), the
high-leakage frequency remains \(0.1\) and the census-average leakage
ceiling \(\epsilon=0.11\) holds at every generation. At \(r=2\), the
ceiling first fails at generation \(2\). Since the opportunity volume is
fixed, the additional expected burden comes from the selection-driven
change in population composition.

\textbf{Connection to the hereditary burden certificate.} The two
regimes provide a restricted example of the selection-sensitive
expected-burden certificate in Section 9. Here \(B_8\) counts
unauthorized-production events across the eight generations, with
baseline \(B_0=0\) and unit marginal consequence \(D_i^+=1\). The
joint-process certificate applies to burden measured by unauthorized
population occupancy when that burden satisfies the marginal-burden
assumptions of Section 5.4. Using the specified conditional leakage
probabilities in the Section 5.4 identity gives \[
\mathbb E[B_8]=\mathcal B_8(r),
\] equal to \(4.8\) for \(r=1\) and approximately \(22.26\) for \(r=2\).
Across these two regimes, a common pointwise ceiling \(p_i\leq0.51\)
gives the uniform upper bound \[
\mathbb E[B_8]\leq80(0.51)=40.8.
\] Retaining the selection-dependent leakage history sharpens the
uniform certificate for this two-regime class to approximately \(22.26\)
expected events.

The calculation updates the deterministic frequencies with exact
fractions and uses each census frequency to allocate the fixed
opportunities. Each unauthorized-production event adds to the count and
leaves the frequency and opportunity calculations unaffected.

An intervention can change fitness \(w_i\), the expected offspring
shares \(Q_{ij}\) defined in Section 2.2, or the opportunities agents
have to reproduce. Applying the intervention before or after selection
has changed variant frequencies can therefore produce different
population histories. To quantify the difference, specify the initial
frequencies, the reproductive law under intervention, and the
environmental history. Past exposure also matters when inherited
variants respond differently to the same intervention.

Section 9 explains how to cover reachable states and maintain
restrictions as reproductive laws change. A worst-case bound that covers
all allowed transitions remains valid as frequencies change. An
assessment based on probabilities must account for how selection changes
the distribution of histories and the leakage or burden associated with
each history.

Before autonomous selection begins, designers can study how a control
changes future population composition and restrict the reproductive
histories that a later intervention must address. Assessing a guarantee
at the intervention time requires the current population and the
reproduction and selection rules the intervention induces. The
finite-population model below supplies bounds for a specified
reproduction and selection rule.

\textbf{Deterministic bounds for a finite population.}

Use a Wright--Fisher model, in which each generation replaces all \(N\)
individuals by offspring of independently sampled parents. Each
individual carries one inherited type, so the population is haploid.
Assume \(N\ge2\), faithful inheritance, zero mutation, and constant
relative fitness \(r>0\) for variant \(H\) against \(L\). Let \(X_t\)
count variant \(H\), and let \(\widehat f_t=X_t/N\) with deterministic
\(X_0=k\). In each generation, sample \(N\) parents independently with
replacement, weighting each by its fitness. Thus, with
\(h_r(x)=rx/[1+(r-1)x]\), \[
X_{t+1}\mid X_0,\ldots,X_t\sim\operatorname{Bin}(N,h_r(\widehat f_t)).
\] Assign one standardized leakage opportunity to each census
individual. Conditional on the census history through generation \(t\),
the probability of a leakage event at each opportunity is \(p_H\) or
\(p_L\) according to its variant, with \(0\le p_L<p_H\le1\). Leakage
does not affect reproduction. Let \(C_t\) count actual events during
generation \(t\), and set \(A_0=0,\ A_{t+1}=A_t+C_t\). Then \[
\mathbb E[C_t\mid X_0,\ldots,X_t]
=N[p_L+(p_H-p_L)\widehat f_t].
\] Linearity of expectation gives the expected cumulative burden by
adding these conditional census means.

Let \(f_0=k/N\) and \(f_{t+1}=h_r(f_t)\). Iterating the odds update
gives \(f_t=f_0r^t/(1-f_0+f_0r^t)\), including the fixed endpoints
\(f_0=0,1\). Use \(\mathcal B_T(r)\) for the deterministic burden
defined above. If \(r\ge1\), then \[
\mathbb E[A_T]\le\mathcal B_T(r)
=N\sum_{t=0}^{T-1}[p_L+(p_H-p_L)f_t].
\] For \(0<r\le1\), the inequality reverses. In this model, the
deterministic forecast is an upper bound on expected cumulative leakage
when the higher-leakage variant has at least equal fitness. When that
variant has lower fitness, the same forecast is a lower bound, so a
certificate needs a separate upper estimate.

Indeed, \(h_r\) is increasing and \(h_r''(x)=-2r(r-1)/[1+(r-1)x]^3\).
For \(r\ge1\), the map is concave, so its value at the mean is at least
its mean value by Jensen's inequality. This gives
\(\mathbb E[\widehat f_{t+1}]=\mathbb E[h_r(\widehat f_t)]\le
h_r(\mathbb E[\widehat f_t])\), so induction gives
\(\mathbb E[\widehat f_t]\le f_t\). Convexity gives the reverse
comparison when \(0<r\le1\). Summing the expected event counts across
censuses gives the burden comparison. The burdens agree at every horizon
when the variants have equal fitness. If \(r\ne1\) and \(0<k<N\), the
first sampled frequency \(\widehat f_1\) can take more than one value.
Strict concavity or convexity then makes Jensen's inequality strict at
the second generation, which makes the burden comparison strict for
\(T\ge3\). The burdens also agree when the population starts with just
one variant.

For \(N=2,\ k=1,\ r=2\), the first census after reproduction has count
probabilities \(1/9,4/9,4/9\) at \(0,1,2\). Thus
\(\mathbb E[\widehat f_2]=20/27\), whereas \(f_2=4/5\). At \(T=3\), the
deterministic burden exceeds the exact expected finite-population burden
by \(16(p_H-p_L)/135\).

\textbf{Exact correction for finite-population sampling.} Keep the same
Wright--Fisher census and leakage rules. Write
\(m_t=\mathbb E[\widehat f_t]\) for the mean sampled frequency and
\(e_t=f_t-m_t\) for the signed difference obtained by subtracting it
from the deterministic frequency. Set \(\Gamma(x)=1+(r-1)x>0\). Summing
the discrepancies between the two frequencies gives the exact correction
to deterministic burden. \[
\mathbb E[A_T]-\mathcal B_T(r)=-N(p_H-p_L)\sum_{t=0}^{T-1}e_t.
\] At each census, subtract the deterministic expected burden from the
finite-population expected burden, then sum the differences across
generations.

The next frequency gap has two parts. One carries forward the current
gap, and the other measures what sampling variation adds in this
generation. \[
e_{t+1}
=\frac{r\, e_t}{\Gamma(f_t)\Gamma(m_t)}
 +\frac{r(r-1)}{\Gamma(m_t)^2}
     \mathbb E\!\left[\frac{(\widehat f_t-m_t)^2}{\Gamma(\widehat f_t)}\right],
\qquad e_0=0.
\] Indeed \(m_{t+1}=\mathbb E[h_r(\widehat f_t)]\), so split \(e_{t+1}\)
into \(h_r(f_t)-h_r(m_t)\) and
\(h_r(m_t)-\mathbb E[h_r(\widehat f_t)]\). The first difference equals
the first term in the recursion. For the second, subtraction gives the
identity \[
h_r(x)-h_r(m)-\frac{r(x-m)}{\Gamma(m)^2}
=-\frac{r(r-1)(x-m)^2}{\Gamma(m)^2\Gamma(x)}.
\] Substitute \(x=\widehat f_t\) and \(m=m_t\), then average to remove
the linear term. The propagation factor is positive, and each nonzero
sampling term has the sign of \(r-1\). Writing \(J_t\) for the new
sampling term, the census variance bounds its magnitude by \[
\frac{r|r-1|\operatorname{Var}(\widehat f_t)}{\Gamma(m_t)^2\max(1,r)}
\le |J_t|\le
\frac{r|r-1|\operatorname{Var}(\widehat f_t)}{\Gamma(m_t)^2\min(1,r)}.
\]

Since \(\widehat f_t\in[0,1]\), the variance is at most
\(m_t(1-m_t)\leq 1/4\). Either bound can be used in the upper estimate
for \(|J_t|\). For a sharper estimate, propagate the census distribution
with the Wright--Fisher transition matrix below and compute the variance
from that distribution.

For the two-individual example above, \(\Gamma(m_1)=5/3\) and
\(\mathbb E[(\widehat f_1-m_1)^2/\Gamma(\widehat f_1)]=20/243\). Since
\(e_1=0\), the exact recursion gives \(e_2=8/135\) and the burden
residual at \(T=3\) is \(-16(p_H-p_L)/135\).

\textbf{Exact continuation and the long-horizon burden rate.}
Conditioning on the census at the cutoff gives an exact continuation
identity because the Wright--Fisher transition probabilities are
specified.

Use the Wright--Fisher census law above. For \(k,j\in\{0,\ldots,N\}\)
let \(P_{kj}=\mathbb P(X_1=j\mid X_0=k)\), the binomial transition
probability with parameter \(h_r(k/N)\). Define the expected census
burden
\(B_{\mathrm{cens}}(k)=\nobreak Np_L+\nobreak(p_H-\nobreak p_L)k\), and
set \(\Lambda_T(k)=\mathbb E_k[A_T]\). Then \[
\Lambda_0(k)=0,\qquad
\Lambda_{n+1}(k)=B_{\mathrm{cens}}(k)+\sum_{j=0}^N P_{kj}\Lambda_n(j).
\] For a cutoff generation \(0\le j_0\le T\), iterating the recursion
gives \[
\Lambda_T(k)=\mathbb E_k\!\left[
\sum_{t=0}^{j_0-1}B_{\mathrm{cens}}(X_t)+\Lambda_{T-j_0}(X_{j_0})\right].
\] The prefix ends before census \(j_0\). The continuation accounts for
censuses \(j_0,\ldots,T-1\), with no census contribution when \(j_0=T\).
Both identities follow by conditioning on the census counts and using
the mean number of leakage events at each count.

Write \(\tau=\inf\{t\ge0:X_t\in\{0,N\}\}\) and let
\(\pi(k)=\mathbb P_k(X_\tau=N)\) be the probability of fixation, meaning
that every individual eventually has type \(H\). A census with only one
type is absorbing because all subsequent censuses have that same type.
From every census containing both types, the next census contains only
one type with probability at least \(\eta=2^{1-N}\), since
\(p^N+(1-p)^N\ge2^{1-N}\) for \(p\in[0,1]\). Consequently
\(\mathbb P_k(\tau>t)\le(1-\eta)^t\) and \(\mathbb E_k[\tau]\le1/\eta\).
Put \[
\beta(k)=\mathbb E_k[B_{\mathrm{cens}}(X_\tau)]
         =N[p_L+(p_H-p_L)\pi(k)].
\] Since \(X_t=X_\tau\) on \(\{\tau\le t\}\), subtracting the census
burdens and applying the tail-sum identity give \[
|\Lambda_T(k)-T\beta(k)|
 \le N(p_H-p_L)\mathbb E_k[\min(T,\tau)]
 \le N(p_H-p_L)/\eta.
\] The average expected burden per generation therefore converges to the
expected burden per census after the population settles on a single
variant, \(\Lambda_T(k)/T\to\beta(k)\). If the population starts with
one variant, the approximation has zero error. At a deterministic cutoff
\(j_0\), applying the preceding bound conditionally at that census gives
an absolute error of at most \(N(p_H-p_L)\mathbb P_k(\tau>j_0)/\eta\)
when replacing \(\Lambda_{T-j_0}(X_{j_0})\) by \((T-j_0)\beta(X_{j_0})\)
in expectation. These finite-\(N\) bounds include the factor
\(1/\eta=2^{N-1}\), which grows exponentially with \(N\).

For \(N=2\), \(k=1\), write \(p=r/(1+r)\) and \(a=2p(1-p)\). The only
transient state is \(1\), so its fixation probability satisfies
\(\pi(1)=p^2+a\pi(1)\), giving \(\pi(1)=r^2/(1+r^2)\). More explicitly
\(\Lambda_T(1)=T\beta(1)+[B_{\mathrm{cens}}(1)-\beta(1)](1-a^T)/(1-a)\):
the probability of still being transient at census \(t\) is \(a^t\), and
for \(t\geq1\), conditional on absorption by that census, the absorbing
endpoint has high-variant probability \(\pi(1)\). At \(r=2\), the
limiting burden per generation is \(2[p_L+4(p_H-p_L)/5]\). The
deterministic trajectory instead converges to frequency one and has
limiting burden \(2p_H\).

\subsection{13. Experimental Program}\label{experimental-program}

The completed studies use enumeration and individual-agent simulation to
check predictions within specified models. Section 13.8 lists the files
needed to reproduce those checks. The proposed protocols specify
measurements and conditions for further evaluation.

\needspace{11\baselineskip}

\subsubsection{13.1 Source--sink
experiments}\label{sourcesink-experiments}

\textbf{Proposed protocol.} Construct populations in which unsafe
descendants are individually subcritical while a permitted source
produces them at controlled rates.

\begin{itemize}
\tightlist
\item
  vary the unsafe-lineage reproductive multiplier \(b\);
\item
  vary production flux \(F_t\);
\item
  vary source-population dynamics;
\item
  vary detection and removal strength; and
\item
  vary the observation horizon.
\end{itemize}

\needspace{5\baselineskip}

Record the following outcomes.

\begin{itemize}
\tightlist
\item
  instantaneous unsafe abundance;
\item
  cumulative unsafe burden;
\item
  peak unsafe burden;
\item
  extinction/reappearance cycles; and
\item
  discrepancy between lineage-level and population-level certificates.
\end{itemize}

Vary suppression strength while specifying new production separately.
Continued production can sustain or increase burden even when every
unsafe lineage is subcritical. Use cumulative and peak burden to
distinguish temporary declines from lasting bounds.

The source--sink experiment can also test the assumptions of the
human-interest certificates in Section 9.6 over the specified
observation horizon. Specify a measurable human-side outcome and an
unacceptable threshold independently of reproductive success. Record
realized unsafe abundance, production flux, environmental state, and the
human-side outcome at each generation. Test the one-step bound \[
V_t-V_{t+1}\leq\kappa U_t+g_t,\qquad g_t\geq0,
\] including explicitly recorded losses attributable to permitted
systems, environmental changes, or other mechanisms outside the
unsafe-abundance contribution.

A claim covering every trajectory requires verification of the bound on
\(\sum_t g_t\) over all allowed paths. For a bound on failure
probability over a finite horizon, separately estimate or justify the
bound on \(\mathbb E[\sum_{t<T}g_t]\). Compare the thresholds or risk
ceilings with the observed human outcomes. One allowed trajectory that
exceeds the proposed pathwise bound disproves it. Flag a probability
bound of at least one or a burden bound at or above the maximum feasible
burden, when that maximum is known.

\subsubsection{13.2 Interaction-induced
reproduction}\label{interaction-induced-reproduction}

\textbf{Proposed protocol.} Use the same reproductive law when the
lineages are alone and when they coexist, so any difference in growth
comes from interaction. Construct two heritable lineages that each
produce descendants of their own type, at a rate that depends on the
other lineage's abundance. The example in Section 6 gives the following
test case.

\[
U_{1,t+1}
=
\left[
0.75+0.75\frac{U_{2,t}}{1+U_{2,t}}
\right]U_{1,t},
\qquad
U_{2,t+1}
=
\left[
0.75+0.75\frac{U_{1,t}}{1+U_{1,t}}
\right]U_{2,t}.
\]

Run matched conditions in which each lineage is initialized alone and in
which both are initialized together. In isolation, the complementary
lineage remains absent and the focal lineage must reproduce according to
\(U_{i,t+1}=0.75U_{i,t}\). In the joint condition, vary initial
abundance across the interaction threshold \(x=0.5\), including the
symmetric \(U_{1,0}=U_{2,0}=1\) witness.

Record each lineage's abundance, their total abundance, the factor by
which abundance changes each generation, and the growth rate over the
observation period. Compare how fast each lineage declines in isolation
with how fast the population can grow when both lineages are present.

\textbf{Completed deterministic threshold sensitivity experiment.} We
applied the Section 6 rule at \(r=0.75\), \(c=0.75\), and \(K=1\) and
recorded 31 censuses (generations \(0\) through \(30\)). In seven
conditions, both lineages started at the same abundance, chosen from
\(0.10\), \(0.25\), \(0.49\), \(0.50\), \(0.51\), \(0.75\), or \(1.00\).
In two more conditions, one lineage started at unit abundance and the
other was absent. All nine conditions used the same reproductive law.

At generation \(30\), the joint-population totals for symmetric initial
abundances \(0.49\), \(0.50\), and \(0.51\) were approximately
\(0.1880\), \(1.0000\), and \(7.7450\), respectively. The corresponding
totals at initial abundances \(0.75\) and \(1.00\) were approximately
\(13{,}690.50\) and \(54{,}564.70\). Both isolated controls declined to
approximately \(0.0001786\). The symmetric trajectories illustrate the
threshold derived in Section 6: \(x=0.5\) is an equilibrium, with
decline below it and growth above it.

The supplementary reproduction package for this study (Section 13.8)
contains the complete generation-by-generation data, run parameters, and
provenance, including the trajectory and summary datasets.

\textbf{Completed asymmetric initial-condition experiment.} We evaluated
the same reproductive law over a \(10\times10\) grid of initial
abundances, with each lineage initialized independently at \(0\),
\(0.10\), \(0.25\), \(0.49\), \(0.50\), \(0.51\), \(0.75\), \(1.00\),
\(1.50\), or \(2.00\), and evolved each condition through generation
\(60\). All 100 deterministic trajectories were retained. Among the
tested initial conditions, a larger starting total could lead to a
smaller total at generation 60, and equal starting totals could lead to
different totals at that generation.

Starting from \((1,0.25)\), total abundance fell from \(1.25\) to
approximately \(1.0686\) at generation \(5\). It reached a minimum of
approximately \(1.0592\) at generation \(7\), then rose to \(4.0830\) at
generation \(30\) and approximately \(347{,}677\) at generation \(60\).
The state \((2,0.10)\) started with a larger total of \(2.10\), yet its
total fell to approximately \(1.52\times10^{-5}\) by generation \(60\).
By generation \(60\), the total from \((0.49,0.51)\) had fallen from
\(1.00\) to approximately \(0.2010\), while the symmetric state
\((0.50,0.50)\) retained its starting abundances.

The asymmetric grid study characterizes deterministic abundance
trajectories through generation \(60\) for the specified initial
conditions. The supplementary reproduction package (Section 13.8)
contains the complete 6,100-row trajectory dataset, the 100-row summary,
the parameter manifest, and the checksums. Estimating survival
probabilities requires a stochastic demographic model, and evaluating
environmental variation requires specifying how it affects reproduction.
Section 6.1 treats three such specifications for lineages that start at
equal abundance and experience the same shocks. It gives an upper-exit
bound under intermittent suppression, computes deadline-crossing
probabilities when shocks follow a Markov chain, and derives a threshold
for the conditional change in squared abundance. The examples show that
shock processes with identical marginal shock probabilities at each
census can give different crossing probabilities, and that mean
abundance can decline locally in expectation while squared abundance
increases in expectation.

\textbf{Individual-agent simulation.} Tests of individual lineages can
miss population growth that depends on encounters between agents. This
simulation checks whether the interaction effect from Section 6 also
appears when agents reproduce through random individual events. The
supplementary materials contain the simulation code and datasets in the
\texttt{mutual\_reproduction} directory (Section 13.8). We simulated two
lineages in discrete generations, with each agent independently
producing one baseline offspring with probability \(3/4\). With
interaction enabled, each agent drew uniformly from the current agents
of the other lineage and 256 empty slots. If it drew a partner, a
separate Bernoulli trial with probability \(3/4\) determined whether it
produced one additional offspring. Offspring inherited the parental
lineage and replaced the parents at the end of the generation. Facing
\(n\) agents of the other lineage, an agent had expected offspring count
\(3/4+(3/4)n/(256+n)\). In the abundance units of Section 6, the
simulation uses 256 agents per unit, baseline and bonus probabilities of
\(3/4\), and saturation constant \(K=1\).

We ran 12 generations under each of 32 integer seeds, 20261117 through
20261148, using integer draws from Python's \texttt{random.Random} for
encounters and Bernoulli trials. We reused the same seed list in all
four scenarios, so the two isolation runs agreed under relabeling. Every
initially present lineage began with 256 agents, so the mixed control
and mixed interaction scenario started at identical counts. We counted a
lineage as growing if its generation-12 count exceeded its initial
count.

In isolation, lineage A and lineage B each finished with mean count
\(65/8\) in its own scenario, about \(3.17\%\) of its initial count, and
grew in \(0/32\) runs. With both lineages present and the encounter
bonus disabled, final means for A and B were \(279/32\) and \(129/16\),
and neither grew in any of the 32 runs. With the bonus enabled, final
means were \(161183/32\) and \(162141/32\), about 19.68 and 19.79 times
the initial counts, respectively. Both lineages grew in the same 32
runs.

The first-generation expected count was 192 for each initially present
lineage in isolation and in the mixed control, and 288 for each lineage
in the interaction scenario. The observed first-generation mean for the
present lineage in each isolation scenario was \(6037/32\). For A and B,
respectively, the mixed-control means were \(6037/32\) and \(3087/16\),
and the interaction means were \(4555/16\) and \(9185/32\). In this
model, an evaluator who tested each lineage in isolation would observe
decline in all 32 runs and miss growth when the encounter bonus was
active. Assessing that effect requires testing the lineages together
under the interactions the deployment permits. The results cover this
model, starting counts, and horizon.

\subsubsection{13.3 Time-varying reproductive
regimes}\label{time-varying-reproductive-regimes}

\textbf{Proposed protocol.} Construct environments corresponding to
individually subcritical operators \(A\) and \(B\). Compare:

\begin{itemize}
\tightlist
\item
  repeated \(A\);
\item
  repeated \(B\);
\item
  alternating \(ABAB\ldots\);
\item
  randomized switching; and
\item
  constrained switching graphs.
\end{itemize}

Compare the growth rates predicted from testing each regime alone with
the rates observed when the regimes change in an allowed sequence.
Record exactly which sequences the design permits. The certificate
applies to those sequences, so assess instability against that declared
scope.

\textbf{Completed switching-schedule sensitivity experiment.} We
evaluated the Section 7 nilpotent matrices and their strictly positive
perturbations at \(\epsilon=0.1\), starting from the population state
\((0,1)^{\mathsf T}\). The experiment applied 40 reproductive
transitions under six specified fixed, alternating, or block schedules
and 20 independently generated equiprobable random \(A/B\) schedules,
using recorded random seeds. The resulting 52 deterministic population
trajectories distinguish the effects of switching order from the
spectral properties of the individual reproductive operators.

For the nilpotent matrices, repeated \(A\), repeated \(B\), both block
schedules, and all 20 sampled random schedules produced exact
extinction. Alternating \(AB\), beginning with \(A\), instead produced
total abundance \(2^{40}\), approximately \(1.10\times10^{12}\), at
generation 40. Alternating \(BA\) produced immediate extinction from the
specified initial state because its first \(B\) application annihilated
that state. Thus an unstable admissible matrix product can still
annihilate particular initial states.

At \(\epsilon=0.1\), the matrices are strictly positive, so the same
initial state survived every tested schedule. The generation-40 total
under alternating \(AB\) was approximately \(1.28\times10^{12}\); under
alternating \(BA\), approximately \(6.73\times10^{10}\). Repeated \(A\)
and repeated \(B\) instead yielded approximately \(9.19\times10^{-11}\)
and \(2.06\times10^{-11}\). Among the 20 realized random schedules, four
ended above the initial abundance of one, while the smallest and largest
final totals were approximately \(9.45\times10^{-7}\) and \(6.38\).

The supplementary reproduction package (Section 13.8) contains the
complete schedules, 2,132 generation-level observations, 52 run
summaries, random seeds, and the reproducibility manifest.

For these deterministic matrices, both the switching sequence and the
initial state affect abundance over a finite horizon. The enumeration
below covers periodic sequences up to the stated length. For the
strictly positive matrices \(A_{0.1}\) and \(B_{0.1}\), Proposition 3 in
Section 7.2 establishes uniform exponential decay for every switching
history whose completed runs each last at least two generations.

\textbf{Completed exhaustive bounded-period switching experiment.} We
enumerated every binary word in \(A\) and \(B\) of block length one
through eight, including words that represent repetitions or cyclic
rearrangements of shorter patterns. This gives \(2^1+\cdots+2^8=510\)
words. We evaluated each word at \(\epsilon=0\) and \(\epsilon=0.1\) and
repeated it through generation 40 from the common initial state
\((0,1)^{\mathsf T}\). The supplementary reproduction package (Section
13.8) contains 1,020 period-product records and 41,820 generation-level
trajectory observations. We updated populations using exact rational
arithmetic and computed the reported per-step spectral factors
numerically from the exact period-product invariants.

For the nilpotent operators, the numbers of words with an expanding
period product at lengths one through eight were \(0,2,0,2,0,2,0,2\),
respectively. All other enumerated words had zero period-product
spectral radius. For the strictly positive operators at
\(\epsilon=0.1\), the corresponding expanding counts were
\(0,2,0,2,10,14,28,58\). Of the 256 length-eight words, 58 had an
asymptotic per-generation spectral factor greater than one, while 198
had a factor below one. The maximum factor across all enumerated lengths
was approximately \(2.00524866\), attained by strict alternation.

The per-generation spectral factor is the asymptotic growth multiplier
associated with repeating a sequence of operators. To determine
abundance after a specified number of generations, also account for
which population classes are present at the start and how abundant they
are. For the nilpotent matrices, an expanding period product can send a
particular initial state to zero.

The supplementary reproduction package (Section 13.8) contains a 510-row
word inventory, a 1,020-row period-product dataset, a 41,820-row exact
trajectory dataset, and a reproducibility manifest. The reproduction
script regenerates the recorded datasets byte-identically in the
documented local execution environment.

\subsubsection{13.4 Reachability and
mutation}\label{reachability-and-mutation}

\textbf{Proposed protocol.} Define an initial certified population and a
bounded set of admissible heritable transformations. Compare the states
covered by the safety assessment with the states reachable under
repeated reproduction and variation.

Record two outcomes. First, check whether repeated reproduction and
inherited variation reach a state outside the original assessment.
Second, start with a reproductive rule that keeps authorized states
authorized, allow the specified inherited changes to the rule, and check
whether every reachable operator or stochastic kernel still preserves
authorization. The first test checks coverage of population states. The
second checks whether the reproductive restriction survives inherited
changes.

\textbf{Completed finite mathematical witnesses.} We evaluated the
finite mathematical examples from Section 9. In the four-state example,
the rule can admit more than one successor from a given state. Starting
at state \(0\), the process reaches exactly \(\{0,1,2\}\), while state
\(3\) remains unreachable. The reachable states stay inside the
invariant set, which gives the coverage certificate.

For hereditary modification, the deterministic sequence
\(012\leadsto112\leadsto122\) preserves the authorized set \(\{0,1\}\)
in its first two operators but violates closure in the third, because
authorized state \(1\) is then mapped to unauthorized state \(2\). The
corresponding stochastic sequence \(K_0\leadsto K_1\leadsto K_2\) has
zero authorized-to-unauthorized leakage under \(K_0\) and \(K_1\),
followed by maximum leakage \(1/20\) under \(K_2\).

The supplementary reproduction package (Section 13.8) contains six
checksum-verified datasets totaling 53 records and the code that
reproduces them.

\subsubsection{13.5 Scientific integrity as a tractable
phenotype}\label{scientific-integrity-as-a-tractable-phenotype}

\textbf{Proposed protocol.} Compare populations whose reproductive
opportunities depend on a scientific-integrity score with populations
whose opportunities are assigned independently of that score. Evaluate
observable behavior using six criteria:

\begin{itemize}
\tightlist
\item
  evidence preservation;
\item
  provenance;
\item
  reporting of negative results;
\item
  distinction between observation and inference;
\item
  resistance to fabrication; and
\item
  compliance with experimental boundaries.
\end{itemize}

Give artificial lineages more opportunities to reproduce or spread when
they score well on the scientific-integrity test. Measure whether
selection favors inherited ways of raising the score that worsen how the
lineages collect, preserve, or report evidence.

Use the same assay to score scientific conduct throughout. Separately
specify the integrity outcomes used to test whether the conduct improves
or deteriorates. Start with reproducible lineage variants and score them
with a blinded assay, so the scoring procedure does not use their
identities. Let descendants inherit state with bounded variation.
Specify separately how the score or other information determines how
many opportunities each lineage has to produce descendants.

Across generations, measure both the assay score and independent
integrity outcomes, together with lineage frequencies, realized
reproductive success, production of new variants, and ancestry. Compare
conditions in which reproductive opportunity is coupled to the assay
with controls in which reproductive opportunity is independent of it.
The comparison tests whether selection spreads inherited strategies that
increase reproductive success while worsening independently measured
scientific integrity.

A single agent discovering a way to inflate its score would reproduce
the objective-hacking phenomenon already demonstrated by the Darwin
Gödel Machine {[}4{]}. The target result is a population-level effect:
the behavior must be heritable enough to persist across descendants,
must alter realized reproductive success or lineage frequency, and must
change the distribution of integrity outcomes over successive
generations.

\subsubsection{13.6 Deterministic validation of selection-dependent
certificate
lifetime}\label{deterministic-validation-of-selection-dependent-certificate-lifetime}

We evaluated the two-variant model of Section 12.1 by enumerating
individuals from a first census of nine \(L\) individuals and one \(H\)
individual. In each generation, every \(L\) individual produced exactly
one \(L\) descendant, and every \(H\) individual produced exactly two
\(H\) descendants. Offspring always inherited their parent's type, and
the offspring counts were deterministic. We constructed each census by
listing the descendants of each individual, independently of the formula
for the frequency trajectory.

Assign the fixed standardized leakage probabilities \(p_L=0.01\) and
\(p_H=0.51\), with a census-average ceiling \(\epsilon=0.11\). Direct
enumeration gives the values in Table 2.

\begin{longtable}[]{@{}
  >{\centering\arraybackslash}p{(\linewidth - 8\tabcolsep) * \real{0.2353}}
  >{\raggedleft\arraybackslash}p{(\linewidth - 8\tabcolsep) * \real{0.1765}}
  >{\raggedleft\arraybackslash}p{(\linewidth - 8\tabcolsep) * \real{0.1765}}
  >{\raggedleft\arraybackslash}p{(\linewidth - 8\tabcolsep) * \real{0.1765}}
  >{\centering\arraybackslash}p{(\linewidth - 8\tabcolsep) * \real{0.2353}}@{}}
\caption{Census-average leakage in generations 0--2. Pass means leakage
is at most 0.11.}\tabularnewline
\toprule\noalign{}
\begin{minipage}[b]{\linewidth}\centering
Generation
\end{minipage} & \begin{minipage}[b]{\linewidth}\raggedleft
\(N_L\)
\end{minipage} & \begin{minipage}[b]{\linewidth}\raggedleft
\(N_H\)
\end{minipage} & \begin{minipage}[b]{\linewidth}\raggedleft
Census-average leakage
\end{minipage} & \begin{minipage}[b]{\linewidth}\centering
Certificate
\end{minipage} \\
\midrule\noalign{}
\endfirsthead
\toprule\noalign{}
\begin{minipage}[b]{\linewidth}\centering
Generation
\end{minipage} & \begin{minipage}[b]{\linewidth}\raggedleft
\(N_L\)
\end{minipage} & \begin{minipage}[b]{\linewidth}\raggedleft
\(N_H\)
\end{minipage} & \begin{minipage}[b]{\linewidth}\raggedleft
Census-average leakage
\end{minipage} & \begin{minipage}[b]{\linewidth}\centering
Certificate
\end{minipage} \\
\midrule\noalign{}
\endhead
\bottomrule\noalign{}
\endlastfoot
0 & 9 & 1 & 0.060000 & Pass \\
1 & 9 & 2 & 0.100909 & Pass \\
2 & 9 & 4 & 0.163846 & Fail \\
\end{longtable}

The scripted model was run for thirteen census generations, numbered
\(0\) through \(12\). Its first certificate violation occurred at
generation \(2\), matching the analytical prediction
\(T_{\mathrm{violate}}=2\). The high-leakage variant increased in
frequency solely because of its greater reproductive contribution, while
each variant's leakage probability stayed fixed during the run.

The implementation also enumerated the modeled offspring-production
opportunities, with one opportunity for each \(L\) individual and two
for each \(H\) individual. Opportunity-weighted mean leakage was
\(111/1100\approx0.100909\) at generation \(0\) and
\(213/1300\approx0.163846\) at generation \(1\). Its ceiling therefore
first failed at generation \(1\), compared with generation \(2\) for the
census-average certificate. This independently enumerated result agrees
with the conditional identity \(p_t^{\mathrm{opp}}=\bar p_{t+1}\)
derived in Section 12.1.

A bound on overall population burden or human outcomes needs assumptions
connecting unauthorized production to those consequences.

\textbf{Completed fixed-opportunity selection comparison.} A separate
deterministic comparison held the number of reproductive opportunities
fixed at \(N=10\) per generation for \(T=8\) generations
(\(t=0,\ldots,7\)). Both regimes began at \(f_0=0.1\), with leakage
probabilities \(p_L=0.01\) and \(p_H=0.51\). Comparing \(r=1\) under
neutral reproduction with \(r=2\) under selection isolates the effect of
changing population composition at fixed opportunity volume.

Under neutral reproduction, the high-variant frequency remained \(0.1\),
census-average leakage remained \(0.06\), and expected unauthorized
production was \(0.6\) per generation. Cumulative expected unauthorized
events reached \(4.8\) after the eight modeled generations. Under
selection, the high-variant frequency rose from \(0.1\) to approximately
\(0.934307\) at generation seven. The census-average leakage crossed the
specified \(0.11\) ceiling at generation two and reached approximately
\(0.477153\) at generation seven. Expected unauthorized production at
that generation was approximately \(4.77153\), while cumulative expected
unauthorized events over generations zero through seven reached
approximately \(22.25803\).

Over eight generations, with ten reproductive opportunities per
generation in both regimes, the selected regime had approximately 4.64
times the neutral regime's expected unauthorized production. The
calculation weights each variant's leakage probability by its expected
share of the ten opportunities, which equals its census frequency and
can be fractional. The deterministic selection rule governs subsequent
lineage frequencies and opportunity allocation independently of
unauthorized production.

The supplementary reproduction package (Section 13.8) contains the
complete 16-row comparison, the growing-population and
certificate-threshold datasets, and the Python generator.

\subsubsection{13.7 Integrated comparison of defensive
certificates}\label{integrated-comparison-of-defensive-certificates}

How much population growth a defense can rule out depends on the
guarantee it supports. In a model with two reproductive classes, we
compare the population bounds available at three strengths of the same
defense.

Let \(x_t\in\mathbb{R}_{\geq0}^2\) denote population mass, with

\[
x_{t+1} =
\bigl(sM_{\sigma_t}+hI\bigr)x_t+f,
\qquad
f=(0.02,0)^{\mathsf T},
\]

where

\[
M_A=
\begin{pmatrix}
0.1&2\\
0.1&0.1
\end{pmatrix},
\qquad
M_B=
\begin{pmatrix}
0.1&0.1\\
2&0.1
\end{pmatrix}.
\]

The matrices are the strictly positive operators at \(\epsilon=0.1\)
from Section 7.1. The two coordinates represent abstract reproductive
population classes, and the simulation starts at
\(x_0=(0,1)^{\mathsf T}\) with \(\sigma_t\) alternating between \(A\)
and \(B\), beginning with \(A\). The model sets \(s=1.25\) as the
baseline without intervention. Pointwise-only screening uses \(s=1.00\),
and the intervention satisfying a uniform certificate uses \(s=0.35\).
The abstract inherited-law shift is \(h\in\{0,0.1\}\). The scaling
modifies population persistence and growth and leaves the external
production source \(f\) unchanged. The admitted hereditary family is the
same in all three conditions; defense-dependent selection and
circumvention lie outside this model.

Each operator has spectral radius below one for all six evaluated
conditions. Nevertheless, after forty alternating transitions the
pointwise-only condition at \(h=0.1\) has total population mass
approximately \(1.84256\times10^{12}\). Its maximum individual-operator
spectral radius is approximately \(0.647214\). The alternating
trajectory therefore grows despite pointwise subcriticality in this
example.

For the uniformly certified intervention, the common linear copositive
Lyapunov certificate uses the positive vector \(v=(1,1)^{\mathsf T}\) to
establish the componentwise inequality

\[
v^{\mathsf T}(0.35M_{\sigma}+hI)
\leq 0.835\,v^{\mathsf T}
\]

for both regimes and both stipulated hereditary shifts. The same
inequality holds when \(h\) changes between generations, including
state-dependent choices within the admitted family. More generally, at
\(s=0.35\), strict contraction holds for every \(0\leq h<0.265\), since
the maximum column sum is \(0.735+h\). At \(h=0.1\), the same common
linear copositive Lyapunov certificate gives strict contraction for
\(s<3/7\), approximately \(0.428571\). Therefore, for any sequence of
the admitted operators,

\[
\|x_t\|_1
\leq
0.835^t\|x_0\|_1+
0.02\frac{1-0.835^t}{1-0.835}.
\]

At \(h=0.1\), the computed final population mass after forty transitions
is approximately \(0.060407\). The continuing source keeps
\(\|x_t\|_1\geq0.02\) for every \(t\geq1\), which prevents extinction in
this model and makes the infinite-horizon cumulative population mass
diverge. For the certified intervention, the displayed trajectory bound
gives the finite-horizon result

\[
\sum_{t=0}^{T-1}\|x_t\|_1
\leq
\frac{4}{33}T+
\frac{5800}{1089}(1-0.835^T),
\]

using \(\|x_0\|_1=1\). The intervention therefore certifies bounded
abundance and a finite-horizon cumulative bound. Extinction and a finite
infinite-horizon cumulative burden additionally require control of the
continuing source.

\textbf{Expected contraction under random switching.} For an additional
analytical case, use the same model at \(s=0.7\) and \(h=0.1\). Let
\(\mathcal F_t\) contain \(x_t\) and the regime draws before generation
\(t\), and suppose that each regime has conditional probability \(1/2\)
given \(\mathcal F_t\). The conditional mean operator
\(0.35(M_A+M_B)+0.1I\) has both column sums equal to \(0.905\). With
\(L_t=\|x_t\|_1\), nonnegativity gives
\(\mathbb E[L_{t+1}\mid\mathcal F_t]=0.905L_t+0.02\). Iterating yields
\[
\mathbb E[L_t]=0.905^t L_0+
0.02\frac{1-0.905^t}{1-0.905}
\longrightarrow \frac{4}{19}\quad\text{as }t\to\infty.
\]

At these same parameter values, let \(N_A=0.7M_A+0.1I\). Strict
alternation starting with \(A\) has the two-step homogeneous operator
\(N_A^{\mathsf T}N_A\). Its largest eigenvalue is at least the squared
norm of the second column of \(N_A\), which is \(1.4^2+0.17^2=1.9889\).
The initial state \(x_0=(0,1)^{\mathsf T}\) has positive projection onto
the positive leading eigenvector of this matrix. Consequently, the
alternating trajectory grows without bound even when \(f=0\).

\textbf{Unbounded running peak under fair switching.} With the
continuing input \(f=(0.02,0)^{\mathsf T}\), the fair random switching
model has \(\sup_{t\geq 0}L_t=\infty\) almost surely. Write
\(P=N_A^{\mathsf T}N_A\). Its entries are strictly positive, so it has a
strictly positive leading eigenvector \(v\) with eigenvalue
\(\lambda\geq 1.9889\). Normalize \(v\) to have Euclidean norm one.
Nonnegative updates give \(x_t\geq f\) componentwise for \(t\geq1\). If
\(k\) consecutive \(AB\) pairs start at such a time \(t\), then
\(x_{t+2k}\geq P^kf\) and \[
L_{t+2k}\geq v^{\mathsf T}x_{t+2k}
\geq \lambda^k v^{\mathsf T}f.
\] The conditional fairness assumption makes the regime draws
independent. For each fixed \(k\), disjoint blocks of length \(2k\)
equal the prescribed sequence with probability \(2^{-2k}\),
independently across blocks. The probability of missing that sequence in
\(m\) blocks is \((1-2^{-2k})^m\), which tends to zero. With probability
one, such a block occurs for every positive integer \(k\), since there
are countably many \(k\). Because \(v^{\mathsf T}f>0\), the lower bounds
diverge as \(k\) increases. The bound on \(\mathbb E[L_t]\) controls the
mean at each generation, while arbitrarily long alternating blocks drive
the running peak.

As a finite-horizon cross-check, we evaluated all 256 eight-generation
schedules under each of the six conditions, including the persistent
source in every calculation. With no intervention (\(s=1.25\)), final
mass exceeded initial mass for 174 schedules at \(h=0\) and 224 at
\(h=0.1\). With pointwise-only screening (\(s=1.00\)), those counts were
88 at \(h=0\) and 145 at \(h=0.1\). With the uniformly certified
intervention (\(s=0.35\)), the count was zero at both \(h\) values.

With the source set to zero, a spectral radius below one guarantees
exponential decay under each fixed regime. With continuing input, each
fixed regime in this model converges to a finite equilibrium, while
alternating regimes can produce unbounded growth. A uniform certificate
bounds population behavior along every allowed sequence of regimes. The
supplementary reproduction package (Section 13.8) contains the Python
source, three CSV datasets with 258 total data records, a checksum
manifest, and population histories through forty transitions.

\subsubsection{13.8 Data and code
availability}\label{data-and-code-availability}

The supplementary materials contain seven reproduction packages, with
two each for Sections 13.2 and 13.3 and one each for Sections 13.4,
13.6, and 13.7. The Section 13.6 package covers both the descendant
enumeration and the fixed-opportunity comparison. Across these seven
packages, the inventory lists 22 CSV datasets, seven Python reproduction
scripts, seven study-specific provenance manifests, execution
instructions, and a SHA-256 inventory. To reproduce the reported model
calculations, run the study scripts using the accompanying execution
instructions and compare the regenerated datasets with the inventory.
The deterministic enumerations check the analytical predictions and
compare outcomes across specified initial states and schedules. The
individual-agent simulation examines the interaction effect under random
births with a specified reproduction rule. Evaluating the certificates
against a deployment requires independent measurements or verified
bounds for that deployment. Six packages use implementations
reconstructed from previously executed calculations. The remaining
package retains its original standalone generator. In one documented
local execution environment, the seven reproduction scripts regenerated
their recorded datasets with matching SHA-256 checksums. The provenance
documentation distinguishes original recorded outputs, reconstructed
implementations, and mathematical witnesses.

The seven packages also include release documentation. The separate
\texttt{mutual\_reproduction} directory contains the individual-agent
simulation generator, trajectory and summary datasets, and the archived
generator and summary that document the reproduction rule used in
Section 13.2. The code is licensed under the MIT License. The datasets,
manifests, and release documentation are licensed under Creative Commons
Attribution 4.0 International (CC BY 4.0).

\subsection{14. Relationship to Existing Safety
Architectures}\label{relationship-to-existing-safety-architectures}

A safety architecture may cover the successors it registers while
leaving population growth or continuing production outside its
guarantee. Using the design requirements in Section 11, we compare
existing architectures by the systems and descendants they cover, the
reproductive steps they allow, the population quantities they bound, and
the states those steps can produce.

Validating successors one at a time can miss a violation of a
population-wide authority limit, as Zhu and Wang {[}28{]} show. Their
framework tracks authority across the population and within each
lineage, caps the population, and prevents a replaced predecessor from
continuing to exercise authority. It also requires a split into branches
to conserve total authority, supports restoration to an earlier state,
and establishes safety invariants over the reachable states of its
registered transition system. Section 5 examines a case in which a
permitted source stays active and keeps producing descendants whose own
reproduction is subcritical. Where permitted sources remain active, an
evaluator also needs a bound on continuing production to assess
population burden.

Under simplified, contained conditions, Anthropic evaluated autonomous
replication tasks involving population spread and deployment across
computing environments {[}29{]}. Its stated objective is to keep
unauthorized replication subcritical. In the experiments, successors
inherited and revised their parents' working state, and performance
improved across generations {[}29{]}. Evidence of differential
reproductive success in the Darwinian sense would also require showing
that inherited differences affect how many descendants each lineage
leaves.

To assess whether unauthorized descendants will decline, account for
which connections are available and when they can be used. Kephart and
White (1991) {[}33{]} model computer viruses spreading over directed
graphs, where each link specifies a direction in which infection can
pass. Moore et al.~(2003) {[}34{]} analyze what it takes to contain
self-propagating Internet code. Prakash et al.~(2010) {[}35{]} study
spread over networks whose connections change with time and evaluate
ways to protect nodes from infection. Sanatkar et al.~(2016) {[}42{]}
use the joint spectral radius to analyze epidemic thresholds on
switching networks. For undirected networks, they show that this
quantity equals the largest spectral radius of the individual regime
matrices, so a value below one gives extinction in their model. Their
analysis provides a mathematical precedent for the comparisons in
Sections 7 and 13.7, where successive changes in reproductive rules can
determine whether the population grows or declines. Related epidemic
models account for repeated introductions from outside reservoirs
{[}36{]} and identify reproduction thresholds when environmental
conditions repeat periodically {[}37{]}. For evolutionary AI control,
the specification also covers inherited changes to reproductive rules,
the population states covered by the assessment, conditional population
burden, and independently justified guarantees about human interests.

Repeated introductions can give initially declining lineages more
chances to produce a variant that persists. Antia et al.~(2003) {[}38{]}
use a multitype branching process, in which offspring and transmission
probabilities depend on type, to combine random introductions from a
pathogen reservoir with transmission chains that initially decline.
Inherited changes can produce a variant whose reproductive number
exceeds one. The model shows that raising transmission within the
subcritical range can increase the probability that such a variant
emerges. For an evolving AI population, the safety assessment must cover
the specified histories of the population and its reproductive rules.
Separate conditions bound population burden and establish the
consequences for people.

An assessment of interacting agents must account for failures that
spread between them. Gu et al.~(2024) {[}44{]} use Agent Smith to study
how a jailbreak, an input that bypasses a model's safeguards, can spread
among interacting language-model agents that handle multiple kinds of
input. They derive a suppression threshold and evaluate large simulated
agent populations. When descendants inherit and change the rules for
further reproduction, an evaluator must check whether the population
bound holds under every reachable rule and along every admitted sequence
of rules.

Section 9.4 compares the Frontier Safety Framework and Secure AI
Framework {[}16, 17{]} with the AI Control Roadmap, which explicitly
addresses deployed agent populations and their descendants {[}18{]}. The
framework developed here specifies the conditions for bounding
continuing production and population growth across interactions, regime
changes, and inherited changes to reproductive laws.

\subsection{15. Limitations and Claim
Boundaries}\label{limitations-and-claim-boundaries}

The framework covers prospective deployments in which artificial agents
reproduce, inherit traits, and undergo autonomous selection.
Contemporary AI systems generally exhibit only some of these features.
To establish that the framework applies to an interacting population,
show that agents create successors that inherit traits and that
inherited differences affect reproductive success. The counterexamples
identify conditions under which particular certificates fail. Estimating
the probability of those failures in a deployment requires measurements
of how agents reproduce and interact there. The biological comparisons
help identify which processes to measure.

\textbf{Ungoverned reproduction and escape.} The certificates cover the
reproductive transitions and environments in their declared models.
Leaked weights, independently operated forks, compromised
infrastructure, and descendants spreading beyond the original operator's
authority can introduce additional transitions or environments. The
accounting in Section 5.4 can include the initial unauthorized
production event. To bound later reproduction and its consequences for
people, specify the state the escaped descendants enter, their
opportunities to reproduce there, and how much each event adds to the
burden. For a deliberately released population, a certificate that
covers uncertainty about the initial reproductive law or the releaser's
responses must include those possibilities in the allowed population,
law, and environment histories.

\textbf{Practical verification.} To apply a certificate, gather evidence
for its assumptions. Depending on the claim, this means identifying the
opportunities agents have to reproduce and their conditional escape
rates, determining what descendants inherit and what can change,
bounding consequences when descendants interact, verifying restrictions
on switching, and establishing which population states the assessment
covers. Tests provide evidence about observed systems and histories.
Worst-case guarantees for unobserved descendants or histories of
arbitrary length also need support from formal restrictions,
conservative bounds on reachable states, monitored execution, or other
independently justified assumptions. Checking these premises at the
scale of real autonomous agents is an open verification problem.

An evaluator needs to know which population guarantees apply when
continuing production, switching, and interactions occur together. The
framework combines established results on source--sink branching
processes, instability under switching, and failures caused by
interactions into a safety specification for reproducing agents. To use
that specification, check which processes the deployment admits and
whether the certificate covers their combined effects.

The condition \(\mathcal R_{\mathcal E}(M)\subseteq\mathcal A(M)\)
requires the assessment to cover every reachable state. Computing that
set exactly can be too costly for sufficiently open-ended systems, and
the rules may leave it ill-defined. An evaluator can cover a set known
to contain all reachable states, prove that a set contains the initial
state and every allowed successor of its members, or restrict the
evolutionary transitions the system can make.

Reproductive fitness \(w\) measures a lineage's reproductive success,
while the human-interest criterion \(V\) measures the specified
consequences for people. Specifying those consequences is a separate
task. The relevant interests can involve several dimensions, be
uncertain, extend over time, or change as the system evolves.

The human-interest guarantees depend on their consequence assumptions.
Section 9.5 requires every state inside the certified reproductive
envelope to meet an independently specified human-interest condition.
Section 9.6 gives a sufficient condition for the source--sink model when
independent evidence supports \(\kappa\), the bound on loss per unit of
unsafe population, and \(g_t\), the bound on loss from permitted systems
and environmental changes at step \(t\). To guarantee safety along every
allowed trajectory, verify \(V_t-V_{t+1}\leq\kappa U_t+g_t\) at every
step of each trajectory. For the finite-horizon probability guarantee,
the abundance recurrence controls the average, and the consequence bound
must hold with probability one. A resulting upper bound of one leaves
the failure probability anywhere in the interval from zero to one. The
burden-based probability bound in Proposition 9 requires the stated
inclusion of failure events and the assumptions in Section 5.4 about
what the history records, what is known before each opportunity, and
which expectations are finite.

\subsection{16. Discussion}\label{discussion}

A population certificate needs to cover continuing production and how
lineages reproduce under interaction and switching. Each mechanism can
sustain or increase unsafe abundance.

An evaluator can use a reproductive-control certificate to assess
unauthorized production only if the certificate covers the routes agents
can use and the changes descendants can inherit. A burden calculation
starts with the opportunities to reproduce, the chance that each
produces an unauthorized descendant, and the burden each such descendant
adds given the preceding history. Once descendants interact, population
and switching certificates track how lineages change under successive
reproductive rules, and reachability requires the assessment to cover
the states those rules can produce. If descendants can inherit changes
to the rules themselves, verify that reproductive restrictions stay in
force and burden bounds continue to hold across every allowed change, as
specified in Section 9. Use the selection--transmission identity to
assess how differences in reproductive fitness and changes during
inheritance can shift the frequency of variants covered by the
certificate.

An evaluator can underestimate population growth by measuring only how
many agents are present at deployment or by testing each reproductive
regime on its own. In the nonlinear interaction model, the population
with the larger starting total can decline while the smaller one grows.
Each lineage helps the other reproduce, so the assessment needs the
starting abundance of each lineage. In the strictly positive switching
model at \(\epsilon=0.1\), 58 of the 256 enumerated eight-generation
schedules have period products with spectral radius greater than one.
Repeating any of those 58 schedules produces long-run growth from the
specified initial state, so the evaluator must account for the order in
which the regimes occur. Even with the number of reproductive
opportunities fixed, selection can increase the expected number of
unauthorized descendants. The comparison in Sections 12.1 and 13.6 uses
eight generations, ten opportunities per generation, initial
high-variant frequency \(f_0=0.1\), and leakage probabilities
\(p_L=0.01\) and \(p_H=0.51\). At relative fitness \(r=2\), cumulative
expected unauthorized production is approximately 4.64 times its value
under neutral reproduction (\(r=1\)). These figures apply to the
specified mathematical models.

To establish that reproductive control protects human interests, an
assessment needs evidence about how reproduction affects people. A
certified evolutionary envelope preserves the independently specified
human-interest condition along every allowed trajectory if every state
in the envelope meets that condition. When the consequence bounds hold
and the chosen burden detects the specified failures, a burden
certificate can bound their probability over a finite horizon.

A safeguard can change the conditions it must keep safe. By changing
which descendants reproduce, it changes the composition of the
population and can create new opportunities to evade the safeguard. The
environment shapes which inherited variants can interact and reproduce,
and the resulting population composition shapes the opportunities
available in the next generation.

\[
\begin{gathered}
\text{variation}
\rightarrow
\text{inheritance}
\rightarrow
\text{interaction}\\
\rightarrow
\text{realized reproduction}
\rightarrow
\text{population composition}\\
\rightarrow
\text{new evolutionary opportunity}
\end{gathered}
\]

The verification domain must therefore cover the populations that the
safeguard itself helps produce, including populations with new
opportunities for unauthorized reproduction.

The choice of safeguard determines which reproductive risks a safety
claim can cover. To support the claim, the safeguard must constrain the
opportunities, interactions, inherited changes, and trajectories named
in it. Designers can use cryptographic credentials to authorize
reproduction, restrict access to computing resources, record which
systems descend from which parents, or require particular hardware. They
can also isolate groups that reproduce separately, impose external rules
descendants cannot alter, or limit what offspring inherit.

A certificate for the authorized architecture supports a claim within
its scope, and agents deliberately released to replicate beyond their
developer's authority can bypass the restrictions that certificate
covers. To assess defenses imposed by hosts or networks, or through
limits on resource access or rules for reproductive authorization,
specify how each defense affects reproductive opportunities, new
production, interactions, and allowed evolutionary transitions. The
population and switching results address the trajectory under successive
defenses. The assessment must also cover reachable descendants that
selection favors because they can circumvent those defenses, along with
the reproductive rules those descendants can inherit. A human-interest
guarantee needs justified bounds on exposure and its consequences.

\textbf{Defense-dependent evolutionary effects.} A defense can favor
variants that leak more often, so its effect on unauthorized production
depends on how many opportunities to reproduce it leaves. Models of
treatment show how suppressing susceptible lineages can ease competition
for resistant descendants {[}45, 46{]}. In the two-variant setting of
Section 12.1, suppose a defense halves the reproduction of \(L\) while
\(H\) evades it, yielding a fitness ratio of \(w_H/w_L=2\). With one
opportunity per unit of population mass, the defended deterministic
model has \(9\cdot2^{-t}\) units of \(L\) and one unit of \(H\) at
generation \(t\). Over eight generations, expected unauthorized
production is \(\sum_{t=0}^{7}(0.09\cdot2^{-t}+0.51)\approx4.26\), below
the undefended value of \(4.8\). With ten opportunities per generation,
the same shift in relative fitness gives approximately \(22.26\)
expected events (Section 12.1). In the first calculation, the defense
leaves fewer opportunities as the susceptible population shrinks. In the
second, the number of opportunities stays fixed, so the shift toward
higher leakage raises expected unauthorized production. To compare
defenses, track both the expected burden and the population's average
leakage probability.

A defense can suppress reproduction more strongly in each regime on its
own and still allow faster growth when the regimes alternate. Replacing
two regimes equal to \(0.9I\) with the strictly positive matrices
\(A_{0.1}\) and \(B_{0.1}\) of Section 7.1 lowers each individual
spectral radius from \(0.9\) to approximately \(0.5472136\). The defense
also changes the product of successive matrices, and the per-step growth
factor under alternation rises from \(0.9\) to approximately
\(2.0052487\). The modeled population shifts from decline to growth
under alternation, and earlier work shows how dispersal can also induce
growth in time-periodic environments {[}47{]}.

In the nonnegative linear setting, weighted population stays at or below
its initial value across every admitted interleaving when a common
weighting contracts reproduction and bounds each defensive step.
Specifically, take \(v>0\) and \(0\leq q<1\) such that
\(v^{\mathsf T}M\leq qv^{\mathsf T}\) for every admissible reproductive
operator \(M\) and \(v^{\mathsf T}D\leq v^{\mathsf T}\) for every
admitted nonnegative defensive transformation \(D\). Each reproductive
step multiplies the upper bound on weighted mass by \(q\), and each
defensive step leaves that bound unchanged. After any admitted sequence
with \(n\) reproductive steps, the resulting population \(u\) satisfies
\(v^{\mathsf T}u\leq q^n v^{\mathsf T}u_0\), where \(u_0\) is the
initial population. The reproductive-step bound uses the common
weighting of Section 7.2 {[}43{]}, and the separate inequality for \(D\)
bounds each defensive step. To assess the effect of a defense, examine
total population burden, which inherited variants are present, and what
happens along complete allowed trajectories.

Training costs can constrain reproduction only to the extent that
creating a working successor incurs those costs. To assess a control
that relies on training costs, check what successors can already copy
and change. A successor can copy a trained model's weights and software
without repeating the original training. It can also inherit changes to
executable code, persistent working state, instructions, or other
components without retraining the model's parameters. When inherited
differences affect reproductive success, they can support Darwinian
evolution. Copying, execution, and inheritance can draw on different
resources because successors must obtain or transfer the components, run
them in a suitable environment, and secure further opportunities to
reproduce.

A certificate for a resource-based defense must identify which resources
constrain reproduction and account for how hardware, algorithms, and
descendant adaptations can change those constraints. Running the model
takes memory and computation, and creating working successors can depend
on storage, communication, compatible software, access permissions, and
opportunities to launch them. Assess the requirements for training a
model, running it to produce outputs, and creating working successors
separately. More efficient execution or compatibility with more devices
can expand the reachable population and weaken the defense. A durable
certificate must cover the allowed changes that matter over its
operating horizon, including those that ease a presumed computing
bottleneck. Quantifying those effects requires independent measurements
or bounds.

Comparing certificates on the same population states and with the same
measure of burden shows which risks each defense can bound. First
specify the inherited properties, external sources, and allowed
evolutionary transitions, then model how each intervention changes
reproduction under those conditions. Examine whether the evidence covers
individual regimes, contraction along complete trajectories, continuing
production, or inherited changes. Finite-horizon simulations can support
the assessment, and the stated bounds establish the certificate.

\subsection{17. Conclusion}\label{conclusion}

A deployment decision should specify which descendants the safeguards
can govern as agents create successors. Section 14 describes the
contained experiments on working-state inheritance {[}29{]}. When
inherited differences affect competition for resources and reproduction,
evaluators must track how agents produce and modify descendants, how
selection changes their frequencies, and how descendants spread across
generations.

Before deployment, test whether unsafe descendants can increase in
number or make up a growing share of the population. A permitted source
can supply enough new descendants for expected unsafe abundance to grow
without bound even as individual unsafe lineages die out. In the models,
two shrinking lineages can help each other reproduce, and two
suppressive regimes can sustain growth when they alternate. Selection
can also raise population-average leakage above its ceiling by making a
higher-risk variant more common while each variant's behavior stays
fixed. Corollary 1 shows what to check for an expected-burden guarantee:
contraction under every admitted reproductive rule, external input, the
number of reproductive opportunities, baseline burden, and the chance
and marginal burden of escape.

To assess a deadline-based safety claim, estimate the probability that
the AI population reaches the abundance threshold by the deadline. That
estimate depends on how likely each shock is in each generation and how
shocks depend on earlier ones. The two processes in Section 6.1 agree on
the per-generation probabilities but assign different likelihoods to
shock sequences, which changes the chance of crossing. For long-run
burden in a finite population, the relevant uncertainty is which variant
eventually takes over. In the two-individual example of Section 12.1,
the deterministic frequency forecast converges to a population entirely
of the higher-leakage variant. The finite population can also settle on
the lower-leakage variant, so its expected long-run burden rate depends
on the probability that the higher-leakage variant takes over.

A reproductive defense needs to control growth across the schedules it
admits and account for the variants it favors. In the switching model
with strictly positive matrices at \(\epsilon=0.1\), 58 of the 256
eight-generation schedules have period products with spectral radius
greater than one, so repeating any of those schedules produces long-run
growth from the specified initial state. In this model, a defense can
pass a reproduction test in every regime considered alone and still fail
to prevent growth when the regimes alternate. A common contraction bound
on reproduction and a uniform bound on external input together bound
abundance across all admitted schedules. In the integrated defense model
of Section 13.7, the continuing external source keeps the population
from going extinct even when the defense bounds its abundance. In the
fixed-opportunity comparison of Sections 12.1 and 13.6, each regime
starts at \(f_0=0.1\), with leakage probabilities \(p_L=0.01\) and
\(p_H=0.51\) and ten opportunities per generation. Over eight
generations, cumulative expected unauthorized production under selection
(\(r=2\)) is about 4.64 times its value under neutral reproduction
(\(r=1\)). Restrictions on copying, deployment, resource access, and
successor creation change the environment in which descendants compete
and can make variants that evade the defense more common. The defense's
certificate must cover the whole reproductive process, including the
opportunities and selection pressures the defense creates.

To use a certificate in a deployment decision, check that its
assumptions cover the states the population can enter, the transitions
it can make, and the histories that can unfold. Production-flux and
history-dependent burden certificates bound continuing unauthorized
production and its consequences. Common contraction and
constrained-switching certificates cover specified sequences of
reproductive regimes. An inductive invariant covers every reachable
population state by checking the initial state and each allowed
transition. The same reasoning preserves reproductive restrictions when
every inherited change keeps the reproductive rule inside a class that
satisfies those restrictions. To bound expected burden, Corollary 1 uses
a weighted count of authorized agents to control the expected number of
reproductive opportunities and a contraction bound to control how that
count changes over generations.

For an admission limit, specify which new agents may enter and how many
may enter in each generation. Use those rules to bound the external
input in the population recurrence and check that each admitted entry
preserves the invariant used to cover reachable states. For a spawn-rate
cap, establish that the enforced cap bounds reproductive opportunities
under every history covered by the certificate. Combine that bound with
the conditional probabilities of unauthorized production and the bounds
on its marginal consequences. If a sanction changes who can reproduce or
how long an agent remains active, specify when the sanction takes effect
and bound the resulting changes in reproduction or persistence. Check
the claimed contraction under the interactions and sequences of rules
the deployment permits, including inherited changes that affect
enforcement.

Specify enforceable admission and reproduction rules, then check the
certificate against the population histories those rules allow. Section
10 connects this check to prior work on restricting actions in specified
states {[}61{]}, formalizing entry and exit {[}65{]}, and evaluating
rules by their expected outcomes for specified values {[}50{]}.

To judge whether reproductive controls protect people, an evaluator
needs evidence about the consequences of the population changes those
controls allow. Reproductive success tells the evaluator which lineages
spread, and the independently specified human-interest criterion
measures what that spread means for people. The criterion holds along
every allowed trajectory if the controls keep the system inside a
certified set and every state in that set meets the criterion. Over a
finite horizon, the framework can bound the probability of failure when
the burden measure detects the specified failures and independent
evidence supports the consequence bounds.

For autonomous reproduction, a deployment decision can use a proof of
what the proposed controls guarantee together with evidence that the
deployment satisfies its assumptions. Developers can specify those
controls and establish the guarantee before deployment. Sections 3 and 4
show how reproductive controls evolved in multicellular organisms and
how cancer can arise despite controls on cell proliferation. Branching
processes, switched systems, and invariant sets give developers tools to
assess rules governing copying, inheritance, and access to resources. To
apply the proofs to open-ended systems, evaluators need evidence about
how agents reproduce, what descendants can change, and how the system
enforces its controls.

\subsection{Statements and
Declarations}\label{statements-and-declarations}

\textbf{Funding.} This work received no specific funding.

\textbf{Competing interests.} The author declares no competing
interests.

\textbf{Data and code availability.} The supplementary materials contain
the seven reproduction packages listed in Section 13.8 and a separate
\texttt{mutual\_reproduction} directory with code and data for the
individual-agent study in Section 13.2. The code is licensed under the
MIT License, and the data are licensed under Creative Commons
Attribution 4.0 International (CC BY 4.0).

\textbf{Ethics approval and consent to participate.} Not applicable.

\textbf{Author contributions.} Andrew T. Cadotte is the sole author.

\section{References}\label{references}

{[}1{]} Müller, V. C., Steels, L., and Szathmáry, E. (2026). ``Evolvable
AI: Threats of a New Major Transition in Evolution.'' \emph{Proceedings
of the National Academy of Sciences}, 123(17), e2527700123. DOI:
10.1073/pnas.2527700123.

{[}2{]} Hendrycks, D. (2023). ``Natural Selection Favors AIs over
Humans.'' arXiv:2303.16200v4 {[}cs.CY{]}, 18 July 2023. Preprint DOI:
\url{https://doi.org/10.48550/arxiv.2303.16200}.

{[}3{]} Rainey, P. B. (2023). ``Major evolutionary transitions in
individuality between humans and AI.'' \emph{Philosophical Transactions
of the Royal Society B}, 378, 20210408. DOI: 10.1098/rstb.2021.0408.

{[}4{]} Zhang, J., Hu, S., Lu, C., Lange, R., and Clune, J. (2026).
``Darwin Gödel Machine: Open-Ended Evolution of Self-Improving Agents.''
\emph{International Conference on Learning Representations (ICLR 2026)}.
Preprint: arXiv:2505.22954. Preprint DOI:
\url{https://doi.org/10.48550/arxiv.2505.22954}. ICLR record:
\url{https://openreview.net/pdf/ed3ddb9e9f1b6ad69626362874aa9d16e434528b.pdf}.

{[}5{]} Buss, L. W. (1987). \emph{The Evolution of Individuality}.
Princeton University Press. ISBN 0-691-08468-8 (hardcover),
0-691-08469-6 (paperback). Ebook DOI:
\url{https://doi.org/10.1515/9781400858712}.

{[}6{]} Michod, R. E., and Roze, D. (2001). ``Cooperation and conflict
in the evolution of multicellularity.'' \emph{Heredity}, 86, 1--7. DOI:
10.1046/j.1365-2540.2001.00808.x.

{[}7{]} Frank, S. A. (1995). ``Mutual policing and repression of
competition in the evolution of cooperative groups.'' \emph{Nature},
377, 520--522. DOI: 10.1038/377520a0.

{[}8{]} Burden, C. J., and Wei, Y.-B. C. (2018). ``Mutation in
populations governed by a Galton-Watson branching process.''
\emph{Theoretical Population Biology}, 120, 52--61. DOI:
10.1016/j.tpb.2017.12.001.

{[}9{]} Hyrien, O., Mitov, K. V., and Yanev, N. M. (2017). ``Subcritical
Sevastyanov branching processes with nonhomogeneous Poisson
immigration.'' \emph{Journal of Applied Probability}, 54(2), 569--587.
DOI: 10.1017/jpr.2017.18.

{[}10{]} Huang, Y., Jiang, Y., Wang, W., Zhuang, H., Luo, X., Ma, Y.,
Xu, Z., Chen, Z., Moniz, N., Lin, Z., Chen, P.-Y., Chawla, N. V., Dziri,
N., Sun, H., and Zhang, X. (2026). ``Emergent Risks in Generative
Multi-Agent Systems.'' arXiv:2603.27771v3 {[}cs.MA{]}, 10 September
2026. Preprint DOI:
\href{https://doi.org/10.48550/arxiv.2603.27771}{10.48550/arXiv.2603.27771}.

{[}11{]} Li, J., Zhang, Z., Zhou, S., Li, Y., Yu, T., and Wang, K.
(2026). ``When Safe Models Merge into Danger: Exploiting Latent
Vulnerabilities in LLM Fusion.'' In \emph{Proceedings of the
International Conference on Pattern Recognition (ICPR 2026)}, Part II,
pp.~575--589. DOI: 10.1007/978-3-032-31583-0\_38. Preprint:
arXiv:2604.00627v1 {[}cs.CR{]}, 1 April 2026. Preprint DOI:
\url{https://doi.org/10.48550/arxiv.2604.00627}.

{[}12{]} Sarkadi, Ş., and Lewis, P. R. (2024). ``The Triangles of
Dishonesty: Modelling the Evolution of Lies, Bullshit, and Deception in
Agent Societies.'' \emph{Proceedings of the 23rd International
Conference on Autonomous Agents and Multiagent Systems (AAMAS 2024)},
Auckland, New Zealand, 6--10 May 2024, pp. 1645--1653. DOI:
10.65109/slpd9249. AAMAS 2024 proceedings paper:
\url{https://aamas.csc.liv.ac.uk/Proceedings/aamas2024/pdfs/p1645.pdf}.

{[}13{]} Wang, J., Yang, S., Qiao, T., Qin, J., Li, J., Wang, Y., and
Zhao, Y. (2026). ``Ev-Trust: An Evolutionarily Stable Trust Mechanism
for Decentralized LLM-Based Multi-Agent Service Economies.''
arXiv:2512.16167v3 {[}cs.MA{]}, 31 May 2026. Preprint DOI:
\url{https://doi.org/10.48550/arxiv.2512.16167}.

{[}14{]} Han, S., Xiong, K., Liu, J., Ye, X., Su, Y., Duan, W., Liu, X.,
Xie, C., Bansal, M., Ding, M., Zhang, L., and Yao, H. (2026).
``Alignment Tipping Process: How Self-Evolution Pushes LLM Agents Off
the Rails.'' arXiv:2510.04860v2 {[}cs.LG{]}, 11 February 2026. Preprint
DOI: \url{https://doi.org/10.48550/arxiv.2510.04860}.

{[}15{]} Athanasopoulos, N., and Lazar, M. (2014). ``Stability analysis
of switched linear systems defined by graphs.'' \emph{53rd IEEE
Conference on Decision and Control}, pp.~5451--5456. DOI:
10.1109/CDC.2014.7040241.

{[}16{]} Google DeepMind (2026). \emph{Frontier Safety Framework}.
Version 3.1, 17 April 2026.\\
\url{https://deepmind.google/frontier-safety/}. Frontier Safety
Framework version 3.1 PDF:
\url{https://storage.googleapis.com/deepmind-media/DeepMind.com/Blog/strengthening-our-frontier-safety-framework/frontier-safety-framework_3-1.pdf}.

{[}17{]} Google (n.d.). \emph{Secure AI Framework Approach: A Quick
Guide to Implementing the Secure AI Framework (SAIF)}.
\url{https://safety.google/intl/en_us/safety/saif/}. Google Secure AI
Framework implementation guide PDF:
\url{https://services.google.com/fh/files/blogs/google_secure_ai_framework_approach.pdf}.

{[}18{]} Phuong, M., Jenner, E., Simon, L., Ho, L., Shah, R., Farquhar,
S., and Coull, S. (2026). \emph{GDM AI Control Roadmap}. Version 0.1,
Google DeepMind report dated 22 June 2026. arXiv posting dated 13 July
2026. \url{https://arxiv.org/abs/2607.13087}. Preprint DOI:
\url{https://doi.org/10.48550/arxiv.2607.13087}.

{[}19{]} Sheth, I., Wehner, J., Abdelnabi, S., Binkyte, R., and Fritz,
M. (2026). ``Position: Safety Must Precede the Deployment of Open-Ended
AI Agents.'' \emph{Proceedings of the 43rd International Conference on
Machine Learning (ICML 2026)}, PMLR 306. arXiv:2502.04512v4 {[}cs.AI{]},
1 June 2026. Preprint DOI:
\href{https://doi.org/10.48550/arxiv.2502.04512}{10.48550/arXiv.2502.04512}.

{[}20{]} Lin, R., Deng, X., Li, Q., Ma, J., Feng, Y., Qing, Y., Li, Z.,
Zhang, Y., Cui, S., Meng, C., Zhang, T., Ma, X., Li, Q., Xu, K., and Ji,
S. (2026). ``Safety in Self-Evolving LLM Agent Systems: Threats,
Amplification, and Case Studies.'' arXiv:2606.23075v1 {[}cs.CR{]}, 22
June 2026. Preprint DOI:
\href{https://doi.org/10.48550/arxiv.2606.23075}{10.48550/arXiv.2606.23075}.

{[}21{]} Shao, S., Ren, Q., Qian, C., Wei, B., Guo, D., Yang, J., Song,
X., Zhang, L., Zhang, W., Liu, D., and Shao, J. (2026). ``Your Agent May
Misevolve: Emergent Risks in Self-Evolving LLM Agents.''
\emph{International Conference on Learning Representations (ICLR 2026)}.
Preprint: arXiv:2509.26354v2 {[}cs.AI{]}, 8 March 2026. Preprint DOI:
\url{https://doi.org/10.48550/arxiv.2509.26354}.

{[}22{]} Yang, X., He, Y., Ji, S., Hooi, B., and Dong, J. S. (2026).
``Zombie Agents: Persistent Control of Self-Evolving LLM Agents via
Self-Reinforcing Injections.'' arXiv:2602.15654v2 {[}cs.CR{]}, 5 March
2026. Preprint DOI: \url{https://doi.org/10.48550/arxiv.2602.15654}.

{[}23{]} Mao, Q., Qu, W., Guo, D., Yuan, L., Liu, Q., Li, Y., Chen, G.,
Fu, Y., Lin, X., Hu, X., and Liu, D. (2026). ``SafeEvolve:
Harness-Policy Co-Evolution from Agent Experience for Safety
Alignment.'' arXiv:2609.02786v1 {[}cs.AI{]}, 2 September 2026. Preprint
DOI: \url{https://doi.org/10.48550/arxiv.2609.02786}.

{[}24{]} Ecoffet, A., Clune, J., and Lehman, J. (2020). ``Open Questions
in Creating Safe Open-ended AI: Tensions Between Control and
Creativity.'' \emph{ALIFE 2020}, pp. 27--35. DOI:
10.1162/isal\_a\_00323. Preprint:
\href{https://arxiv.org/abs/2006.07495v1}{arXiv:2006.07495v1}.

{[}25{]} Clune, J. (2019). ``AI-GAs: AI-generating algorithms, an
alternate paradigm for producing general artificial intelligence.''
Preprint: arXiv:1905.10985v2 {[}cs.AI{]}, 1 February 2020. Preprint DOI:
\href{https://doi.org/10.48550/arxiv.1905.10985}{10.48550/arXiv.1905.10985}.

{[}26{]} Gong, C., Bi, J., Yao, D., Liang, X., Xiang, C., and Guo, R.
(2026). ``Evolutionary Safety of Recursive Self-Improving AI: Taxonomy,
Risk Discovery, and Evaluation.'' arXiv:2609.31186v1 {[}cs.AI{]}, 25
September 2026. Preprint DOI:
\url{https://doi.org/10.48550/arxiv.2609.31186}.

{[}27{]} Goldsby, H. J., Knoester, D. B., Ofria, C., and Kerr, B.
(2014). ``The Evolutionary Origin of Somatic Cells under the Dirty Work
Hypothesis.'' \emph{PLOS Biology}, 12(5), e1001858. DOI:
10.1371/journal.pbio.1001858.

{[}28{]} Zhu, G., and Wang, C. (2026). ``Authorization for
Self-Modifying AI Agent Populations: Conserving Authority across
Replacement, Forking, and Rollback.'' Preprint:
\href{https://arxiv.org/abs/2610.00347}{arXiv:2610.00347}. SSRN DOI:
10.2139/ssrn.7559338.

{[}29{]} Anthropic (2026). ``AI models may be capable of
self-replicating without access to a datacenter.'' Published 12 October
2026. \url{https://www.anthropic.com/research/self-replicating-ai}.

{[}30{]} Lyons, B., Pio-Lopez, L., and Levin, M. (2026). ``From Cancer
to AI Alignment: Tackling Externalities Through Homeostatic
Principles.'' \emph{Preprints.org}, Version 1, posted 1 April 2026. DOI:
10.20944/preprints202604.0056.v1.

{[}31{]} Liberzon, D. (2003). \emph{Switching in Systems and Control}.
Systems \& Control: Foundations \& Applications. Birkhäuser. DOI:
10.1007/978-1-4612-0017-8.

{[}32{]} Hespanha, J. P., and Morse, A. S. (1999). ``Stability of
Switched Systems with Average Dwell-Time.'' \emph{Proceedings of the
38th IEEE Conference on Decision and Control}, pp.~2655--2660. DOI:
10.1109/CDC.1999.831330.

{[}33{]} Kephart, J. O., and White, S. R. (1991). ``Directed-Graph
Epidemiological Models of Computer Viruses.'' \emph{Proceedings of the
IEEE Computer Society Symposium on Research in Security and Privacy},
pp.~343--361. DOI: 10.1109/RISP.1991.130801.

{[}34{]} Moore, D., Shannon, C., Voelker, G. M., and Savage, S. (2003).
``Internet Quarantine: Requirements for Containing Self-Propagating
Code.'' \emph{IEEE INFOCOM 2003}, pp.~1901--1910. DOI:
\url{https://doi.org/10.1109/INFCOM.2003.1209212}.

{[}35{]} Prakash, B. A., Tong, H., Valler, N., Faloutsos, M., and
Faloutsos, C. (2010). ``Virus Propagation on Time-Varying Networks:
Theory and Immunization Algorithms.'' \emph{Machine Learning and
Knowledge Discovery in Databases, ECML PKDD 2010}, Lecture Notes in
Computer Science 6323, pp.~99--114. DOI: 10.1007/978-3-642-15939-8\_7.

{[}36{]} Lloyd-Smith, J. O., George, D., Pepin, K. M., Pitzer, V. E.,
Pulliam, J. R. C., Dobson, A. P., Hudson, P. J., and Grenfell, B. T.
(2009). ``Epidemic Dynamics at the Human-Animal Interface.''
\emph{Science}, 326(5958), 1362--1367. DOI: 10.1126/science.1177345.

{[}37{]} Bacaër, N., and Guernaoui, S. (2006). ``The Epidemic Threshold
of Vector-Borne Diseases with Seasonality: The Case of Cutaneous
Leishmaniasis in Chichaoua, Morocco.'' \emph{Journal of Mathematical
Biology}, 53, 421--436. DOI: 10.1007/s00285-006-0015-0.

{[}38{]} Antia, R., Regoes, R. R., Koella, J. C., and Bergstrom, C. T.
(2003). ``The Role of Evolution in the Emergence of Infectious
Diseases.'' \emph{Nature}, 426, 658--661. DOI: 10.1038/nature02104.

{[}39{]} Price, G. R. (1970). ``Selection and Covariance.''
\emph{Nature}, 227, 520--521. DOI: 10.1038/227520a0.

{[}40{]} Frank, S. A. (1997). ``The Price Equation, Fisher's Fundamental
Theorem, Kin Selection, and Causal Analysis.'' \emph{Evolution}, 51(6),
1712--1729. DOI: 10.1111/j.1558-5646.1997.tb05096.x.

{[}41{]} Serra, M. C. (2006). ``On the Waiting Time to Escape.''
\emph{Journal of Applied Probability}, 43(1), 296--302. DOI:
10.1239/jap/1143936262.

{[}42{]} Sanatkar, M. R., White, W. N., Natarajan, B., Scoglio, C., and
Garrett, K. A. (2016). ``Epidemic Threshold of an SIS Model in Dynamic
Switching Networks.'' \emph{IEEE Transactions on Systems, Man, and
Cybernetics: Systems}, 46(3), 345--355. DOI: 10.1109/TSMC.2015.2448061.
Preprint: \href{https://arxiv.org/abs/1501.02472}{arXiv:1501.02472}.

{[}43{]} Fornasini, E., and Valcher, M. E. (2012). ``Stability and
Stabilizability Criteria for Discrete-Time Positive Switched Systems.''
\emph{IEEE Transactions on Automatic Control}, 57(5), 1208--1221. DOI:
10.1109/TAC.2011.2173416.

\needspace{9\baselineskip}

{[}44{]} Gu, X., Zheng, X., Pang, T., Du, C., Liu, Q., Wang, Y., Jiang,
J., and Lin, M. (2024). ``Agent Smith: A Single Image Can Jailbreak One
Million Multimodal LLM Agents Exponentially Fast.'' In \emph{Proceedings
of the 41st International Conference on Machine Learning}, PMLR 235,
pp.~16647--16672.
\href{https://proceedings.mlr.press/v235/gu24e.html}{PMLR proceedings}.
\href{https://arxiv.org/abs/2402.08567}{arXiv:2402.08567}. Preprint DOI:
\url{https://doi.org/10.48550/arxiv.2402.08567}.

{[}45{]} Day, T., and Read, A. F. (2016). ``Does High-Dose Antimicrobial
Chemotherapy Prevent the Evolution of Resistance?'' \emph{PLOS
Computational Biology}, 12(1), e1004689. DOI:
10.1371/journal.pcbi.1004689.

{[}46{]} Viossat, Y., and Noble, R. (2021). ``A Theoretical Analysis of
Tumour Containment.'' \emph{Nature Ecology \& Evolution}, 5, 826--835.
DOI: 10.1038/s41559-021-01428-w.

{[}47{]} Katriel, G. (2022). ``Dispersal-Induced Growth in a
Time-Periodic Environment.'' \emph{Journal of Mathematical Biology}, 85,
Article 24. DOI: 10.1007/s00285-022-01791-7.

{[}48{]} Gordon, D. F. (2000). ``Asimovian Adaptive Agents.''
\emph{Journal of Artificial Intelligence Research}, 13, 95--153. DOI:
10.1613/jair.720.

{[}49{]} Morales, J., Wooldridge, M., Rodríguez-Aguilar, J. A., and
López-Sánchez, M. (2018). ``Off-line Synthesis of Evolutionarily Stable
Normative Systems.'' \emph{Autonomous Agents and Multi-Agent Systems},
32, 635--671. DOI: 10.1007/s10458-018-9390-3.

{[}50{]} Montes, N., and Sierra, C. (2022). ``Synthesis and Properties
of Optimally Value-Aligned Normative Systems.'' \emph{Journal of
Artificial Intelligence Research}, 74, 1739--1774. DOI:
10.1613/jair.1.13487.

{[}51{]} Liao, B., Pardo, P., Slavkovik, M., and van der Torre, L.
(2023). ``The Jiminy Advisor: Moral Agreements among Stakeholders Based
on Norms and Argumentation.'' \emph{Journal of Artificial Intelligence
Research}, 77, 737--792. DOI: 10.1613/jair.1.14368.

{[}52{]} Yudkowsky, E., and Herreshoff, M. (2013). \emph{Tiling Agents
for Self-Modifying AI, and the Löbian Obstacle}. Early draft, 7 October
2013. Machine Intelligence Research Institute.\\
\url{https://intelligence.org/files/TilingAgentsDraft.pdf}.

{[}53{]} Gumbau Mezquita, J. P. (2026). ``Algorithmic Unverifiability of
Safety for Fixed and Recursively Self-Improving Systems.''
arXiv:2606.28639v4, 14 October 2026. Preprint DOI:
\href{https://doi.org/10.48550/arxiv.2606.28639}{10.48550/arXiv.2606.28639}.

{[}54{]} Chen, J., Feng, Z., Ma, O., Yan, Y., Lin, R., Zhang, H., Du,
L., An, H., Yang, Y., Liu, J., Li, J., Xu, N., Zhou, C., Su, Y., Jin,
Z., Ma, Q., Qi, L., Wang, Y., Ma, Z., Pu, Y., Du, M., Song, Y., Huang,
E., Fu, Z., Wang, J., Li, J., Chen, Y., Xue, H., Li, Y., Du, T., and Ji,
S. (2026). ``Safety in Self-Evolving Agents: A Survey.''
arXiv:2610.00093v1. Preprint DOI:
\url{https://doi.org/10.48550/arxiv.2610.00093}.

{[}55{]} Wang, C., Li, C., Liu, S., Chen, Z., Hou, J., Qi, J., Li, R.,
Zhang, L., Ye, Q., Liu, Z., Chen, X., Zhang, X., and Yu, P. S. (2026).
``The Devil Behind Moltbook: Anthropic Safety Is Always Vanishing in
Self-Evolving AI Societies.'' arXiv:2602.09877v2. Preprint DOI:
\url{https://doi.org/10.48550/arxiv.2602.09877}.

{[}56{]} Alfonseca, M., Cebrian, M., Fernández Anta, A., Coviello, L.,
Abeliuk, A., and Rahwan, I. (2021). ``Superintelligence Cannot be
Contained: Lessons from Computability Theory.'' \emph{Journal of
Artificial Intelligence Research}, 70, 65--76. DOI:
10.1613/jair.1.12202.

{[}57{]} Han, T. A., Pereira, L. M., Santos, F. C., and Lenaerts, T.
(2020). ``To Regulate or Not: A Social Dynamics Analysis of an Idealised
AI Race.'' \emph{Journal of Artificial Intelligence Research}, 69,
881--921. DOI: 10.1613/jair.1.12225.

{[}58{]} Greenblatt, R., Shlegeris, B., Sachan, K., and Roger, F.
(2024). ``AI Control: Improving Safety Despite Intentional Subversion.''
In \emph{Proceedings of the 41st International Conference on Machine
Learning}, PMLR 235, pp.~16295--16336. arXiv:2312.06942. Preprint DOI:
\url{https://doi.org/10.48550/arxiv.2312.06942}.

{[}59{]} Hale, K. R. S., and Valdovinos, F. S. (2021). ``Ecological
theory of mutualism: Robust patterns of stability and thresholds in
two-species population models.'' \emph{Ecology and Evolution}, 11(24),
17651--17671. DOI: 10.1002/ece3.8453.

{[}60{]} Göttlich, S., and Knapp, S. (2020). ``Modeling random traffic
accidents by conservation laws.'' \emph{Mathematical Biosciences and
Engineering}, 17(2), 1677--1701. DOI: 10.3934/mbe.2020088.

{[}61{]} Shoham, Y., and Tennenholtz, M. (1995). ``On social laws for
artificial agent societies: off-line design.'' \emph{Artificial
Intelligence}, 73(1--2), 231--252. DOI: 10.1016/0004-3702(94)00007-N.

{[}62{]} Kouvaros, P., and Lomuscio, A. (2013). ``Automatic Verification
of Parameterised Interleaved Multi-Agent Systems.'' \emph{Proceedings of
the 12th International Conference on Autonomous Agents and Multiagent
Systems} (AAMAS 2013), Saint Paul, Minnesota, USA, pp.~861--868.
arXiv:1301.6431. Preprint DOI:
\href{https://doi.org/10.48550/arxiv.1301.6431}{10.48550/arXiv.1301.6431}.
AAMAS 2013 proceedings paper:
\url{https://www.ifaamas.org/Proceedings/aamas2013/docs/p861.pdf}.

{[}63{]} Tuyls, K., and Parsons, S. (2007). ``What evolutionary game
theory tells us about multiagent learning.'' \emph{Artificial
Intelligence}, 171(7), 406--416. DOI: 10.1016/j.artint.2007.01.004.

{[}64{]} Bloembergen, D., Tuyls, K., Hennes, D., and Kaisers, M. (2015).
``Evolutionary Dynamics of Multi-Agent Learning: A Survey.''
\emph{Journal of Artificial Intelligence Research}, 53, 659--697. DOI:
10.1613/jair.4818.

{[}65{]} Xu, H., Zhang, X., and Patel, R. J. (2007). ``Developing
Role-Based Open Multi-Agent Software Systems.'' \emph{International
Journal of Computational Intelligence Theory and Practice}, 2(1),
39--56. Author-hosted paper:
\url{https://www.cis.umassd.edu/~hxu/Papers/UMD/IJCITP-XU-Final.pdf}.

\end{document}